\documentclass[copyright,creativecommons]{eptcs}

\usepackage[latin1]{inputenc}

\usepackage[T1]{fontenc}

\usepackage[english]{babel}

\usepackage[font=small,labelfont=bf]{caption}

\usepackage{amsthm}

\usepackage{hyperref}

\usepackage{cite}

\usepackage[final]{microtype}

\hypersetup
  {
  pdfsubject    = {},
  pdfkeywords   = {},
  pdfproducer   = {},
  pdfcreator    = {},
  pdfpagemode   = {UseNone},
  pdfstartview  = {FitH}
  }

\usepackage{etoolbox}

\AtEndPreamble
  {

  \usepackage{fixltx2e}

  }

\usepackage{xargs}
\usepackage{xspace}
\usepackage{xstring}
\usepackage{boolexpr}

\usepackage[cmex10]{mathtools}
\usepackage{amsxtra}
\usepackage{amssymb}
\usepackage{amsfonts}
\usepackage{mathrsfs}
\usepackage{mathdots}
\usepackage{stmaryrd}
\usepackage{latexsym}
\usepackage{textcomp}
\usepackage{wasysym}
\usepackage{pifont}

\newcommand{\argemp}[2]
	{\if&#1&\else#2\fi}

\newcommand{\argdef}[2]
	{\if&#1&#2\else#1\fi}

\newcommand{\argint}[3]
	{\if&#2&\else#1#2#3\fi}

\newcommand{\argext}[3]
	{\if&#1&#3\else#1\if&#3&\else#2#3\fi\fi}

\newcommandx{\argsubsup}[3][2=, 3=]
	{\def\argsubscript{{#2}}\def\argsuperscript{{#3}}#1}

\newcommandx{\argind}[9][2=, 3=, 4=, 5=, 6=, 7=, 8=, 9=]
	{%
	\switch[#1=]%
		\case{0}#2%
		\case{1}#3%
		\case{2}#4%
		\case{3}#5%
		\case{4}#6%
		\case{5}#7%
		\case{6}#8%
		\case{7}#9%
		\otherwise\ensuremath{\clubsuit}%
	\endswitch%
	}

\newcommand{\arga}[1]
	{#1}
\newcommand{\argb}[2]
	{\argext{\arga{#1}}{, \allowbreak}{#2}}
\newcommand{\argc}[3]
	{\argext{\argb{#1}{#2}}{, \allowbreak}{#3}}
\newcommand{\argd}[4]
	{\argext{\argc{#1}{#2}{#3}}{, \allowbreak}{#4}}
\newcommand{\arge}[5]
	{\argext{\argd{#1}{#2}{#3}{#4}}{, \allowbreak}{#5}}
\newcommand{\argf}[6]
	{\argext{\arge{#1}{#2}{#3}{#4}{#5}}{, \allowbreak}{#6}}
\newcommand{\argg}[7]
	{\argext{\argf{#1}{#2}{#3}{#4}{#5}{#6}}{, \allowbreak}{#7}}
\newcommand{\argh}[8]
	{\argext{\argg{#1}{#2}{#3}{#4}{#5}{#6}{#7}}{, \allowbreak}{#8}}
\newcommand{\argi}[9]
	{\argext{\argh{#1}{#2}{#3}{#4}{#5}{#6}{#7}{#8}}{, \allowbreak}{#9}}

\newcommand{\argj}[9]
	{%
	\def\valarga{#1}%
	\def\valargb{#2}%
	\def\valargc{#3}%
	\def\valargd{#4}%
	\def\valarge{#5}%
	\def\valargf{#6}%
	\def\valargg{#7}%
	\def\valargh{#8}%
	\def\valargi{#9}%
	\argauxj%
	}
\newcommand{\argk}[9]
	{%
	\def\valarga{#1}%
	\def\valargb{#2}%
	\def\valargc{#3}%
	\def\valargd{#4}%
	\def\valarge{#5}%
	\def\valargf{#6}%
	\def\valargg{#7}%
	\def\valargh{#8}%
	\def\valargi{#9}%
	\argauxk%
	}
\newcommand{\argl}[9]
	{%
	\def\valarga{#1}%
	\def\valargb{#2}%
	\def\valargc{#3}%
	\def\valargd{#4}%
	\def\valarge{#5}%
	\def\valargf{#6}%
	\def\valargg{#7}%
	\def\valargh{#8}%
	\def\valargi{#9}%
	\argauxl%
	}
\newcommand{\argm}[9]
	{%
	\def\valarga{#1}%
	\def\valargb{#2}%
	\def\valargc{#3}%
	\def\valargd{#4}%
	\def\valarge{#5}%
	\def\valargf{#6}%
	\def\valargg{#7}%
	\def\valargh{#8}%
	\def\valargi{#9}%
	\argauxm%
	}
\newcommand{\argn}[9]
	{%
	\def\valarga{#1}%
	\def\valargb{#2}%
	\def\valargc{#3}%
	\def\valargd{#4}%
	\def\valarge{#5}%
	\def\valargf{#6}%
	\def\valargg{#7}%
	\def\valargh{#8}%
	\def\valargi{#9}%
	\argauxn%
	}
\newcommand{\argo}[9]
	{%
	\def\valarga{#1}%
	\def\valargb{#2}%
	\def\valargc{#3}%
	\def\valargd{#4}%
	\def\valarge{#5}%
	\def\valargf{#6}%
	\def\valargg{#7}%
	\def\valargh{#8}%
	\def\valargi{#9}%
	\argauxo%
	}
\newcommand{\argp}[9]
	{%
	\def\valarga{#1}%
	\def\valargb{#2}%
	\def\valargc{#3}%
	\def\valargd{#4}%
	\def\valarge{#5}%
	\def\valargf{#6}%
	\def\valargg{#7}%
	\def\valargh{#8}%
	\def\valargi{#9}%
	\argauxp%
	}
\newcommand{\argq}[9]
	{%
	\def\valarga{#1}%
	\def\valargb{#2}%
	\def\valargc{#3}%
	\def\valargd{#4}%
	\def\valarge{#5}%
	\def\valargf{#6}%
	\def\valargg{#7}%
	\def\valargh{#8}%
	\def\valargi{#9}%
	\argauxq%
	}
\newcommand{\argr}[9]
	{%
	\def\valarga{#1}%
	\def\valargb{#2}%
	\def\valargc{#3}%
	\def\valargd{#4}%
	\def\valarge{#5}%
	\def\valargf{#6}%
	\def\valargg{#7}%
	\def\valargh{#8}%
	\def\valargi{#9}%
	\argauxr%
	}

\newcommand{\argauxj}[1]
	{%
	\argext%
		{\argi{\valarga}{\valargb}{\valargc}{\valargd}{\valarge}{\valargf}{\valargg}
			{\valargh}{\valargi}}
		{, \allowbreak}{#1}%
	}
\newcommand{\argauxk}[2]
	{\argext{\argauxj{#1}}{, \allowbreak}{#2}}
\newcommand{\argauxl}[3]
	{\argext{\argauxk{#1}{#2}}{, \allowbreak}{#3}}
\newcommand{\argauxm}[4]
	{\argext{\argauxl{#1}{#2}{#3}}{, \allowbreak}{#4}}
\newcommand{\argauxn}[5]
	{\argext{\argauxm{#1}{#2}{#3}{#4}}{, \allowbreak}{#5}}
\newcommand{\argauxo}[6]
	{\argext{\argauxn{#1}{#2}{#3}{#4}{#5}}{, \allowbreak}{#6}}
\newcommand{\argauxp}[7]
	{\argext{\argauxo{#1}{#2}{#3}{#4}{#5}{#6}}{, \allowbreak}{#7}}
\newcommand{\argauxq}[8]
	{\argext{\argauxp{#1}{#2}{#3}{#4}{#5}{#6}{#7}}{, \allowbreak}{#8}}
\newcommand{\argauxr}[9]
	{\argext{\argauxq{#1}{#2}{#3}{#4}{#5}{#6}{#7}{#8}}{, \allowbreak}{#9}}

\newcommand{\txtfnt}[2][]
	{{%
	\IfStrEq{#1}{}
		{#2}
		{%
		\StrLeft{#1}{2}[\optbgn]%
		\StrGobbleLeft{#1}{2}[\optend]%
		\IfStrEqCase{\optbgn}
			{%
			{Rm}{\rmfamily\txtfnt[\optend]{#2}}%
			{Sf}{\sffamily\txtfnt[\optend]{#2}}%
			{Tt}{\ttfamily\txtfnt[\optend]{#2}}%
			{Up}{\upshape\txtfnt[\optend]{#2}}%
			{It}{\itshape\txtfnt[\optend]{#2}}%
			{Sl}{\slshape\txtfnt[\optend]{#2}}%
			{Sc}{\scshape\txtfnt[\optend]{#2}}%
			{Md}{\mdseries\txtfnt[\optend]{#2}}%
			{Bf}{\bfseries\txtfnt[\optend]{#2}}%
			{Em}{\emph{\txtfnt[\optend]{#2}}}%
			}
			[\ensuremath{\clubsuit}]%
		}%
	}}

\newcommand{\txtsub}[2][]
	{\argemp{#2}{\ensuremath{_{\text{\txtfnt[#1]{#2}}}}}}

\newcommand{\txtsup}[2][]
	{\argemp{#2}{\ensuremath{^{\text{\txtfnt[#1]{#2}}}}}}

\newcommandx{\txt}[4][1=, 3=, 4=]
	{\text{\txtfnt[#1]{#2}\ensuremath{\txtsub[#1]{#3}\txtsup[#1]{#4}}}}

\newcommandx{\txtarg}[5][1=, 3=, 4=]
	{{\txt[#1]{#2}[#3][#4]\argint{(}{#5}{)}}}

\newcommand{\txtstyname}{RmScMd}
\newcommand{\txtname}[1][]
	{\txt[\argdef{#1}{\txtstyname}]}
\newcommand{\txtargname}[1][]
	{\txtarg[\argdef{#1}{\txtstyname}]}

\newcommand{\txtstyabr}{Em}
\newcommand{\txtabr}[1][]
	{\txt[\argdef{#1}{\txtstyabr}]}

\newcommandx{\mthfnt}[3][1=, 2=0]
	{{%
	\IfStrEqCase{#1}
		{%
		{}%
			{#3}%
		{Name}%
			{%
			\IfStrEqCase{#2}
				{%
				{0}{\mathcal{#3}}%
				{1}{\mathscr{#3}}%
				{2}{\mathfrak{#3}}%
				{3}{\mathbf{#3}}%
				}
				[\ensuremath{\clubsuit}]%
			}%
		{Set}%
			{%
			\IfStrEqCase{#2}
				{%
				{0}{\mathrm{#3}}%
				{1}{\mathsf{#3}}%
				{2}{\mathbb{#3}}%
				{3}{\mathtt{#3}}%
				}
				[\ensuremath{\clubsuit}]%
			}%
		{Fun}%
			{%
			\IfStrEqCase{#2}
				{%
				{0}{\mathsf{#3}}%
				{1}{\mathrm{#3}}%
				}
				[\ensuremath{\clubsuit}]%
			}%
		{Rel}%
			{%
			\IfStrEqCase{#2}
				{%
				{0}{\mathit{#3}}%
				{1}{\mathtt{#3}}%
				}
				[\ensuremath{\clubsuit}]%
			}%
		{Sym}%
			{%
			\IfStrEqCase{#2}
				{%
				{0}{\mathtt{#3}}%
				{1}{\mathbf{#3}}%
				}
				[\ensuremath{\clubsuit}]%
			}%
		{Elm}%
			{\mathnormal{#3}}
		}
		[\ensuremath{\clubsuit}]%
	}}

\newcommand{\mthsub}[1]
	{\argemp{#1}{\ensuremath{_{\mathnormal{#1}}}}}

\newcommand{\mthsup}[1]
	{\argemp{#1}{\ensuremath{^{\mathnormal{#1}}}}}

\newcommandx{\mth}[5][1=, 2=0, 4=, 5=]
	{{\ensuremath{\mthfnt[#1][#2]{#3}\mthsub{#4}\mthsup{#5}}}}

\newcommandx{\mtharg}[6][1=, 2=0, 4=, 5=]
	{{\mth[#1][#2]{#3}[#4][#5]\ensuremath{\argint{\!\left(}{#6}{\right)}}}}

\newcommand{\mthempty}
	{\mth[][]}

\newcommand{\mthstyname}{0}
\newcommand{\mthname}[1][]
	{\mth[Name][\argdef{#1}{\mthstyname}]}

\newcommand{\mthstyset}{0}
\newcommand{\mthset}[1][]
	{\mth[Set][\argdef{#1}{\mthstyset}]}
\newcommand{\mthargset}[1][]
	{\mtharg[Set][\argdef{#1}{\mthstyset}]}

\newcommand{\mthstyfun}{0}
\newcommand{\mthfun}[1][]
	{\mth[Fun][\argdef{#1}{\mthstyfun}]}
\newcommand{\mthargfun}[1][]
	{\mtharg[Fun][\argdef{#1}{\mthstyfun}]}

\newcommand{\mthstyrel}{0}
\newcommand{\mthrel}[1][]
	{\mth[Rel][\argdef{#1}{\mthstyrel}]}

\newcommand{\mthstysym}{0}
\newcommand{\mthsym}[1][]
	{\mth[Sym][\argdef{#1}{\mthstysym}]}

\newcommand{\mthstyelm}{0}
\newcommand{\mthelm}[1][]
	{\mth[Elm][\argdef{#1}{\mthstyelm}]}

\newcommandx{\AName}[4][1=, 2=, 3=, 4=]{\mthname[#4]{A#3}[#1][#2]}
\newcommandx{\BName}[4][1=, 2=, 3=, 4=]{\mthname[#4]{B#3}[#1][#2]}
\newcommandx{\CName}[4][1=, 2=, 3=, 4=]{\mthname[#4]{C#3}[#1][#2]}
\newcommandx{\DName}[4][1=, 2=, 3=, 4=]{\mthname[#4]{D#3}[#1][#2]}
\newcommandx{\EName}[4][1=, 2=, 3=, 4=]{\mthname[#4]{E#3}[#1][#2]}
\newcommandx{\FName}[4][1=, 2=, 3=, 4=]{\mthname[#4]{F#3}[#1][#2]}
\newcommandx{\GName}[4][1=, 2=, 3=, 4=]{\mthname[#4]{G#3}[#1][#2]}
\newcommandx{\HName}[4][1=, 2=, 3=, 4=]{\mthname[#4]{H#3}[#1][#2]}
\newcommandx{\IName}[4][1=, 2=, 3=, 4=]{\mthname[#4]{I#3}[#1][#2]}
\newcommandx{\JName}[4][1=, 2=, 3=, 4=]{\mthname[#4]{J#3}[#1][#2]}
\newcommandx{\KName}[4][1=, 2=, 3=, 4=]{\mthname[#4]{K#3}[#1][#2]}
\newcommandx{\LName}[4][1=, 2=, 3=, 4=]{\mthname[#4]{L#3}[#1][#2]}
\newcommandx{\MName}[4][1=, 2=, 3=, 4=]{\mthname[#4]{M#3}[#1][#2]}
\newcommandx{\NName}[4][1=, 2=, 3=, 4=]{\mthname[#4]{N#3}[#1][#2]}
\newcommandx{\OName}[4][1=, 2=, 3=, 4=]{\mthname[#4]{O#3}[#1][#2]}
\newcommandx{\PName}[4][1=, 2=, 3=, 4=]{\mthname[#4]{P#3}[#1][#2]}
\newcommandx{\QName}[4][1=, 2=, 3=, 4=]{\mthname[#4]{Q#3}[#1][#2]}
\newcommandx{\RName}[4][1=, 2=, 3=, 4=]{\mthname[#4]{R#3}[#1][#2]}
\newcommandx{\SName}[4][1=, 2=, 3=, 4=]{\mthname[#4]{S#3}[#1][#2]}
\newcommandx{\TName}[4][1=, 2=, 3=, 4=]{\mthname[#4]{T#3}[#1][#2]}
\newcommandx{\UName}[4][1=, 2=, 3=, 4=]{\mthname[#4]{U#3}[#1][#2]}
\newcommandx{\VName}[4][1=, 2=, 3=, 4=]{\mthname[#4]{V#3}[#1][#2]}
\newcommandx{\WName}[4][1=, 2=, 3=, 4=]{\mthname[#4]{W#3}[#1][#2]}
\newcommandx{\XName}[4][1=, 2=, 3=, 4=]{\mthname[#4]{X#3}[#1][#2]}
\newcommandx{\YName}[4][1=, 2=, 3=, 4=]{\mthname[#4]{Y#3}[#1][#2]}
\newcommandx{\ZName}[4][1=, 2=, 3=, 4=]{\mthname[#4]{Z#3}[#1][#2]}

\newcommandx{\ASet}[4][1=, 2=, 3=, 4=]{\mthset[#4]{A#3}[#1][#2]}
\newcommandx{\BSet}[4][1=, 2=, 3=, 4=]{\mthset[#4]{B#3}[#1][#2]}
\newcommandx{\CSet}[4][1=, 2=, 3=, 4=]{\mthset[#4]{C#3}[#1][#2]}
\newcommandx{\DSet}[4][1=, 2=, 3=, 4=]{\mthset[#4]{D#3}[#1][#2]}
\newcommandx{\ESet}[4][1=, 2=, 3=, 4=]{\mthset[#4]{E#3}[#1][#2]}
\newcommandx{\FSet}[4][1=, 2=, 3=, 4=]{\mthset[#4]{F#3}[#1][#2]}
\newcommandx{\GSet}[4][1=, 2=, 3=, 4=]{\mthset[#4]{G#3}[#1][#2]}
\newcommandx{\HSet}[4][1=, 2=, 3=, 4=]{\mthset[#4]{H#3}[#1][#2]}
\newcommandx{\ISet}[4][1=, 2=, 3=, 4=]{\mthset[#4]{I#3}[#1][#2]}
\newcommandx{\JSet}[4][1=, 2=, 3=, 4=]{\mthset[#4]{J#3}[#1][#2]}
\newcommandx{\KSet}[4][1=, 2=, 3=, 4=]{\mthset[#4]{K#3}[#1][#2]}
\newcommandx{\LSet}[4][1=, 2=, 3=, 4=]{\mthset[#4]{L#3}[#1][#2]}
\newcommandx{\MSet}[4][1=, 2=, 3=, 4=]{\mthset[#4]{M#3}[#1][#2]}
\newcommandx{\NSet}[4][1=, 2=, 3=, 4=]{\mthset[#4]{N#3}[#1][#2]}
\newcommandx{\OSet}[4][1=, 2=, 3=, 4=]{\mthset[#4]{O#3}[#1][#2]}
\newcommandx{\PSet}[4][1=, 2=, 3=, 4=]{\mthset[#4]{P#3}[#1][#2]}
\newcommandx{\QSet}[4][1=, 2=, 3=, 4=]{\mthset[#4]{Q#3}[#1][#2]}
\newcommandx{\RSet}[4][1=, 2=, 3=, 4=]{\mthset[#4]{R#3}[#1][#2]}
\newcommandx{\SSet}[4][1=, 2=, 3=, 4=]{\mthset[#4]{S#3}[#1][#2]}
\newcommandx{\TSet}[4][1=, 2=, 3=, 4=]{\mthset[#4]{T#3}[#1][#2]}
\newcommandx{\USet}[4][1=, 2=, 3=, 4=]{\mthset[#4]{U#3}[#1][#2]}
\newcommandx{\VSet}[4][1=, 2=, 3=, 4=]{\mthset[#4]{V#3}[#1][#2]}
\newcommandx{\WSet}[4][1=, 2=, 3=, 4=]{\mthset[#4]{W#3}[#1][#2]}
\newcommandx{\XSet}[4][1=, 2=, 3=, 4=]{\mthset[#4]{X#3}[#1][#2]}
\newcommandx{\YSet}[4][1=, 2=, 3=, 4=]{\mthset[#4]{Y#3}[#1][#2]}
\newcommandx{\ZSet}[4][1=, 2=, 3=, 4=]{\mthset[#4]{Z#3}[#1][#2]}

\newcommandx{\aSet}[4][1=, 2=, 3=, 4=]{\mthset[#4]{a#3}[#1][#2]}
\newcommandx{\bSet}[4][1=, 2=, 3=, 4=]{\mthset[#4]{b#3}[#1][#2]}
\newcommandx{\cSet}[4][1=, 2=, 3=, 4=]{\mthset[#4]{c#3}[#1][#2]}
\newcommandx{\dSet}[4][1=, 2=, 3=, 4=]{\mthset[#4]{d#3}[#1][#2]}
\newcommandx{\eSet}[4][1=, 2=, 3=, 4=]{\mthset[#4]{e#3}[#1][#2]}
\newcommandx{\fSet}[4][1=, 2=, 3=, 4=]{\mthset[#4]{f#3}[#1][#2]}
\newcommandx{\gSet}[4][1=, 2=, 3=, 4=]{\mthset[#4]{g#3}[#1][#2]}
\newcommandx{\hSet}[4][1=, 2=, 3=, 4=]{\mthset[#4]{h#3}[#1][#2]}
\newcommandx{\iSet}[4][1=, 2=, 3=, 4=]{\mthset[#4]{i#3}[#1][#2]}
\newcommandx{\jSet}[4][1=, 2=, 3=, 4=]{\mthset[#4]{j#3}[#1][#2]}
\newcommandx{\kSet}[4][1=, 2=, 3=, 4=]{\mthset[#4]{k#3}[#1][#2]}
\newcommandx{\lSet}[4][1=, 2=, 3=, 4=]{\mthset[#4]{l#3}[#1][#2]}
\newcommandx{\mSet}[4][1=, 2=, 3=, 4=]{\mthset[#4]{m#3}[#1][#2]}
\newcommandx{\nSet}[4][1=, 2=, 3=, 4=]{\mthset[#4]{n#3}[#1][#2]}
\newcommandx{\oSet}[4][1=, 2=, 3=, 4=]{\mthset[#4]{o#3}[#1][#2]}
\newcommandx{\pSet}[4][1=, 2=, 3=, 4=]{\mthset[#4]{p#3}[#1][#2]}
\newcommandx{\qSet}[4][1=, 2=, 3=, 4=]{\mthset[#4]{q#3}[#1][#2]}
\newcommandx{\rSet}[4][1=, 2=, 3=, 4=]{\mthset[#4]{r#3}[#1][#2]}
\newcommandx{\sSet}[4][1=, 2=, 3=, 4=]{\mthset[#4]{s#3}[#1][#2]}
\newcommandx{\tSet}[4][1=, 2=, 3=, 4=]{\mthset[#4]{t#3}[#1][#2]}
\newcommandx{\uSet}[4][1=, 2=, 3=, 4=]{\mthset[#4]{u#3}[#1][#2]}
\newcommandx{\vSet}[4][1=, 2=, 3=, 4=]{\mthset[#4]{v#3}[#1][#2]}
\newcommandx{\wSet}[4][1=, 2=, 3=, 4=]{\mthset[#4]{w#3}[#1][#2]}
\newcommandx{\xSet}[4][1=, 2=, 3=, 4=]{\mthset[#4]{x#3}[#1][#2]}
\newcommandx{\ySet}[4][1=, 2=, 3=, 4=]{\mthset[#4]{y#3}[#1][#2]}
\newcommandx{\zSet}[4][1=, 2=, 3=, 4=]{\mthset[#4]{z#3}[#1][#2]}

\newcommandx{\AFun}[4][1=, 2=, 3=, 4=]{\mthfun[#4]{A#3}[#1][#2]}
\newcommandx{\BFun}[4][1=, 2=, 3=, 4=]{\mthfun[#4]{B#3}[#1][#2]}
\newcommandx{\CFun}[4][1=, 2=, 3=, 4=]{\mthfun[#4]{C#3}[#1][#2]}
\newcommandx{\DFun}[4][1=, 2=, 3=, 4=]{\mthfun[#4]{D#3}[#1][#2]}
\newcommandx{\EFun}[4][1=, 2=, 3=, 4=]{\mthfun[#4]{E#3}[#1][#2]}
\newcommandx{\FFun}[4][1=, 2=, 3=, 4=]{\mthfun[#4]{F#3}[#1][#2]}
\newcommandx{\GFun}[4][1=, 2=, 3=, 4=]{\mthfun[#4]{G#3}[#1][#2]}
\newcommandx{\HFun}[4][1=, 2=, 3=, 4=]{\mthfun[#4]{H#3}[#1][#2]}
\newcommandx{\IFun}[4][1=, 2=, 3=, 4=]{\mthfun[#4]{I#3}[#1][#2]}
\newcommandx{\JFun}[4][1=, 2=, 3=, 4=]{\mthfun[#4]{J#3}[#1][#2]}
\newcommandx{\KFun}[4][1=, 2=, 3=, 4=]{\mthfun[#4]{K#3}[#1][#2]}
\newcommandx{\LFun}[4][1=, 2=, 3=, 4=]{\mthfun[#4]{L#3}[#1][#2]}
\newcommandx{\MFun}[4][1=, 2=, 3=, 4=]{\mthfun[#4]{M#3}[#1][#2]}
\newcommandx{\NFun}[4][1=, 2=, 3=, 4=]{\mthfun[#4]{N#3}[#1][#2]}
\newcommandx{\OFun}[4][1=, 2=, 3=, 4=]{\mthfun[#4]{O#3}[#1][#2]}
\newcommandx{\PFun}[4][1=, 2=, 3=, 4=]{\mthfun[#4]{P#3}[#1][#2]}
\newcommandx{\QFun}[4][1=, 2=, 3=, 4=]{\mthfun[#4]{Q#3}[#1][#2]}
\newcommandx{\RFun}[4][1=, 2=, 3=, 4=]{\mthfun[#4]{R#3}[#1][#2]}
\newcommandx{\SFun}[4][1=, 2=, 3=, 4=]{\mthfun[#4]{S#3}[#1][#2]}
\newcommandx{\TFun}[4][1=, 2=, 3=, 4=]{\mthfun[#4]{T#3}[#1][#2]}
\newcommandx{\UFun}[4][1=, 2=, 3=, 4=]{\mthfun[#4]{U#3}[#1][#2]}
\newcommandx{\VFun}[4][1=, 2=, 3=, 4=]{\mthfun[#4]{V#3}[#1][#2]}
\newcommandx{\WFun}[4][1=, 2=, 3=, 4=]{\mthfun[#4]{W#3}[#1][#2]}
\newcommandx{\XFun}[4][1=, 2=, 3=, 4=]{\mthfun[#4]{X#3}[#1][#2]}
\newcommandx{\YFun}[4][1=, 2=, 3=, 4=]{\mthfun[#4]{Y#3}[#1][#2]}
\newcommandx{\ZFun}[4][1=, 2=, 3=, 4=]{\mthfun[#4]{Z#3}[#1][#2]}

\newcommandx{\aFun}[4][1=, 2=, 3=, 4=]{\mthfun[#4]{a#3}[#1][#2]}
\newcommandx{\bFun}[4][1=, 2=, 3=, 4=]{\mthfun[#4]{b#3}[#1][#2]}
\newcommandx{\cFun}[4][1=, 2=, 3=, 4=]{\mthfun[#4]{c#3}[#1][#2]}
\newcommandx{\dFun}[4][1=, 2=, 3=, 4=]{\mthfun[#4]{d#3}[#1][#2]}
\newcommandx{\eFun}[4][1=, 2=, 3=, 4=]{\mthfun[#4]{e#3}[#1][#2]}
\newcommandx{\fFun}[4][1=, 2=, 3=, 4=]{\mthfun[#4]{f#3}[#1][#2]}
\newcommandx{\gFun}[4][1=, 2=, 3=, 4=]{\mthfun[#4]{g#3}[#1][#2]}
\newcommandx{\hFun}[4][1=, 2=, 3=, 4=]{\mthfun[#4]{h#3}[#1][#2]}
\newcommandx{\iFun}[4][1=, 2=, 3=, 4=]{\mthfun[#4]{i#3}[#1][#2]}
\newcommandx{\jFun}[4][1=, 2=, 3=, 4=]{\mthfun[#4]{j#3}[#1][#2]}
\newcommandx{\kFun}[4][1=, 2=, 3=, 4=]{\mthfun[#4]{k#3}[#1][#2]}
\newcommandx{\lFun}[4][1=, 2=, 3=, 4=]{\mthfun[#4]{l#3}[#1][#2]}
\newcommandx{\mFun}[4][1=, 2=, 3=, 4=]{\mthfun[#4]{m#3}[#1][#2]}
\newcommandx{\nFun}[4][1=, 2=, 3=, 4=]{\mthfun[#4]{n#3}[#1][#2]}
\newcommandx{\oFun}[4][1=, 2=, 3=, 4=]{\mthfun[#4]{o#3}[#1][#2]}
\newcommandx{\pFun}[4][1=, 2=, 3=, 4=]{\mthfun[#4]{p#3}[#1][#2]}
\newcommandx{\qFun}[4][1=, 2=, 3=, 4=]{\mthfun[#4]{q#3}[#1][#2]}
\newcommandx{\rFun}[4][1=, 2=, 3=, 4=]{\mthfun[#4]{r#3}[#1][#2]}
\newcommandx{\sFun}[4][1=, 2=, 3=, 4=]{\mthfun[#4]{s#3}[#1][#2]}
\newcommandx{\tFun}[4][1=, 2=, 3=, 4=]{\mthfun[#4]{t#3}[#1][#2]}
\newcommandx{\uFun}[4][1=, 2=, 3=, 4=]{\mthfun[#4]{u#3}[#1][#2]}
\newcommandx{\vFun}[4][1=, 2=, 3=, 4=]{\mthfun[#4]{v#3}[#1][#2]}
\newcommandx{\wFun}[4][1=, 2=, 3=, 4=]{\mthfun[#4]{w#3}[#1][#2]}
\newcommandx{\xFun}[4][1=, 2=, 3=, 4=]{\mthfun[#4]{x#3}[#1][#2]}
\newcommandx{\yFun}[4][1=, 2=, 3=, 4=]{\mthfun[#4]{y#3}[#1][#2]}
\newcommandx{\zFun}[4][1=, 2=, 3=, 4=]{\mthfun[#4]{z#3}[#1][#2]}

\newcommandx{\ARel}[4][1=, 2=, 3=, 4=]{\mthrel[#4]{A#3}[#1][#2]}
\newcommandx{\BRel}[4][1=, 2=, 3=, 4=]{\mthrel[#4]{B#3}[#1][#2]}
\newcommandx{\CRel}[4][1=, 2=, 3=, 4=]{\mthrel[#4]{C#3}[#1][#2]}
\newcommandx{\DRel}[4][1=, 2=, 3=, 4=]{\mthrel[#4]{D#3}[#1][#2]}
\newcommandx{\ERel}[4][1=, 2=, 3=, 4=]{\mthrel[#4]{E#3}[#1][#2]}
\newcommandx{\FRel}[4][1=, 2=, 3=, 4=]{\mthrel[#4]{F#3}[#1][#2]}
\newcommandx{\GRel}[4][1=, 2=, 3=, 4=]{\mthrel[#4]{G#3}[#1][#2]}
\newcommandx{\HRel}[4][1=, 2=, 3=, 4=]{\mthrel[#4]{H#3}[#1][#2]}
\newcommandx{\IRel}[4][1=, 2=, 3=, 4=]{\mthrel[#4]{I#3}[#1][#2]}
\newcommandx{\JRel}[4][1=, 2=, 3=, 4=]{\mthrel[#4]{J#3}[#1][#2]}
\newcommandx{\KRel}[4][1=, 2=, 3=, 4=]{\mthrel[#4]{K#3}[#1][#2]}
\newcommandx{\LRel}[4][1=, 2=, 3=, 4=]{\mthrel[#4]{L#3}[#1][#2]}
\newcommandx{\MRel}[4][1=, 2=, 3=, 4=]{\mthrel[#4]{M#3}[#1][#2]}
\newcommandx{\NRel}[4][1=, 2=, 3=, 4=]{\mthrel[#4]{N#3}[#1][#2]}
\newcommandx{\ORel}[4][1=, 2=, 3=, 4=]{\mthrel[#4]{O#3}[#1][#2]}
\newcommandx{\PRel}[4][1=, 2=, 3=, 4=]{\mthrel[#4]{P#3}[#1][#2]}
\newcommandx{\QRel}[4][1=, 2=, 3=, 4=]{\mthrel[#4]{Q#3}[#1][#2]}
\newcommandx{\RRel}[4][1=, 2=, 3=, 4=]{\mthrel[#4]{R#3}[#1][#2]}
\newcommandx{\SRel}[4][1=, 2=, 3=, 4=]{\mthrel[#4]{S#3}[#1][#2]}
\newcommandx{\TRel}[4][1=, 2=, 3=, 4=]{\mthrel[#4]{T#3}[#1][#2]}
\newcommandx{\URel}[4][1=, 2=, 3=, 4=]{\mthrel[#4]{U#3}[#1][#2]}
\newcommandx{\VRel}[4][1=, 2=, 3=, 4=]{\mthrel[#4]{V#3}[#1][#2]}
\newcommandx{\WRel}[4][1=, 2=, 3=, 4=]{\mthrel[#4]{W#3}[#1][#2]}
\newcommandx{\XRel}[4][1=, 2=, 3=, 4=]{\mthrel[#4]{X#3}[#1][#2]}
\newcommandx{\YRel}[4][1=, 2=, 3=, 4=]{\mthrel[#4]{Y#3}[#1][#2]}
\newcommandx{\ZRel}[4][1=, 2=, 3=, 4=]{\mthrel[#4]{Z#3}[#1][#2]}

\newcommandx{\aRel}[4][1=, 2=, 3=, 4=]{\mthrel[#4]{a#3}[#1][#2]}
\newcommandx{\bRel}[4][1=, 2=, 3=, 4=]{\mthrel[#4]{b#3}[#1][#2]}
\newcommandx{\cRel}[4][1=, 2=, 3=, 4=]{\mthrel[#4]{c#3}[#1][#2]}
\newcommandx{\dRel}[4][1=, 2=, 3=, 4=]{\mthrel[#4]{d#3}[#1][#2]}
\newcommandx{\eRel}[4][1=, 2=, 3=, 4=]{\mthrel[#4]{e#3}[#1][#2]}
\newcommandx{\fRel}[4][1=, 2=, 3=, 4=]{\mthrel[#4]{f#3}[#1][#2]}
\newcommandx{\gRel}[4][1=, 2=, 3=, 4=]{\mthrel[#4]{g#3}[#1][#2]}
\newcommandx{\hRel}[4][1=, 2=, 3=, 4=]{\mthrel[#4]{h#3}[#1][#2]}
\newcommandx{\iRel}[4][1=, 2=, 3=, 4=]{\mthrel[#4]{i#3}[#1][#2]}
\newcommandx{\jRel}[4][1=, 2=, 3=, 4=]{\mthrel[#4]{j#3}[#1][#2]}
\newcommandx{\kRel}[4][1=, 2=, 3=, 4=]{\mthrel[#4]{k#3}[#1][#2]}
\newcommandx{\lRel}[4][1=, 2=, 3=, 4=]{\mthrel[#4]{l#3}[#1][#2]}
\newcommandx{\mRel}[4][1=, 2=, 3=, 4=]{\mthrel[#4]{m#3}[#1][#2]}
\newcommandx{\nRel}[4][1=, 2=, 3=, 4=]{\mthrel[#4]{n#3}[#1][#2]}
\newcommandx{\oRel}[4][1=, 2=, 3=, 4=]{\mthrel[#4]{o#3}[#1][#2]}
\newcommandx{\pRel}[4][1=, 2=, 3=, 4=]{\mthrel[#4]{p#3}[#1][#2]}
\newcommandx{\qRel}[4][1=, 2=, 3=, 4=]{\mthrel[#4]{q#3}[#1][#2]}
\newcommandx{\rRel}[4][1=, 2=, 3=, 4=]{\mthrel[#4]{r#3}[#1][#2]}
\newcommandx{\sRel}[4][1=, 2=, 3=, 4=]{\mthrel[#4]{s#3}[#1][#2]}
\newcommandx{\tRel}[4][1=, 2=, 3=, 4=]{\mthrel[#4]{t#3}[#1][#2]}
\newcommandx{\uRel}[4][1=, 2=, 3=, 4=]{\mthrel[#4]{u#3}[#1][#2]}
\newcommandx{\vRel}[4][1=, 2=, 3=, 4=]{\mthrel[#4]{v#3}[#1][#2]}
\newcommandx{\wRel}[4][1=, 2=, 3=, 4=]{\mthrel[#4]{w#3}[#1][#2]}
\newcommandx{\xRel}[4][1=, 2=, 3=, 4=]{\mthrel[#4]{x#3}[#1][#2]}
\newcommandx{\yRel}[4][1=, 2=, 3=, 4=]{\mthrel[#4]{y#3}[#1][#2]}
\newcommandx{\zRel}[4][1=, 2=, 3=, 4=]{\mthrel[#4]{z#3}[#1][#2]}

\newcommandx{\ASym}[4][1=, 2=, 3=, 4=]{\mthsym[#4]{A#3}[#1][#2]}
\newcommandx{\BSym}[4][1=, 2=, 3=, 4=]{\mthsym[#4]{B#3}[#1][#2]}
\newcommandx{\CSym}[4][1=, 2=, 3=, 4=]{\mthsym[#4]{C#3}[#1][#2]}
\newcommandx{\DSym}[4][1=, 2=, 3=, 4=]{\mthsym[#4]{D#3}[#1][#2]}
\newcommandx{\ESym}[4][1=, 2=, 3=, 4=]{\mthsym[#4]{E#3}[#1][#2]}
\newcommandx{\FSym}[4][1=, 2=, 3=, 4=]{\mthsym[#4]{F#3}[#1][#2]}
\newcommandx{\GSym}[4][1=, 2=, 3=, 4=]{\mthsym[#4]{G#3}[#1][#2]}
\newcommandx{\HSym}[4][1=, 2=, 3=, 4=]{\mthsym[#4]{H#3}[#1][#2]}
\newcommandx{\ISym}[4][1=, 2=, 3=, 4=]{\mthsym[#4]{I#3}[#1][#2]}
\newcommandx{\JSym}[4][1=, 2=, 3=, 4=]{\mthsym[#4]{J#3}[#1][#2]}
\newcommandx{\KSym}[4][1=, 2=, 3=, 4=]{\mthsym[#4]{K#3}[#1][#2]}
\newcommandx{\LSym}[4][1=, 2=, 3=, 4=]{\mthsym[#4]{L#3}[#1][#2]}
\newcommandx{\MSym}[4][1=, 2=, 3=, 4=]{\mthsym[#4]{M#3}[#1][#2]}
\newcommandx{\NSym}[4][1=, 2=, 3=, 4=]{\mthsym[#4]{N#3}[#1][#2]}
\newcommandx{\OSym}[4][1=, 2=, 3=, 4=]{\mthsym[#4]{O#3}[#1][#2]}
\newcommandx{\PSym}[4][1=, 2=, 3=, 4=]{\mthsym[#4]{P#3}[#1][#2]}
\newcommandx{\QSym}[4][1=, 2=, 3=, 4=]{\mthsym[#4]{Q#3}[#1][#2]}
\newcommandx{\RSym}[4][1=, 2=, 3=, 4=]{\mthsym[#4]{R#3}[#1][#2]}
\newcommandx{\SSym}[4][1=, 2=, 3=, 4=]{\mthsym[#4]{S#3}[#1][#2]}
\newcommandx{\TSym}[4][1=, 2=, 3=, 4=]{\mthsym[#4]{T#3}[#1][#2]}
\newcommandx{\USym}[4][1=, 2=, 3=, 4=]{\mthsym[#4]{U#3}[#1][#2]}
\newcommandx{\VSym}[4][1=, 2=, 3=, 4=]{\mthsym[#4]{V#3}[#1][#2]}
\newcommandx{\WSym}[4][1=, 2=, 3=, 4=]{\mthsym[#4]{W#3}[#1][#2]}
\newcommandx{\XSym}[4][1=, 2=, 3=, 4=]{\mthsym[#4]{X#3}[#1][#2]}
\newcommandx{\YSym}[4][1=, 2=, 3=, 4=]{\mthsym[#4]{Y#3}[#1][#2]}
\newcommandx{\ZSym}[4][1=, 2=, 3=, 4=]{\mthsym[#4]{Z#3}[#1][#2]}

\newcommandx{\aSym}[4][1=, 2=, 3=, 4=]{\mthsym[#4]{a#3}[#1][#2]}
\newcommandx{\bSym}[4][1=, 2=, 3=, 4=]{\mthsym[#4]{b#3}[#1][#2]}
\newcommandx{\cSym}[4][1=, 2=, 3=, 4=]{\mthsym[#4]{c#3}[#1][#2]}
\newcommandx{\dSym}[4][1=, 2=, 3=, 4=]{\mthsym[#4]{d#3}[#1][#2]}
\newcommandx{\eSym}[4][1=, 2=, 3=, 4=]{\mthsym[#4]{e#3}[#1][#2]}
\newcommandx{\fSym}[4][1=, 2=, 3=, 4=]{\mthsym[#4]{f#3}[#1][#2]}
\newcommandx{\gSym}[4][1=, 2=, 3=, 4=]{\mthsym[#4]{g#3}[#1][#2]}
\newcommandx{\hSym}[4][1=, 2=, 3=, 4=]{\mthsym[#4]{h#3}[#1][#2]}
\newcommandx{\iSym}[4][1=, 2=, 3=, 4=]{\mthsym[#4]{i#3}[#1][#2]}
\newcommandx{\jSym}[4][1=, 2=, 3=, 4=]{\mthsym[#4]{j#3}[#1][#2]}
\newcommandx{\kSym}[4][1=, 2=, 3=, 4=]{\mthsym[#4]{k#3}[#1][#2]}
\newcommandx{\lSym}[4][1=, 2=, 3=, 4=]{\mthsym[#4]{l#3}[#1][#2]}
\newcommandx{\mSym}[4][1=, 2=, 3=, 4=]{\mthsym[#4]{m#3}[#1][#2]}
\newcommandx{\nSym}[4][1=, 2=, 3=, 4=]{\mthsym[#4]{n#3}[#1][#2]}
\newcommandx{\oSym}[4][1=, 2=, 3=, 4=]{\mthsym[#4]{o#3}[#1][#2]}
\newcommandx{\pSym}[4][1=, 2=, 3=, 4=]{\mthsym[#4]{p#3}[#1][#2]}
\newcommandx{\qSym}[4][1=, 2=, 3=, 4=]{\mthsym[#4]{q#3}[#1][#2]}
\newcommandx{\rSym}[4][1=, 2=, 3=, 4=]{\mthsym[#4]{r#3}[#1][#2]}
\newcommandx{\sSym}[4][1=, 2=, 3=, 4=]{\mthsym[#4]{s#3}[#1][#2]}
\newcommandx{\tSym}[4][1=, 2=, 3=, 4=]{\mthsym[#4]{t#3}[#1][#2]}
\newcommandx{\uSym}[4][1=, 2=, 3=, 4=]{\mthsym[#4]{u#3}[#1][#2]}
\newcommandx{\vSym}[4][1=, 2=, 3=, 4=]{\mthsym[#4]{v#3}[#1][#2]}
\newcommandx{\wSym}[4][1=, 2=, 3=, 4=]{\mthsym[#4]{w#3}[#1][#2]}
\newcommandx{\xSym}[4][1=, 2=, 3=, 4=]{\mthsym[#4]{x#3}[#1][#2]}
\newcommandx{\ySym}[4][1=, 2=, 3=, 4=]{\mthsym[#4]{y#3}[#1][#2]}
\newcommandx{\zSym}[4][1=, 2=, 3=, 4=]{\mthsym[#4]{z#3}[#1][#2]}

\newcommandx{\AElm}[4][1=, 2=, 3=, 4=]{\mthelm[#4]{A#3}[#1][#2]}
\newcommandx{\BElm}[4][1=, 2=, 3=, 4=]{\mthelm[#4]{B#3}[#1][#2]}
\newcommandx{\CElm}[4][1=, 2=, 3=, 4=]{\mthelm[#4]{C#3}[#1][#2]}
\newcommandx{\DElm}[4][1=, 2=, 3=, 4=]{\mthelm[#4]{D#3}[#1][#2]}
\newcommandx{\EElm}[4][1=, 2=, 3=, 4=]{\mthelm[#4]{E#3}[#1][#2]}
\newcommandx{\FElm}[4][1=, 2=, 3=, 4=]{\mthelm[#4]{F#3}[#1][#2]}
\newcommandx{\GElm}[4][1=, 2=, 3=, 4=]{\mthelm[#4]{G#3}[#1][#2]}
\newcommandx{\HElm}[4][1=, 2=, 3=, 4=]{\mthelm[#4]{H#3}[#1][#2]}
\newcommandx{\IElm}[4][1=, 2=, 3=, 4=]{\mthelm[#4]{I#3}[#1][#2]}
\newcommandx{\JElm}[4][1=, 2=, 3=, 4=]{\mthelm[#4]{J#3}[#1][#2]}
\newcommandx{\KElm}[4][1=, 2=, 3=, 4=]{\mthelm[#4]{K#3}[#1][#2]}
\newcommandx{\LElm}[4][1=, 2=, 3=, 4=]{\mthelm[#4]{L#3}[#1][#2]}
\newcommandx{\MElm}[4][1=, 2=, 3=, 4=]{\mthelm[#4]{M#3}[#1][#2]}
\newcommandx{\NElm}[4][1=, 2=, 3=, 4=]{\mthelm[#4]{N#3}[#1][#2]}
\newcommandx{\OElm}[4][1=, 2=, 3=, 4=]{\mthelm[#4]{O#3}[#1][#2]}
\newcommandx{\PElm}[4][1=, 2=, 3=, 4=]{\mthelm[#4]{P#3}[#1][#2]}
\newcommandx{\QElm}[4][1=, 2=, 3=, 4=]{\mthelm[#4]{Q#3}[#1][#2]}
\newcommandx{\RElm}[4][1=, 2=, 3=, 4=]{\mthelm[#4]{R#3}[#1][#2]}
\newcommandx{\SElm}[4][1=, 2=, 3=, 4=]{\mthelm[#4]{S#3}[#1][#2]}
\newcommandx{\TElm}[4][1=, 2=, 3=, 4=]{\mthelm[#4]{T#3}[#1][#2]}
\newcommandx{\UElm}[4][1=, 2=, 3=, 4=]{\mthelm[#4]{U#3}[#1][#2]}
\newcommandx{\VElm}[4][1=, 2=, 3=, 4=]{\mthelm[#4]{V#3}[#1][#2]}
\newcommandx{\WElm}[4][1=, 2=, 3=, 4=]{\mthelm[#4]{W#3}[#1][#2]}
\newcommandx{\XElm}[4][1=, 2=, 3=, 4=]{\mthelm[#4]{X#3}[#1][#2]}
\newcommandx{\YElm}[4][1=, 2=, 3=, 4=]{\mthelm[#4]{Y#3}[#1][#2]}
\newcommandx{\ZElm}[4][1=, 2=, 3=, 4=]{\mthelm[#4]{Z#3}[#1][#2]}

\newcommandx{\aElm}[4][1=, 2=, 3=, 4=]{\mthelm[#4]{a#3}[#1][#2]}
\newcommandx{\bElm}[4][1=, 2=, 3=, 4=]{\mthelm[#4]{b#3}[#1][#2]}
\newcommandx{\cElm}[4][1=, 2=, 3=, 4=]{\mthelm[#4]{c#3}[#1][#2]}
\newcommandx{\dElm}[4][1=, 2=, 3=, 4=]{\mthelm[#4]{d#3}[#1][#2]}
\newcommandx{\eElm}[4][1=, 2=, 3=, 4=]{\mthelm[#4]{e#3}[#1][#2]}
\newcommandx{\fElm}[4][1=, 2=, 3=, 4=]{\mthelm[#4]{f#3}[#1][#2]}
\newcommandx{\gElm}[4][1=, 2=, 3=, 4=]{\mthelm[#4]{g#3}[#1][#2]}
\newcommandx{\hElm}[4][1=, 2=, 3=, 4=]{\mthelm[#4]{h#3}[#1][#2]}
\newcommandx{\iElm}[4][1=, 2=, 3=, 4=]{\mthelm[#4]{i#3}[#1][#2]}
\newcommandx{\jElm}[4][1=, 2=, 3=, 4=]{\mthelm[#4]{j#3}[#1][#2]}
\newcommandx{\kElm}[4][1=, 2=, 3=, 4=]{\mthelm[#4]{k#3}[#1][#2]}
\newcommandx{\lElm}[4][1=, 2=, 3=, 4=]{\mthelm[#4]{l#3}[#1][#2]}
\newcommandx{\mElm}[4][1=, 2=, 3=, 4=]{\mthelm[#4]{m#3}[#1][#2]}
\newcommandx{\nElm}[4][1=, 2=, 3=, 4=]{\mthelm[#4]{n#3}[#1][#2]}
\newcommandx{\oElm}[4][1=, 2=, 3=, 4=]{\mthelm[#4]{o#3}[#1][#2]}
\newcommandx{\pElm}[4][1=, 2=, 3=, 4=]{\mthelm[#4]{p#3}[#1][#2]}
\newcommandx{\qElm}[4][1=, 2=, 3=, 4=]{\mthelm[#4]{q#3}[#1][#2]}
\newcommandx{\rElm}[4][1=, 2=, 3=, 4=]{\mthelm[#4]{r#3}[#1][#2]}
\newcommandx{\sElm}[4][1=, 2=, 3=, 4=]{\mthelm[#4]{s#3}[#1][#2]}
\newcommandx{\tElm}[4][1=, 2=, 3=, 4=]{\mthelm[#4]{t#3}[#1][#2]}
\newcommandx{\uElm}[4][1=, 2=, 3=, 4=]{\mthelm[#4]{u#3}[#1][#2]}
\newcommandx{\vElm}[4][1=, 2=, 3=, 4=]{\mthelm[#4]{v#3}[#1][#2]}
\newcommandx{\wElm}[4][1=, 2=, 3=, 4=]{\mthelm[#4]{w#3}[#1][#2]}
\newcommandx{\xElm}[4][1=, 2=, 3=, 4=]{\mthelm[#4]{x#3}[#1][#2]}
\newcommandx{\yElm}[4][1=, 2=, 3=, 4=]{\mthelm[#4]{y#3}[#1][#2]}
\newcommandx{\zElm}[4][1=, 2=, 3=, 4=]{\mthelm[#4]{z#3}[#1][#2]}

\newcommand{\aka}
	{\txtabr{a.k.a.}\xspace}

\newcommand{\divideetimpera}
	{\txtabr{divide et impera}\xspace}

\newcommand{\eg}
	{\txtabr{e.g.}\xspace}

\newcommand{\ie}
	{\txtabr{i.e.}\xspace}

\newcommand{\resp}
	{\txtabr{resp.}\xspace}

\newcommand{\wrt}
	{\txtabr{w.r.t.}\xspace}

\newcommand{\defeq}
	{\ensuremath{\triangleq}}

\newcommand{\lst}
	{\mthargfun{lst}}

\newcommand{\dual}[1]
	{\mthempty{\overline{#1}}}
\newcommand{\adj}[1]
	{\mthempty{\widetilde{#1}}}
\newcommand{\der}[1]
	{\mthempty{\widehat{#1}}}

\newcommand{\tuple}[1]
	{\ensuremath{\!\argint{\langle}{#1}{\rangle}}}

\newcommand{\tupleb}[2]
	{\tuple{\argb{#1}{#2}}}
\newcommand{\tuplec}[3]
	{\tuple{\argc{#1}{#2}{#3}}}
\newcommand{\tupled}[4]
	{\tuple{\argd{#1}{#2}{#3}{#4}}}
\newcommand{\tuplee}[5]
	{\tuple{\arge{#1}{#2}{#3}{#4}{#5}}}
\newcommand{\tuplef}[6]
	{\tuple{\argf{#1}{#2}{#3}{#4}{#5}{#6}}}
\newcommand{\tupleg}[7]
	{\tuple{\argg{#1}{#2}{#3}{#4}{#5}{#6}{#7}}}
\newcommand{\tupleh}[8]
	{\tuple{\argh{#1}{#2}{#3}{#4}{#5}{#6}{#7}{#8}}}
\newcommand{\tuplei}[9]
	{\tuple{\argi{#1}{#2}{#3}{#4}{#5}{#6}{#7}{#8}{#9}}}

\newcommand{\tuplej}[9]
	{%
	\def\defarga{#1}%
	\def\defargb{#2}%
	\def\defargc{#3}%
	\def\defargd{#4}%
	\def\defarge{#5}%
	\def\defargf{#6}%
	\def\defargg{#7}%
	\def\defargh{#8}%
	\def\defargi{#9}%
	\tupleauxj%
	}
\newcommand{\tuplek}[9]
	{%
	\def\defarga{#1}%
	\def\defargb{#2}%
	\def\defargc{#3}%
	\def\defargd{#4}%
	\def\defarge{#5}%
	\def\defargf{#6}%
	\def\defargg{#7}%
	\def\defargh{#8}%
	\def\defargi{#9}%
	\tupleauxk%
	}
\newcommand{\tuplel}[9]
	{%
	\def\defarga{#1}%
	\def\defargb{#2}%
	\def\defargc{#3}%
	\def\defargd{#4}%
	\def\defarge{#5}%
	\def\defargf{#6}%
	\def\defargg{#7}%
	\def\defargh{#8}%
	\def\defargi{#9}%
	\tupleauxl%
	}
\newcommand{\tuplem}[9]
	{%
	\def\defarga{#1}%
	\def\defargb{#2}%
	\def\defargc{#3}%
	\def\defargd{#4}%
	\def\defarge{#5}%
	\def\defargf{#6}%
	\def\defargg{#7}%
	\def\defargh{#8}%
	\def\defargi{#9}%
	\tupleauxm%
	}
\newcommand{\tuplen}[9]
	{%
	\def\defarga{#1}%
	\def\defargb{#2}%
	\def\defargc{#3}%
	\def\defargd{#4}%
	\def\defarge{#5}%
	\def\defargf{#6}%
	\def\defargg{#7}%
	\def\defargh{#8}%
	\def\defargi{#9}%
	\tupleauxn%
	}
\newcommand{\tupleo}[9]
	{%
	\def\defarga{#1}%
	\def\defargb{#2}%
	\def\defargc{#3}%
	\def\defargd{#4}%
	\def\defarge{#5}%
	\def\defargf{#6}%
	\def\defargg{#7}%
	\def\defargh{#8}%
	\def\defargi{#9}%
	\tupleauxo%
	}
\newcommand{\tuplep}[9]
	{%
	\def\defarga{#1}%
	\def\defargb{#2}%
	\def\defargc{#3}%
	\def\defargd{#4}%
	\def\defarge{#5}%
	\def\defargf{#6}%
	\def\defargg{#7}%
	\def\defargh{#8}%
	\def\defargi{#9}%
	\tupleauxp%
	}
\newcommand{\tupleq}[9]
	{%
	\def\defarga{#1}%
	\def\defargb{#2}%
	\def\defargc{#3}%
	\def\defargd{#4}%
	\def\defarge{#5}%
	\def\defargf{#6}%
	\def\defargg{#7}%
	\def\defargh{#8}%
	\def\defargi{#9}%
	\tupleauxq%
	}
\newcommand{\tupler}[9]
	{%
	\def\defarga{#1}%
	\def\defargb{#2}%
	\def\defargc{#3}%
	\def\defargd{#4}%
	\def\defarge{#5}%
	\def\defargf{#6}%
	\def\defargg{#7}%
	\def\defargh{#8}%
	\def\defargi{#9}%
	\tupleauxr%
	}

\newcommand{\tupleauxj}[1]
	{%
	\tuple{\argj{\defarga}{\defargb}{\defargc}{\defargd}{\defarge}{\defargf}%
		{\defargg}{\defargh}{\defargi}{#1}}%
	}
\newcommand{\tupleauxk}[2]
	{%
	\tuple{\argk{\defarga}{\defargb}{\defargc}{\defargd}{\defarge}{\defargf}%
		{\defargg}{\defargh}{\defargi}{#1}{#2}}%
	}
\newcommand{\tupleauxl}[3]
	{%
	\tuple{\argl{\defarga}{\defargb}{\defargc}{\defargd}{\defarge}{\defargf}%
		{\defargg}{\defargh}{\defargi}{#1}{#2}{#3}}%
	}
\newcommand{\tupleauxm}[4]
	{%
	\tuple{\argm{\defarga}{\defargb}{\defargc}{\defargd}{\defarge}{\defargf}%
		{\defargg}{\defargh}{\defargi}{#1}{#2}{#3}{#4}}%
	}
\newcommand{\tupleauxn}[5]
	{%
	\tuple{\argn{\defarga}{\defargb}{\defargc}{\defargd}{\defarge}{\defargf}%
		{\defargg}{\defargh}{\defargi}{#1}{#2}{#3}{#4}{#5}}%
	}
\newcommand{\tupleauxo}[6]
	{%
	\tuple{\argo{\defarga}{\defargb}{\defargc}{\defargd}{\defarge}{\defargf}%
		{\defargg}{\defargh}{\defargi}{#1}{#2}{#3}{#4}{#5}{#6}}%
	}
\newcommand{\tupleauxp}[7]
	{%
	\tuple{\argp{\defarga}{\defargb}{\defargc}{\defargd}{\defarge}{\defargf}%
		{\defargg}{\defargh}{\defargi}{#1}{#2}{#3}{#4}{#5}{#6}{#7}}%
	}
\newcommand{\tupleauxq}[8]
	{%
	\tuple{\argq{\defarga}{\defargb}{\defargc}{\defargd}{\defarge}{\defargf}%
		{\defargg}{\defargh}{\defargi}{#1}{#2}{#3}{#4}{#5}{#6}{#7}{#8}}%
	}
\newcommand{\tupleauxr}[9]
	{%
	\tuple{\argr{\defarga}{\defargb}{\defargc}{\defargd}{\defarge}{\defargf}%
		{\defargg}{\defargh}{\defargi}{#1}{#2}{#3}{#4}{#5}{#6}{#7}{#8}{#9}}%
	}

\newcommand{\tuplecx}[3]
	{%
	\def\defarga{#1}%
	\def\defargb{#2}%
	\def\defargc{#3}%
	\argsubsup{\tupleauxcx}%
	}
\newcommand{\tupledx}[4]
	{%
	\def\defarga{#1}%
	\def\defargb{#2}%
	\def\defargc{#3}%
	\def\defargd{#4}%
	\argsubsup{\tupleauxdx}%
	}
\newcommand{\tupleex}[5]
	{%
	\def\defarga{#1}%
	\def\defargb{#2}%
	\def\defargc{#3}%
	\def\defargd{#4}%
	\def\defarge{#5}%
	\argsubsup{\tupleauxex}%
	}
\newcommand{\tuplefx}[6]
	{%
	\def\defarga{#1}%
	\def\defargb{#2}%
	\def\defargc{#3}%
	\def\defargd{#4}%
	\def\defarge{#5}%
	\def\defargf{#6}%
	\argsubsup{\tupleauxfx}%
	}
\newcommand{\tuplegx}[7]
	{%
	\def\defarga{#1}%
	\def\defargb{#2}%
	\def\defargc{#3}%
	\def\defargd{#4}%
	\def\defarge{#5}%
	\def\defargf{#6}%
	\def\defargg{#7}%
	\argsubsup{\tupleauxgx}%
	}
\newcommand{\tuplehx}[8]
	{%
	\def\defarga{#1}%
	\def\defargb{#2}%
	\def\defargc{#3}%
	\def\defargd{#4}%
	\def\defarge{#5}%
	\def\defargf{#6}%
	\def\defargg{#7}%
	\def\defargh{#8}%
	\argsubsup{\tupleauxhx}%
	}
\newcommand{\tupleix}[9]
	{%
	\def\defarga{#1}%
	\def\defargb{#2}%
	\def\defargc{#3}%
	\def\defargd{#4}%
	\def\defarge{#5}%
	\def\defargf{#6}%
	\def\defargg{#7}%
	\def\defargh{#8}%
	\def\defargi{#9}%
	\argsubsup{\tupleauxix}%
	}

\newcommandx{\tupleauxbx}[2][1=, 2=]
	{%
	\tupleb
		{\argdef{#1}{\defarga[\argsubscript][\argsuperscript]}}
		{\argdef{#2}{\defargb[\argsubscript][\argsuperscript]}}%
	}
\newcommandx{\tupleauxcx}[3][1=, 2=, 3=]
	{%
	\tuplec
		{\argdef{#1}{\defarga[\argsubscript][\argsuperscript]}}
		{\argdef{#2}{\defargb[\argsubscript][\argsuperscript]}}
		{\argdef{#3}{\defargc[\argsubscript][\argsuperscript]}}%
	}
\newcommandx{\tupleauxdx}[4][1=, 2=, 3=, 4=]
	{%
	\tupled
		{\argdef{#1}{\defarga[\argsubscript][\argsuperscript]}}
		{\argdef{#2}{\defargb[\argsubscript][\argsuperscript]}}
		{\argdef{#3}{\defargc[\argsubscript][\argsuperscript]}}
		{\argdef{#4}{\defargd[\argsubscript][\argsuperscript]}}%
	}
\newcommandx{\tupleauxex}[5][1=, 2=, 3=, 4=, 5=]
	{%
	\tuplee
		{\argdef{#1}{\defarga[\argsubscript][\argsuperscript]}}
		{\argdef{#2}{\defargb[\argsubscript][\argsuperscript]}}
		{\argdef{#3}{\defargc[\argsubscript][\argsuperscript]}}
		{\argdef{#4}{\defargd[\argsubscript][\argsuperscript]}}
		{\argdef{#5}{\defarge[\argsubscript][\argsuperscript]}}%
	}
\newcommandx{\tupleauxfx}[6][1=, 2=, 3=, 4=, 5=, 6=]
	{%
	\tuplef
		{\argdef{#1}{\defarga[\argsubscript][\argsuperscript]}}
		{\argdef{#2}{\defargb[\argsubscript][\argsuperscript]}}
		{\argdef{#3}{\defargc[\argsubscript][\argsuperscript]}}
		{\argdef{#4}{\defargd[\argsubscript][\argsuperscript]}}
		{\argdef{#5}{\defarge[\argsubscript][\argsuperscript]}}
		{\argdef{#6}{\defargf[\argsubscript][\argsuperscript]}}%
	}
\newcommandx{\tupleauxgx}[7][1=, 2=, 3=, 4=, 5=, 6=, 7=]
	{%
	\tupleg
		{\argdef{#1}{\defarga[\argsubscript][\argsuperscript]}}
		{\argdef{#2}{\defargb[\argsubscript][\argsuperscript]}}
		{\argdef{#3}{\defargc[\argsubscript][\argsuperscript]}}
		{\argdef{#4}{\defargd[\argsubscript][\argsuperscript]}}
		{\argdef{#5}{\defarge[\argsubscript][\argsuperscript]}}
		{\argdef{#6}{\defargf[\argsubscript][\argsuperscript]}}
		{\argdef{#7}{\defargg[\argsubscript][\argsuperscript]}}%
	}
\newcommandx{\tupleauxhx}[8][1=, 2=, 3=, 4=, 5=, 6=, 7=, 8=]
	{%
	\tupleh
		{\argdef{#1}{\defarga[\argsubscript][\argsuperscript]}}
		{\argdef{#2}{\defargb[\argsubscript][\argsuperscript]}}
		{\argdef{#3}{\defargc[\argsubscript][\argsuperscript]}}
		{\argdef{#4}{\defargd[\argsubscript][\argsuperscript]}}
		{\argdef{#5}{\defarge[\argsubscript][\argsuperscript]}}
		{\argdef{#6}{\defargf[\argsubscript][\argsuperscript]}}
		{\argdef{#7}{\defargg[\argsubscript][\argsuperscript]}}
		{\argdef{#8}{\defargh[\argsubscript][\argsuperscript]}}%
	}
\newcommandx{\tupleauxix}[9][1=, 2=, 3=, 4=, 5=, 6=, 7=, 8=, 9=]
	{%
	\tuplei
		{\argdef{#1}{\defarga[\argsubscript][\argsuperscript]}}
		{\argdef{#2}{\defargb[\argsubscript][\argsuperscript]}}
		{\argdef{#3}{\defargc[\argsubscript][\argsuperscript]}}
		{\argdef{#4}{\defargd[\argsubscript][\argsuperscript]}}
		{\argdef{#5}{\defarge[\argsubscript][\argsuperscript]}}
		{\argdef{#6}{\defargf[\argsubscript][\argsuperscript]}}
		{\argdef{#7}{\defargg[\argsubscript][\argsuperscript]}}
		{\argdef{#8}{\defargh[\argsubscript][\argsuperscript]}}
		{\argdef{#9}{\defargi[\argsubscript][\argsuperscript]}}%
	}

\newcommand{\tuplejx}[9]
	{%
	\def\tuplearga{#1}%
	\def\tupleargb{#2}%
	\def\tupleargc{#3}%
	\def\tupleargd{#4}%
	\def\tuplearge{#5}%
	\def\tupleargf{#6}%
	\def\tupleargg{#7}%
	\def\tupleargh{#8}%
	\def\tupleargi{#9}%
	\argsubsup{\tupleauxjx}%
	}

\newcommand{\tupleauxjx}[1]
	{%
	\def\tupleargj{#1}%
	\argsubsup{\tupleauxxjx}%
	}

\newcommandx{\tupleauxxjx}[9][1=, 2=, 3=, 4=, 5=, 6=, 7=, 8=, 9=]
	{%
	\def\optarga{#1}%
	\def\optargb{#2}%
	\def\optargc{#3}%
	\def\optargd{#4}%
	\def\optarge{#5}%
	\def\optargf{#6}%
	\def\optargg{#7}%
	\def\optargh{#8}%
	\def\optargi{#9}%
	\tupleauxxxjx%
	}
\newcommandx{\tupleauxxkx}[9][1=, 2=, 3=, 4=, 5=, 6=, 7=, 8=, 9=]
	{%
	\def\optarga{#1}%
	\def\optargb{#2}%
	\def\optargc{#3}%
	\def\optargd{#4}%
	\def\optarge{#5}%
	\def\optargf{#6}%
	\def\optargg{#7}%
	\def\optargh{#8}%
	\def\optargi{#9}%
	\tupleauxxxkx%
	}
\newcommandx{\tupleauxxlx}[9][1=, 2=, 3=, 4=, 5=, 6=, 7=, 8=, 9=]
	{%
	\def\optarga{#1}%
	\def\optargb{#2}%
	\def\optargc{#3}%
	\def\optargd{#4}%
	\def\optarge{#5}%
	\def\optargf{#6}%
	\def\optargg{#7}%
	\def\optargh{#8}%
	\def\optargi{#9}%
	\tupleauxxxlx%
	}
\newcommandx{\tupleauxxmx}[9][1=, 2=, 3=, 4=, 5=, 6=, 7=, 8=, 9=]
	{%
	\def\optarga{#1}%
	\def\optargb{#2}%
	\def\optargc{#3}%
	\def\optargd{#4}%
	\def\optarge{#5}%
	\def\optargf{#6}%
	\def\optargg{#7}%
	\def\optargh{#8}%
	\def\optargi{#9}%
	\tupleauxxxmx%
	}
\newcommandx{\tupleauxxnx}[9][1=, 2=, 3=, 4=, 5=, 6=, 7=, 8=, 9=]
	{%
	\def\optarga{#1}%
	\def\optargb{#2}%
	\def\optargc{#3}%
	\def\optargd{#4}%
	\def\optarge{#5}%
	\def\optargf{#6}%
	\def\optargg{#7}%
	\def\optargh{#8}%
	\def\optargi{#9}%
	\tupleauxxxnx%
	}
\newcommandx{\tupleauxxox}[9][1=, 2=, 3=, 4=, 5=, 6=, 7=, 8=, 9=]
	{%
	\def\optarga{#1}%
	\def\optargb{#2}%
	\def\optargc{#3}%
	\def\optargd{#4}%
	\def\optarge{#5}%
	\def\optargf{#6}%
	\def\optargg{#7}%
	\def\optargh{#8}%
	\def\optargi{#9}%
	\tupleauxxxox%
	}
\newcommandx{\tupleauxxpx}[9][1=, 2=, 3=, 4=, 5=, 6=, 7=, 8=, 9=]
	{%
	\def\optarga{#1}%
	\def\optargb{#2}%
	\def\optargc{#3}%
	\def\optargd{#4}%
	\def\optarge{#5}%
	\def\optargf{#6}%
	\def\optargg{#7}%
	\def\optargh{#8}%
	\def\optargi{#9}%
	\tupleauxxxpx%
	}
\newcommandx{\tupleauxxqx}[9][1=, 2=, 3=, 4=, 5=, 6=, 7=, 8=, 9=]
	{%
	\def\optarga{#1}%
	\def\optargb{#2}%
	\def\optargc{#3}%
	\def\optargd{#4}%
	\def\optarge{#5}%
	\def\optargf{#6}%
	\def\optargg{#7}%
	\def\optargh{#8}%
	\def\optargi{#9}%
	\tupleauxxxqx%
	}
\newcommandx{\tupleauxxrx}[9][1=, 2=, 3=, 4=, 5=, 6=, 7=, 8=, 9=]
	{%
	\def\optarga{#1}%
	\def\optargb{#2}%
	\def\optargc{#3}%
	\def\optargd{#4}%
	\def\optarge{#5}%
	\def\optargf{#6}%
	\def\optargg{#7}%
	\def\optargh{#8}%
	\def\optargi{#9}%
	\tupleauxxxrx%
	}

\newcommandx{\tupleauxxxjx}[1][1=]
	{%
	\tuplej
		{\argdef{\optarga}{\tuplearga[\argsubscript][\argsuperscript]}}
		{\argdef{\optargb}{\tupleargb[\argsubscript][\argsuperscript]}}
		{\argdef{\optargc}{\tupleargc[\argsubscript][\argsuperscript]}}
		{\argdef{\optargd}{\tupleargd[\argsubscript][\argsuperscript]}}
		{\argdef{\optarge}{\tuplearge[\argsubscript][\argsuperscript]}}
		{\argdef{\optargf}{\tupleargf[\argsubscript][\argsuperscript]}}
		{\argdef{\optargg}{\tupleargg[\argsubscript][\argsuperscript]}}
		{\argdef{\optargh}{\tupleargh[\argsubscript][\argsuperscript]}}
		{\argdef{\optargi}{\tupleargi[\argsubscript][\argsuperscript]}}
		{\argdef{#1}{\tupleargj[\argsubscript][\argsuperscript]}}%
	}
\newcommandx{\tupleauxxxkx}[2][1=, 2=]
	{%
	\tuplek
		{\argdef{\optarga}{\tuplearga[\argsubscript][\argsuperscript]}}
		{\argdef{\optargb}{\tupleargb[\argsubscript][\argsuperscript]}}
		{\argdef{\optargc}{\tupleargc[\argsubscript][\argsuperscript]}}
		{\argdef{\optargd}{\tupleargd[\argsubscript][\argsuperscript]}}
		{\argdef{\optarge}{\tuplearge[\argsubscript][\argsuperscript]}}
		{\argdef{\optargf}{\tupleargf[\argsubscript][\argsuperscript]}}
		{\argdef{\optargg}{\tupleargg[\argsubscript][\argsuperscript]}}
		{\argdef{\optargh}{\tupleargh[\argsubscript][\argsuperscript]}}
		{\argdef{\optargi}{\tupleargi[\argsubscript][\argsuperscript]}}
		{\argdef{#1}{\tupleargj[\argsubscript][\argsuperscript]}}
		{\argdef{#2}{\tupleargk[\argsubscript][\argsuperscript]}}
	}
\newcommandx{\tupleauxxxlx}[3][1=, 2=, 3=]
	{%
	\tuplel
		{\argdef{\optarga}{\tuplearga[\argsubscript][\argsuperscript]}}
		{\argdef{\optargb}{\tupleargb[\argsubscript][\argsuperscript]}}
		{\argdef{\optargc}{\tupleargc[\argsubscript][\argsuperscript]}}
		{\argdef{\optargd}{\tupleargd[\argsubscript][\argsuperscript]}}
		{\argdef{\optarge}{\tuplearge[\argsubscript][\argsuperscript]}}
		{\argdef{\optargf}{\tupleargf[\argsubscript][\argsuperscript]}}
		{\argdef{\optargg}{\tupleargg[\argsubscript][\argsuperscript]}}
		{\argdef{\optargh}{\tupleargh[\argsubscript][\argsuperscript]}}
		{\argdef{\optargi}{\tupleargi[\argsubscript][\argsuperscript]}}
		{\argdef{#1}{\tupleargj[\argsubscript][\argsuperscript]}}
		{\argdef{#2}{\tupleargk[\argsubscript][\argsuperscript]}}
		{\argdef{#3}{\tupleargl[\argsubscript][\argsuperscript]}}
	}
\newcommandx{\tupleauxxxmx}[4][1=, 2=, 3=, 4=]
	{%
	\tuplem
		{\argdef{\optarga}{\tuplearga[\argsubscript][\argsuperscript]}}
		{\argdef{\optargb}{\tupleargb[\argsubscript][\argsuperscript]}}
		{\argdef{\optargc}{\tupleargc[\argsubscript][\argsuperscript]}}
		{\argdef{\optargd}{\tupleargd[\argsubscript][\argsuperscript]}}
		{\argdef{\optarge}{\tuplearge[\argsubscript][\argsuperscript]}}
		{\argdef{\optargf}{\tupleargf[\argsubscript][\argsuperscript]}}
		{\argdef{\optargg}{\tupleargg[\argsubscript][\argsuperscript]}}
		{\argdef{\optargh}{\tupleargh[\argsubscript][\argsuperscript]}}
		{\argdef{\optargi}{\tupleargi[\argsubscript][\argsuperscript]}}
		{\argdef{#1}{\tupleargj[\argsubscript][\argsuperscript]}}
		{\argdef{#2}{\tupleargk[\argsubscript][\argsuperscript]}}
		{\argdef{#3}{\tupleargl[\argsubscript][\argsuperscript]}}
		{\argdef{#4}{\tupleargm[\argsubscript][\argsuperscript]}}
	}
\newcommandx{\tupleauxxxnx}[5][1=, 2=, 3=, 4=, 5=]
	{%
	\tuplen
		{\argdef{\optarga}{\tuplearga[\argsubscript][\argsuperscript]}}
		{\argdef{\optargb}{\tupleargb[\argsubscript][\argsuperscript]}}
		{\argdef{\optargc}{\tupleargc[\argsubscript][\argsuperscript]}}
		{\argdef{\optargd}{\tupleargd[\argsubscript][\argsuperscript]}}
		{\argdef{\optarge}{\tuplearge[\argsubscript][\argsuperscript]}}
		{\argdef{\optargf}{\tupleargf[\argsubscript][\argsuperscript]}}
		{\argdef{\optargg}{\tupleargg[\argsubscript][\argsuperscript]}}
		{\argdef{\optargh}{\tupleargh[\argsubscript][\argsuperscript]}}
		{\argdef{\optargi}{\tupleargi[\argsubscript][\argsuperscript]}}
		{\argdef{#1}{\tupleargj[\argsubscript][\argsuperscript]}}
		{\argdef{#2}{\tupleargk[\argsubscript][\argsuperscript]}}
		{\argdef{#3}{\tupleargl[\argsubscript][\argsuperscript]}}
		{\argdef{#4}{\tupleargm[\argsubscript][\argsuperscript]}}
		{\argdef{#5}{\tupleargn[\argsubscript][\argsuperscript]}}
	}
\newcommandx{\tupleauxxxox}[6][1=, 2=, 3=, 4=, 5=, 6=]
	{%
	\tupleo
		{\argdef{\optarga}{\tuplearga[\argsubscript][\argsuperscript]}}
		{\argdef{\optargb}{\tupleargb[\argsubscript][\argsuperscript]}}
		{\argdef{\optargc}{\tupleargc[\argsubscript][\argsuperscript]}}
		{\argdef{\optargd}{\tupleargd[\argsubscript][\argsuperscript]}}
		{\argdef{\optarge}{\tuplearge[\argsubscript][\argsuperscript]}}
		{\argdef{\optargf}{\tupleargf[\argsubscript][\argsuperscript]}}
		{\argdef{\optargg}{\tupleargg[\argsubscript][\argsuperscript]}}
		{\argdef{\optargh}{\tupleargh[\argsubscript][\argsuperscript]}}
		{\argdef{\optargi}{\tupleargi[\argsubscript][\argsuperscript]}}
		{\argdef{#1}{\tupleargj[\argsubscript][\argsuperscript]}}
		{\argdef{#2}{\tupleargk[\argsubscript][\argsuperscript]}}
		{\argdef{#3}{\tupleargl[\argsubscript][\argsuperscript]}}
		{\argdef{#4}{\tupleargm[\argsubscript][\argsuperscript]}}
		{\argdef{#5}{\tupleargn[\argsubscript][\argsuperscript]}}
		{\argdef{#6}{\tupleargo[\argsubscript][\argsuperscript]}}
	}
\newcommandx{\tupleauxxxpx}[7][1=, 2=, 3=, 4=, 5=, 6=, 7=]
	{%
	\tuplep
		{\argdef{\optarga}{\tuplearga[\argsubscript][\argsuperscript]}}
		{\argdef{\optargb}{\tupleargb[\argsubscript][\argsuperscript]}}
		{\argdef{\optargc}{\tupleargc[\argsubscript][\argsuperscript]}}
		{\argdef{\optargd}{\tupleargd[\argsubscript][\argsuperscript]}}
		{\argdef{\optarge}{\tuplearge[\argsubscript][\argsuperscript]}}
		{\argdef{\optargf}{\tupleargf[\argsubscript][\argsuperscript]}}
		{\argdef{\optargg}{\tupleargg[\argsubscript][\argsuperscript]}}
		{\argdef{\optargh}{\tupleargh[\argsubscript][\argsuperscript]}}
		{\argdef{\optargi}{\tupleargi[\argsubscript][\argsuperscript]}}
		{\argdef{#1}{\tupleargj[\argsubscript][\argsuperscript]}}
		{\argdef{#2}{\tupleargk[\argsubscript][\argsuperscript]}}
		{\argdef{#3}{\tupleargl[\argsubscript][\argsuperscript]}}
		{\argdef{#4}{\tupleargm[\argsubscript][\argsuperscript]}}
		{\argdef{#5}{\tupleargn[\argsubscript][\argsuperscript]}}
		{\argdef{#6}{\tupleargo[\argsubscript][\argsuperscript]}}
		{\argdef{#7}{\tupleargp[\argsubscript][\argsuperscript]}}
	}
\newcommandx{\tupleauxxxqx}[8][1=, 2=, 3=, 4=, 5=, 6=, 7=, 8=]
	{%
	\tupleq
		{\argdef{\optarga}{\tuplearga[\argsubscript][\argsuperscript]}}
		{\argdef{\optargb}{\tupleargb[\argsubscript][\argsuperscript]}}
		{\argdef{\optargc}{\tupleargc[\argsubscript][\argsuperscript]}}
		{\argdef{\optargd}{\tupleargd[\argsubscript][\argsuperscript]}}
		{\argdef{\optarge}{\tuplearge[\argsubscript][\argsuperscript]}}
		{\argdef{\optargf}{\tupleargf[\argsubscript][\argsuperscript]}}
		{\argdef{\optargg}{\tupleargg[\argsubscript][\argsuperscript]}}
		{\argdef{\optargh}{\tupleargh[\argsubscript][\argsuperscript]}}
		{\argdef{\optargi}{\tupleargi[\argsubscript][\argsuperscript]}}
		{\argdef{#1}{\tupleargj[\argsubscript][\argsuperscript]}}
		{\argdef{#2}{\tupleargk[\argsubscript][\argsuperscript]}}
		{\argdef{#3}{\tupleargl[\argsubscript][\argsuperscript]}}
		{\argdef{#4}{\tupleargm[\argsubscript][\argsuperscript]}}
		{\argdef{#5}{\tupleargn[\argsubscript][\argsuperscript]}}
		{\argdef{#6}{\tupleargo[\argsubscript][\argsuperscript]}}
		{\argdef{#7}{\tupleargp[\argsubscript][\argsuperscript]}}
		{\argdef{#8}{\tupleargq[\argsubscript][\argsuperscript]}}
	}
\newcommandx{\tupleauxxxrx}[9][1=, 2=, 3=, 4=, 5=, 6=, 7=, 8=, 9=]
	{%
	\tupler
		{\argdef{\optarga}{\tuplearga[\argsubscript][\argsuperscript]}}
		{\argdef{\optargb}{\tupleargb[\argsubscript][\argsuperscript]}}
		{\argdef{\optargc}{\tupleargc[\argsubscript][\argsuperscript]}}
		{\argdef{\optargd}{\tupleargd[\argsubscript][\argsuperscript]}}
		{\argdef{\optarge}{\tuplearge[\argsubscript][\argsuperscript]}}
		{\argdef{\optargf}{\tupleargf[\argsubscript][\argsuperscript]}}
		{\argdef{\optargg}{\tupleargg[\argsubscript][\argsuperscript]}}
		{\argdef{\optargh}{\tupleargh[\argsubscript][\argsuperscript]}}
		{\argdef{\optargi}{\tupleargi[\argsubscript][\argsuperscript]}}
		{\argdef{#1}{\tupleargj[\argsubscript][\argsuperscript]}}
		{\argdef{#2}{\tupleargk[\argsubscript][\argsuperscript]}}
		{\argdef{#3}{\tupleargl[\argsubscript][\argsuperscript]}}
		{\argdef{#4}{\tupleargm[\argsubscript][\argsuperscript]}}
		{\argdef{#5}{\tupleargn[\argsubscript][\argsuperscript]}}
		{\argdef{#6}{\tupleargo[\argsubscript][\argsuperscript]}}
		{\argdef{#7}{\tupleargp[\argsubscript][\argsuperscript]}}
		{\argdef{#8}{\tupleargq[\argsubscript][\argsuperscript]}}
		{\argdef{#9}{\tupleargr[\argsubscript][\argsuperscript]}}%
	}

\newcommand{\set}[2]
	{\ensuremath{\argint{\{}{\argext{#1}{\allowbreak:\allowbreak}{#2}}{\}}}}

\newcommand{\pow}[1]
	{\ensuremath{2^{#1}}}

\newcommand{\card}[1]
	{\mthempty{\argint{\vert}{#1}{\vert}}}

\newcommand{\dom}
	{\mthargfun{dom}}

\newcommand{\rng}
	{\mthargfun{rng}}

\newcommand{\rst}
	{\mthempty{\upharpoonright}}

\newcommandx{\pto}[2][1=, 2=]
	{\ensuremath{\rightharpoonup}}

\newcommandx{\cto}[2][1=, 2=]
	{\:\mthempty{\to}[#1][#2]\:}
\newcommandx{\cpto}[2][1=, 2=]
	{\:\mthempty{\pto}[#1][#2]\:}

\newcommand{\emptyfun}
	{\mthempty{\varnothing}}

\newcommand{\ATheta}
	{\mthargset{\Theta}}

\newcommand{\AOmicron}
	{\mthargset{O}}

\newcommand{\SetN}
	{\mthset[2]{N}}

\newcommand{\SetR}
	{\mthset[2]{R}}

\newcommand{\numcc}[2]
	{\mthempty{[\argb{#1}{#2}]}}
\newcommand{\numco}[2]
	{\mthempty{[\argb{#1}{#2}[\:\!}}

\newcommand{\floor}[1]
	{\mthempty{\left\lfloor{#1}\right\rfloor}}

\DeclareRobustCommand{\min}
	{\mthfun{min}}
\DeclareRobustCommand{\max}
	{\mthfun{max}}

\newcommand{\argset}{Ar}
\newcommandx{\ArgSet}[3][1=, 2=, 3=]
	{\mthset{\argset#3}[#1][#2]}

\newcommand{\argsym}{a}
\newcommandx{\argSym}[3][1=, 2=, 3=]
	{\mthsym{\argsym#3}[#1][#2]}

\newcommand{\argelm}{a}
\newcommandx{\argElm}[3][1=, 2=, 3=]
	{\mthelm{\argelm#3}[#1][#2]}

\newcommand{\relset}{Rl}
\newcommandx{\RelSet}[3][1=, 2=, 3=]
	{\mthset{\relset#3}[#1][#2]}

\newcommand{\relsym}{r}
\newcommandx{\relSym}[3][1=, 2=, 3=]
	{\mthsym{\relsym#3}[#1][#2]}

\newcommand{\relelm}{r}
\newcommandx{\relElm}[3][1=, 2=, 3=]
	{\mthelm{\relelm#3}[#1][#2]}

\newcommand{\argfun}{ar}
\newcommandx{\argFun}[4][1=, 2=, 3=, 4=]
	{\mthargfun{\argfun#4}[#1][#2]{#3}}

\newcommand{\lansig}{LS}
\newcommandx{\LanSig}[5][1=, 2=, 3=, 4=, 5=]
	{\txtargname{\lansig#5{\small\argint{$[$}{#1}{$]$}}}[#2][#3]{#4}\xspace}

\newcommand{\lansigcls}{LS}
\newcommandx{\LanSigCls}[5][1=, 2=, 3=, 4=, 5=]
	{\mthset[#5]{\lansigcls#4\text{\txtname{\small\argint{$[$}{#1}{$]$}}}}[#2]%
	[#3]}

\newcommand{\domset}{Dm}
\newcommandx{\DomSet}[3][1=, 2=, 3=]
	{\mthset{\domset#3}[#1][#2]}

\newcommand{\domsym}{d}
\newcommandx{\domSym}[3][1=, 2=, 3=]
	{\mthsym{\domsym#3}[#1][#2]}

\newcommand{\domelm}{d}
\newcommandx{\domElm}[3][1=, 2=, 3=]
	{\mthelm{\domelm#3}[#1][#2]}

\newcommand{\relfun}{rl}
\newcommandx{\relFun}[4][1=, 2=, 3=, 4=]
	{\mthargfun{\relfun#4}[#1][#2]{#3}}

\newcommand{\relstr}{RS}
\newcommandx{\RelStr}[5][1=, 2=, 3=, 4=, 5=]
	{\txtargname{\relstr#5{\small\argint{$[$}{#1}{$]$}}}[#2][#3]{#4}\xspace}

\newcommand{\relstrcls}{RS}
\newcommandx{\RelStrCls}[5][1=, 2=, 3=, 4=, 5=]
	{\mthset[#5]{\relstrcls#4\text{\txtname{\small\argint{$[$}{#1}{$]$}}}}[#2]%
	[#3]}

\newcommandx{\ordFun}[3][1=, 2=, 3=]
	{\mthempty{\argint{\left\vert}{#3}{\right\vert}}[#1][#2]}

\newcommandx{\sizFun}[3][1=, 2=, 3=]
	{\mthempty{\argint{\left\Vert}{#3}{\right\Vert}}[#1][#2]}

\newcommand{\verset}{Vr}
\newcommandx{\VerSet}[3][1=, 2=, 3=]
	{\mthset{\verset#3}[#1][#2]}

\newcommand{\versym}{v}
\newcommandx{\verSym}[3][1=, 2=, 3=]
	{\mthsym{\versym#3}[#1][#2]}

\newcommand{\verelm}{v}
\newcommandx{\verElm}[3][1=, 2=, 3=]
	{\mthelm{\verelm#3}[#1][#2]}

\newcommand{\edgrel}{Ed}
\newcommandx{\EdgRel}[3][1=, 2=, 3=]
	{\mthrel{\edgrel#3}[#1][#2]}

\newcommand{\edgsym}{e}
\newcommandx{\edgSym}[3][1=, 2=, 3=]
	{\mthsym{\edgsym#3}[#1][#2]}

\newcommand{\edgelm}{e}
\newcommandx{\edgElm}[3][1=, 2=, 3=]
	{\mthelm{\edgelm#3}[#1][#2]}

\newcommand{\orgfun}{or}
\newcommandx{\orgFun}[4][1=, 2=, 3=, 4=]
	{\mthargfun{\orgfun#4}[#1][#2]{#3}}

\newcommand{\desfun}{ds}
\newcommandx{\desFun}[4][1=, 2=, 3=, 4=]
	{\mthargfun{\desfun#4}[#1][#2]{#3}}

\newcommand{\grp}{Gr}
\newcommandx{\Grp}[5][1=, 2=, 3=, 4=, 5=]
	{\txtargname{\grp#5{\small\argint{$[$}{#1}{$]$}}}[#2][#3]{#4}\xspace}

\newcommand{\grpcls}{Gr}
\newcommandx{\GrpCls}[5][1=, 2=, 3=, 4=, 5=]
	{\mthset[#5]{\grpcls#4\text{\small\txtname{\argint{$[$}{#1}{$]$}}}}[#2][#3]}

\newcommand{\pthset}{Pth}
\newcommandx{\PthSet}[3][1=, 2=, 3=]
	{\mthset{\pthset#3}[#1][#2]}

\newcommand{\pthsym}{\pi}
\newcommandx{\pthSym}[3][1=, 2=, 3=]
	{\mthsym{\pthsym#3}[#1][#2]}

\newcommand{\pthelm}{\pi}
\newcommandx{\pthElm}[3][1=, 2=, 3=]
	{\mthelm{\pthelm#3}[#1][#2]}

\newcommand{\apset}{AP}
\newcommandx{\APSet}[3][1=, 2=, 3=]
	{\mthset{\apset#3}[#1][#2]}

\newcommand{\apsym}{p}
\newcommandx{\apSym}[3][1=, 2=, 3=]
	{\mthsym{\apsym#3}[#1][#2]}

\newcommand{\apelm}{p}
\newcommandx{\apElm}[3][1=, 2=, 3=]
	{\mthelm{\apelm#3}[#1][#2]}

\newcommand{\apfun}{ap}
\newcommandx{\apFun}[4][1=, 2=, 3=, 4=]
	{\mthargfun{\apfun#4}[#1][#2]{#3}}

\newcommand{\labgrp}{L\grp}
\newcommandx{\LabGrp}[5][1=, 2=, 3=, 4=, 5=]
	{\txtargname{\labgrp#5{\small\argint{$[$}{#1}{$]$}}}[#2][#3]{#4}\xspace}

\newcommand{\labgrpcls}{L\grpcls}
\newcommandx{\LabGrpCls}[5][1=, 2=, 3=, 4=, 5=]
	{\mthset[#5]{\labgrpcls#4\text{\small\txtname{\argint{$[$}{#1}{$]$}}}}[#2]%
	[#3]}

\newcommand{\trcset}{Trc}
\newcommandx{\TrcSet}[3][1=, 2=, 3=]
	{\mthset{\trcset#3}[#1][#2]}

\newcommand{\trcsym}{\varrho}
\newcommandx{\trcSym}[3][1=, 2=, 3=]
	{\mthsym{\trcsym#3}[#1][#2]}

\newcommand{\trcelm}{\varrho}
\newcommandx{\trcElm}[3][1=, 2=, 3=]
	{\mthelm{\trcelm#3}[#1][#2]}

\newcommand{\colset}{Cl}
\newcommandx{\ColSet}[3][1=, 2=, 3=]
	{\mthset{\colset#3}[#1][#2]}

\newcommand{\colsym}{c}
\newcommandx{\colSym}[3][1=, 2=, 3=]
	{\mthsym{\colsym#3}[#1][#2]}

\newcommand{\colelm}{c}
\newcommandx{\colElm}[3][1=, 2=, 3=]
	{\mthelm{\colelm#3}[#1][#2]}

\newcommand{\colfun}{cl}
\newcommandx{\colFun}[4][1=, 2=, 3=, 4=]
	{\mthargfun{\colfun#4}[#1][#2]{#3}}

\newcommand{\colgrp}{C\grp}
\newcommandx{\ColGrp}[5][1=, 2=, 3=, 4=, 5=]
	{\txtargname{\colgrp#5{\small\argint{$[$}{#1}{$]$}}}[#2][#3]{#4}\xspace}

\newcommand{\colgrpcls}{C\grpcls}
\newcommandx{\ColGrpCls}[5][1=, 2=, 3=, 4=, 5=]
	{\mthset[#5]{\colgrpcls#4\text{\small\txtname{\argint{$[$}{#1}{$]$}}}}[#2]%
	[#3]}

\newcommand{\wghset}{Wg}
\newcommandx{\WghSet}[3][1=, 2=, 3=]
	{\mthset{\wghset#3}[#1][#2]}

\newcommand{\wghsym}{w}
\newcommandx{\wghSym}[3][1=, 2=, 3=]
	{\mthsym{\wghsym#3}[#1][#2]}

\newcommand{\wghelm}{w}
\newcommandx{\wghElm}[3][1=, 2=, 3=]
	{\mthelm{\wghelm#3}[#1][#2]}

\newcommand{\wghfun}{wg}
\newcommandx{\wghFun}[4][1=, 2=, 3=, 4=]
	{\mthargfun{\wghfun#4}[#1][#2]{#3}}

\newcommand{\wghgrp}{W\grp}
\newcommandx{\WghGrp}[5][1=, 2=, 3=, 4=, 5=]
	{\txtargname{\wghgrp#5{\small\argint{$[$}{#1}{$]$}}}[#2][#3]{#4}\xspace}

\newcommand{\wghgrpcls}{W\grpcls}
\newcommandx{\WghGrpCls}[5][1=, 2=, 3=, 4=, 5=]
	{\mthset[#5]{\wghgrpcls#4\text{\small\txtname{\argint{$[$}{#1}{$]$}}}}[#2]%
	[#3]}

\newcommand{\gamkin}{2PT}

\newcommand{\plrset}{Pl}
\newcommandx{\PlrSet}[3][1=, 2=, 3=]
	{\mthset{\plrset#3}[#1][#2]}

\newcommand{\plrsym}{p}
\newcommandx{\plrSym}[3][1=, 2=, 3=]
	{\mthsym{\plrsym#3}[#1][#2]}

\newcommand{\plrelm}{p}
\newcommandx{\plrElm}[3][1=, 2=, 3=]
	{\mthelm{\plrelm#3}[#1][#2]}

\newcommand{\agnset}{Ag}
\newcommandx{\AgnSet}[3][1=, 2=, 3=]
	{\mthset{\agnset#3}[#1][#2]}

\newcommand{\agnsym}{a}
\newcommandx{\agnSym}[3][1=, 2=, 3=]
	{\mthsym{\agnsym#3}[#1][#2]}

\newcommand{\agnelm}{a}
\newcommandx{\agnElm}[3][1=, 2=, 3=]
	{\mthelm{\agnelm#3}[#1][#2]}

\newcommand{\movset}{Mv}
\newcommandx{\MovSet}[3][1=, 2=, 3=]
	{\mthset{\movset#3}[#1][#2]}

\newcommand{\movrel}{Mv}
\newcommandx{\MovRel}[3][1=, 2=, 3=]
	{\mthrel{\movrel#3}[#1][#2]}

\newcommand{\movsym}{m}
\newcommandx{\movSym}[3][1=, 2=, 3=]
	{\mthsym{\movsym#3}[#1][#2]}

\newcommand{\movelm}{m}
\newcommandx{\movElm}[3][1=, 2=, 3=]
	{\mthelm{\movelm#3}[#1][#2]}

\newcommand{\actset}{Ac}
\newcommandx{\ActSet}[3][1=, 2=, 3=]
	{\mthset{\actset#3}[#1][#2]}

\newcommand{\actrel}{Ac}
\newcommandx{\ActRel}[3][1=, 2=, 3=]
	{\mthrel{\actrel#3}[#1][#2]}

\newcommand{\actsym}{c}
\newcommandx{\actSym}[3][1=, 2=, 3=]
	{\mthsym{\actsym#3}[#1][#2]}

\newcommand{\actelm}{c}
\newcommandx{\actElm}[3][1=, 2=, 3=]
	{\mthelm{\actelm#3}[#1][#2]}

\newcommand{\decset}{Dc}
\newcommandx{\DecSet}[3][1=, 2=, 3=]
	{\mthset{\decset#3}[#1][#2]}

\newcommand{\decsym}{\delta}
\newcommandx{\decSym}[4][1=, 2=, 3=, 4=]
	{\mthargfun{\decsym#4}[#1][#2]{#3}}

\newcommand{\decelm}{\delta}
\newcommandx{\decElm}[4][1=, 2=, 3=, 4=]
	{\mthargfun{\decelm#4}[#1][#2]{#3}}

\newcommand{\posset}{Ps}
\newcommandx{\PosSet}[3][1=, 2=, 3=]
	{\mthset{\posset#3}[#1][#2]}
\newcommand{\fpossub}{0}
\newcommandx{\FPosSet}[3][1=, 2=, 3=]
	{\mthset{\posset#3}[\fpossub#1][#2]}
\newcommand{\spossub}{1}
\newcommandx{\SPosSet}[3][1=, 2=, 3=]
	{\mthset{\posset#3}[\spossub#1][#2]}

\newcommand{\possym}{v}
\newcommandx{\posSym}[3][1=, 2=, 3=]
	{\mthsym{\possym#3}[#1][#2]}
\newcommandx{\fposSym}[1][1=]
	{\posSym[\fpossub#1]}
\newcommandx{\sposSym}[1][1=]
	{\posSym[\spossub#1]}
\newcommand{\ipossub}{I}
\newcommandx{\iposSym}[1][1=]
	{\posSym[\ipossub#1]}

\newcommand{\poselm}{v}
\newcommandx{\posElm}[3][1=, 2=, 3=]
	{\mthelm{\poselm#3}[#1][#2]}
\newcommandx{\fposElm}[1][1=]
	{\posElm[\fpossub#1]}
\newcommandx{\sposElm}[1][1=]
	{\posElm[\spossub#1]}
\newcommandx{\iposElm}[1][1=]
	{\posElm[\ipossub#1]}

\newcommand{\sttset}{St}
\newcommandx{\SttSet}[3][1=, 2=, 3=]
	{\mthset{\sttset#3}[#1][#2]}
\newcommand{\fsttsub}{0}
\newcommandx{\FSttSet}[3][1=, 2=, 3=]
	{\mthset{\sttset#3}[\fsttsub#1][#2]}
\newcommand{\ssttsub}{1}
\newcommandx{\SSttSet}[3][1=, 2=, 3=]
	{\mthset{\sttset#3}[\ssttsub#1][#2]}

\newcommand{\sttsym}{s}
\newcommandx{\sttSym}[3][1=, 2=, 3=]
	{\mthsym{\sttsym#3}[#1][#2]}
\newcommandx{\fsttSym}[1][1=]
	{\sttSym[\fsttsub#1]}
\newcommandx{\ssttSym}[1][1=]
	{\sttSym[\ssttsub#1]}
\newcommand{\isttsub}{I}
\newcommandx{\isttSym}[1][1=]
	{\sttSym[\isttsub#1]}

\newcommand{\sttelm}{s}
\newcommandx{\sttElm}[3][1=, 2=, 3=]
	{\mthelm{\sttelm#3}[#1][#2]}
\newcommandx{\fsttElm}[1][1=]
	{\sttElm[\fsttsub#1]}
\newcommandx{\ssttElm}[1][1=]
	{\sttElm[\ssttsub#1]}
\newcommandx{\isttElm}[1][1=]
	{\sttElm[\isttsub#1]}

\newcommand{\plrfun}{pl}
\newcommandx{\plrFun}[4][1=, 2=, 3=, 4=]
	{\mthargfun{\plrfun#4}[#1][#2]{#3}}

\newcommand{\agnfun}{ag}
\newcommandx{\agnFun}[4][1=, 2=, 3=, 4=]
	{\mthargfun{\agnfun#4}[#1][#2]{#3}}

\newcommand{\movfun}{mv}
\newcommandx{\movFun}[4][1=, 2=, 3=, 4=]
	{\mthargfun{\movfun#4}[#1][#2]{#3}}

\newcommand{\actfun}{ac}
\newcommandx{\actFun}[4][1=, 2=, 3=, 4=]
	{\mthargfun{\actfun#4}[#1][#2]{#3}}

\newcommand{\decfun}{dc}
\newcommandx{\decFun}[4][1=, 2=, 3=, 4=]
	{\mthargfun{\decfun#4}[#1][#2]{#3}}

\newcommand{\trnfun}{tr}
\newcommandx{\trnFun}[4][1=, 2=, 3=, 4=]
	{\mthargfun{\trnfun#4}[#1][#2]{#3}}

\newcommand{\arn}{Ar}
\newcommandx{\Arn}[5][1=, 2=, 3=, 4=, 5=]
	{\txtargname{\arn#5{\small\argint{$[$}{#1}{$]$}}}[#2][#3]{#4}\xspace}

\newcommand{\arnname}{A}
\newcommand{\ArnName}
	{\mthname{\arnname}}

\newcommand{\arncls}{Ar}
\newcommandx{\ArnCls}[5][1=, 2=, 3=, 4=, 5=]
	{\mthset[#5]{\arncls#4\text{\small\txtname{\argint{$[$}{#1}{$]$}}}}[#2][#3]}

\newcommand{\ArnStr}[1][]
	{%
	\IfStrEqCase{\argdef{#1}{\gamkin}}
		{%
		{2PT}
			{\tuplecx{\FPosSet}{\SPosSet}{\MovRel}}%
		{MPC0}
			{\tupledx{\PlrSet}{\MovSet}{\PosSet}{\trnFun}}%
		{MPC1}
			{\tupleex{\PlrSet}{\MovSet}{\PosSet}{\decFun}{\trnFun}}%
		{MPC2}
			{\tuplefx{\PlrSet}{\MovSet}{\PosSet}{\plrFun}{\movFun}{\trnFun}}%
		{MPC3}
			{\tuplegx{\PlrSet}{\MovSet}{\PosSet}{\plrFun}{\movFun}{\decFun}{\trnFun}}%
		{2AT}
			{\tuplecx{\FSttSet}{\SSttSet}{\ActRel}}%
		{MAC0}
			{\tupledx{\AgnSet}{\ActSet}{\SttSet}{\trnFun}}%
		{MAC1}
			{\tupleex{\AgnSet}{\ActSet}{\SttSet}{\decFun}{\trnFun}}%
		{MAC2}
			{\tuplefx{\AgnSet}{\ActSet}{\SttSet}{\agnFun}{\actFun}{\trnFun}}%
		{MAC3}
			{\tuplegx{\AgnSet}{\ActSet}{\SttSet}{\agnFun}{\actFun}{\decFun}{\trnFun}}%
		}
		[\ensuremath{\clubsuit}]%
	}

\newcommand{\hstset}{Hst}
\newcommandx{\HstSet}[3][1=, 2=, 3=]
	{\mthset{\hstset#3}[#1][#2]}

\newcommand{\hstsym}{\rho}
\newcommandx{\hstSym}[3][1=, 2=, 3=]
	{\mthsym{\hstsym#3}[#1][#2]}

\newcommand{\hstelm}{\rho}
\newcommandx{\hstElm}[3][1=, 2=, 3=]
	{\mthelm{\hstelm#3}[#1][#2]}

\newcommand{\strset}{Str}
\newcommandx{\StrSet}[3][1=, 2=, 3=]
	{\mthset{\strset#3}[#1][#2]}

\newcommand{\strsym}{\sigma}
\newcommandx{\strSym}[4][1=, 2=, 3=, 4=]
	{\mthargfun{\strsym#4}[#1][#2]{#3}}

\newcommand{\strelm}{\sigma}
\newcommandx{\strElm}[4][1=, 2=, 3=, 4=]
	{\mthargfun{\strelm#4}[#1][#2]{#3}}

\newcommand{\prfset}{Prf}
\newcommandx{\PrfSet}[3][1=, 2=, 3=]
	{\mthset{\prfset#3}[#1][#2]}

\newcommand{\prfsym}{\xi}
\newcommandx{\prfSym}[4][1=, 2=, 3=, 4=]
	{\mthargfun{\prfsym#4}[#1][#2]{#3}}

\newcommandx{\prfElm}[4][1=, 2=, 3=, 4=]
	{\mthargfun{\prfsym#4}[#1][#2]{#3}}

\newcommand{\playfun}{play}
\newcommandx{\playFun}[4][1=, 2=, 3=, 4=]
	{\mthargfun{\playfun#4}[#1][#2]{#3}}

\newcommand{\labarn}{L\arn}
\newcommandx{\LabArn}[5][1=, 2=, 3=, 4=, 5=]
	{\txtargname{\labarn#5{\small\argint{$[$}{#1}{$]$}}}[#2][#3]{#4}\xspace}

\newcommand{\labarncls}{L\arncls}
\newcommandx{\LabArnCls}[5][1=, 2=, 3=, 4=, 5=]
	{\mthset[#5]{\labarncls#4\text{\small\txtname{\argint{$[$}{#1}{$]$}}}}[#2]%
	[#3]}

\newcommand{\colarn}{C\arn}
\newcommandx{\ColArn}[5][1=, 2=, 3=, 4=, 5=]
	{\txtargname{\colarn#5{\small\argint{$[$}{#1}{$]$}}}[#2][#3]{#4}\xspace}

\newcommand{\colarncls}{C\arncls}
\newcommandx{\ColArnCls}[5][1=, 2=, 3=, 4=, 5=]
	{\mthset[#5]{\colarncls#4\text{\small\txtname{\argint{$[$}{#1}{$]$}}}}[#2]%
	[#3]}

\newcommand{\wgharn}{W\arn}
\newcommandx{\WghArn}[5][1=, 2=, 3=, 4=, 5=]
	{\txtargname{\wgharn#5{\small\argint{$[$}{#1}{$]$}}}[#2][#3]{#4}\xspace}

\newcommand{\wgharncls}{W\arncls}
\newcommandx{\WghArnCls}[5][1=, 2=, 3=, 4=, 5=]
	{\mthset[#5]{\wgharncls#4\text{\small\txtname{\argint{$[$}{#1}{$]$}}}}[#2]%
	[#3]}

\newcommand{\winset}{Wn}
\newcommandx{\WinSet}[3][1=, 2=, 3=]
	{\mthset{\winset#3}[#1][#2]}

\newcommand{\prdset}{Pr}
\newcommandx{\PrdSet}[3][1=, 2=, 3=]
	{\mthset{\prdset#3}[#1][#2]}

\newcommand{\prdsym}{p}
\newcommandx{\prdSym}[3][1=, 2=, 3=]
	{\mthsym{\prdsym#3}[#1][#2]}

\newcommand{\prdelm}{p}
\newcommandx{\prdElm}[3][1=, 2=, 3=]
	{\mthelm{\prdelm#3}[#1][#2]}

\newcommand{\prdfun}{pr}
\newcommandx{\prdFun}[4][1=, 2=, 3=, 4=]
	{\mthargfun{\prdfun#4}[#1][#2]{#3}}

\newcommand{\extname}{E}
\newcommand{\ExtName}
	{\mthname{\extname}}

\newcommand{\extcls}{Ex}
\newcommandx{\ExtCls}[5][1=, 2=, 3=, 4=, 5=]
	{\mthset[#5]{\extcls#4\text{\small\txtname{\argint{$[$}{#1}{$]$}}}}[#2][#3]}

\newcommand{\conset}{Cn}
\newcommandx{\ConSet}[3][1=, 2=, 3=]
	{\mthset{\conset#3}[#1][#2]}

\newcommand{\consym}{\varphi}
\newcommandx{\conSym}[3][1=, 2=, 3=]
	{\mthsym{\consym#3}[#1][#2]}

\newcommand{\conelm}{\varphi}
\newcommandx{\conElm}[3][1=, 2=, 3=]
	{\mthelm{\conelm#3}[#1][#2]}

\newcommand{\schrel}{\models}
\newcommandx{\schRel}[4][1=, 2=, 3=, 4=]
	{\mthrel{\schrel#3}[#1][#2]}

\newcommand{\schcls}{Sc}
\newcommandx{\SchCls}[5][1=, 2=, 3=, 4=, 5=]
	{\mthset[#5]{\schcls#4\text{\small\txtname{\argint{$[$}{#1}{$]$}}}}[#2][#3]}

\newcommand{\gamname}{\Game}
\newcommand{\GamName}
	{\mthname{\gamname}}

\newcommand{\gamcls}{Gm}
\newcommandx{\GamCls}[5][1=, 2=, 3=, 4=, 5=]
	{\mthset[#5]{\gamcls#4\text{\small\txtname{\argint{$[$}{#1}{$]$}}}}[#2][#3]}

\newcommandx{\GamStr}[1][1=]
	{%
	\StrLeft{\argdef{#1}{\gamkin}}{2}[\optgamkin]%
	\IfStrEqCase{\optgamkin}
		{%
		{2P}
			{\gamstrauxtp}%
		{MP}
			{\gamstrauxmp}%
		{2A}
			{\gamstrauxta}%
		{MA}
			{\gamstrauxma}%
		}
		[\ensuremath{\clubsuit}]%
	}
\newcommandx{\gamstrauxtp}[5][1=, 2=, 3=, 4=, 5=]
	{\tuplecx{\ArnName}{\iposElm}{\WinSet}[#3][#4][#5][#1][#2]}
\newcommandx{\gamstrauxmp}[5][1=, 2=, 3=, 4=, 5=]
	{\tuplecx{\ExtName}{\iposElm}{\conElm}[#3][#4][#5][#1][#2]}
\newcommandx{\gamstrauxta}[5][1=, 2=, 3=, 4=, 5=]
	{\tuplecx{\ArnName}{\isttElm}{\WinSet}[#3][#4][#5][#1][#2]}
\newcommandx{\gamstrauxma}[5][1=, 2=, 3=, 4=, 5=]
	{\tuplecx{\ExtName}{\isttElm}{\conElm}[#3][#4][#5][#1][#2]}

\newcommand{\worset}{W}
\newcommandx{\WorSet}[3][1=, 2=, 3=]
	{\mthset{\worset#3}[#1][#2]}

\newcommand{\worsym}{w}
\newcommandx{\worSym}[3][1=, 2=, 3=]
	{\mthsym{\worsym#3}[#1][#2]}

\newcommand{\worelm}{w}
\newcommandx{\worElm}[3][1=, 2=, 3=]
	{\mthelm{\worelm#3}[#1][#2]}

\newcommand{\trnrel}{R}
\newcommandx{\TrnRel}[3][1=, 2=, 3=]
	{\mthrel{\trnrel#3}[#1][#2]}

\newcommand{\trnsym}{r}
\newcommandx{\trnSym}[3][1=, 2=, 3=]
	{\mthsym{\trnsym#3}[#1][#2]}

\newcommand{\trnelm}{r}
\newcommandx{\trnElm}[3][1=, 2=, 3=]
	{\mthelm{\trnelm#3}[#1][#2]}

\newcommand{\labfun}{L}
\newcommandx{\labFun}[4][1=, 2=, 3=, 4=]
	{\mthargfun{\labfun#4}[#1][#2]{#3}}

\newcommand{\krpstr}{KS}
\newcommandx{\KrpStr}[5][1=, 2=, 3=, 4=, 5=]
	{\txtargname{\krpstr#5{\small\argint{$[$}{#1}{$]$}}}[#2][#3]{#4}\xspace}

\newcommand{\krpstrcls}{KS}
\newcommandx{\KrpStrCls}[5][1=, 2=, 3=, 4=, 5=]
	{\mthset[#5]{\krpstrcls#4\text{\small\txtname{\argint{$[$}{#1}{$]$}}}}[#2]%
	[#3]}

\newcommand{\trkset}{Trk}
\newcommandx{\TrkSet}[3][1=, 2=, 3=]
	{\mthset{\trkset#3}[#1][#2]}

\newcommand{\trksym}{\rho}
\newcommandx{\trkSym}[3][1=, 2=, 3=]
	{\mthsym{\trksym#3}[#1][#2]}

\newcommand{\trkelm}{\rho}
\newcommandx{\trkElm}[3][1=, 2=, 3=]
	{\mthelm{\trkelm#3}[#1][#2]}

\newcommand{\krptree}{KT}
\newcommandx{\KrpTree}[5][1=, 2=, 3=, 4=, 5=]
	{\txtargname{\krptree#5{\small\argint{$[$}{#1}{$]$}}}[#2][#3]{#4}\xspace}

\newcommand{\krptreecls}{KT}
\newcommandx{\KrpTreeCls}[5][1=, 2=, 3=, 4=, 5=]
	{\mthset[#5]{\krptreecls#4\text{\small\txtname{\argint{$[$}{#1}{$]$}}}}[#2]%
	[#3]}

\newcommand{\dirset}{Dir}
\newcommandx{\DirSet}[3][1=, 2=, 3=]
	{\mthset{\dirset#3}[#1][#2]}

\newcommand{\dirsym}{d}
\newcommandx{\dirSym}[3][1=, 2=, 3=]
	{\mthsym{\dirsym#3}[#1][#2]}

\newcommand{\direlm}{d}
\newcommandx{\dirElm}[3][1=, 2=, 3=]
	{\mthelm{\direlm#3}[#1][#2]}

\newcommand{\unwfun}{unw}
\newcommandx{\unwFun}[4][1=, 2=, 3=, 4=]
	{\mthargfun{\unwfun#4}[#1][#2]{#3}}

\newcommand{\congamstrkin}{MAC0}

\newcommand{\congamstr}{CGS}
\newcommandx{\ConGamStr}[5][1=, 2=, 3=, 4=, 5=]
	{\txtargname{\congamstr#5{\small\argint{$[$}{#1}{$]$}}}[#2][#3]{#4}\xspace}

\newcommandx{\ConGamStrCls}[5][1=, 2=, 3=, 4=, 5=]
	{\mthset[#5]{\arncls#4\text{\small\txtname{\argint{$[$}{#1}{$]$}}}}[#2][#3]}

\newcommandx{\ConGamStrStr}[1][1=]
	{%
	\IfStrEqCase{\argdef{#1}{\congamstrkin}}
		{%
		{IP}
			{\congamstrstrauxip}%
		{2PT}
			{\congamstrstrauxpt}%
		{MPC0}
			{\congamstrstrauxpca}%
		{MPC1}
			{\congamstrstrauxpcb}%
		{MPC2}
			{\congamstrstrauxpcc}%
		{MPC3}
			{\congamstrstrauxpcd}%
		{IA}
			{\congamstrstrauxia}%
		{2AT}
			{\congamstrstrauxat}%
		{MAC0}
			{\congamstrstrauxaca}%
		{MAC1}
			{\congamstrstrauxacb}%
		{MAC2}
			{\congamstrstrauxacc}%
		{MAC3}
			{\congamstrstrauxacd}%
		}
		[\ensuremath{\clubsuit}]%
	}
\newcommandx{\congamstrstrauxip}[3][1=, 2=, 3=]
	{%
	\def\defini{#1}%
	\def\defsubscr{#2}%
	\def\defsupscr{#3}%
	\congamstrstrauxxip%
	}
\newcommandx{\congamstrstrauxxip}[3][1=, 2=, 3=]
	{%
	\tupledx{\ArnName}{\APSet}{\apFun}{\iposElm}%
		[\defsubscr][\defsupscr][#1][#2][#3][\defini]%
	}
\newcommandx{\congamstrstrauxpt}[3][1=, 2=, 3=]
	{%
	\def\defini{#1}%
	\def\defsubscr{#2}%
	\def\defsupscr{#3}%
	\congamstrstrauxxpt%
	}
\newcommandx{\congamstrstrauxxpt}[5][1=, 2=, 3=, 4=, 5=]
	{%
	\tuplefx{\APSet}{\FPosSet}{\SPosSet}{\MovRel}{\apFun}{\iposElm}%
		[\defsubscr][\defsupscr][#1][#2][#3][#4][#5][\defini]%
	}
\newcommandx{\congamstrstrauxpca}[3][1=, 2=, 3=]
	{%
	\def\defini{#1}%
	\def\defsubscr{#2}%
	\def\defsupscr{#3}%
	\congamstrstrauxxpca%
	}
\newcommandx{\congamstrstrauxxpca}[6][1=, 2=, 3=, 4=, 5=, 6=]
	{%
	\tuplegx{\APSet}{\PlrSet}{\MovSet}{\PosSet}{\trnFun}{\apFun}{\iposElm}%
		[\defsubscr][\defsupscr][#1][#2][#3][#4][#5][#6][\defini]%
	}
\newcommandx{\congamstrstrauxpcb}[3][1=, 2=, 3=]
	{%
	\def\defini{#1}%
	\def\defsubscr{#2}%
	\def\defsupscr{#3}%
	\congamstrstrauxxpcb%
	}
\newcommandx{\congamstrstrauxxpcb}[7][1=, 2=, 3=, 4=, 5=, 6=, 7=]
	{%
	\tuplehx{\APSet}{\PlrSet}{\MovSet}{\PosSet}{\decFun}{\trnFun}{\apFun}%
		{\iposElm}%
		[\defsubscr][\defsupscr][#1][#2][#3][#4][#5][#6][#7][\defini]%
	}
\newcommandx{\congamstrstrauxpcc}[3][1=, 2=, 3=]
	{%
	\def\defini{#1}%
	\def\defsubscr{#2}%
	\def\defsupscr{#3}%
	\congamstrstrauxxpcc%
	}
\newcommandx{\congamstrstrauxxpcc}[8][1=, 2=, 3=, 4=, 5=, 6=, 7=, 8=]
	{%
	\tupleix{\APSet}{\PlrSet}{\MovSet}{\PosSet}{\plrFun}{\movFun}{\trnFun}%
		{\apFun}{\iposElm}%
		[\defsubscr][\defsupscr][#1][#2][#3][#4][#5][#6][#7][#8][\defini]%
	}
\newcommandx{\congamstrstrauxpcd}[3][1=, 2=, 3=]
	{%
	\def\defini{#1}%
	\def\defsubscr{#2}%
	\def\defsupscr{#3}%
	\congamstrstrauxxpcd%
	}
\newcommandx{\congamstrstrauxxpcd}[9][1=, 2=, 3=, 4=, 5=, 6=, 7=, 8=, 9=]
	{%
	\tuplejx{\APSet}{\PlrSet}{\MovSet}{\PosSet}{\plrFun}{\movFun}{\decFun}%
	{\trnFun}{\apFun}{\iposElm}%
		[\defsubscr][\defsupscr][#1][#2][#3][#4][#5][#6][#7][#8][#9][\defini]%
	}
\newcommandx{\congamstrstrauxia}[3][1=, 2=, 3=]
	{%
	\def\defini{#1}%
	\def\defsubscr{#2}%
	\def\defsupscr{#3}%
	\congamstrstrauxxia%
	}
\newcommandx{\congamstrstrauxxia}[3][1=, 2=, 3=]
	{%
	\tupledx{\ArnName}{\APSet}{\apFun}{\isttElm}%
		[\defsubscr][\defsupscr][#1][#2][#3][\defini]%
	}
\newcommandx{\congamstrstrauxat}[3][1=, 2=, 3=]
	{%
	\def\defini{#1}%
	\def\defsubscr{#2}%
	\def\defsupscr{#3}%
	\congamstrstrauxxat%
	}
\newcommandx{\congamstrstrauxxat}[5][1=, 2=, 3=, 4=, 5=]
	{%
	\tuplefx{\APSet}{\FSttSet}{\SSttSet}{\ActRel}{\apFun}{\isttElm}%
		[\defsubscr][\defsupscr][#1][#2][#3][#4][#5][\defini]%
	}
\newcommandx{\congamstrstrauxaca}[3][1=, 2=, 3=]
	{%
	\def\defini{#1}%
	\def\defsubscr{#2}%
	\def\defsupscr{#3}%
	\congamstrstrauxxaca%
	}
\newcommandx{\congamstrstrauxxaca}[6][1=, 2=, 3=, 4=, 5=, 6=]
	{%
	\tuplegx{\APSet}{\AgnSet}{\ActSet}{\SttSet}{\trnFun}{\apFun}{\isttElm}%
		[\defsubscr][\defsupscr][#1][#2][#3][#4][#5][#6][\defini]%
	}
\newcommandx{\congamstrstrauxacb}[3][1=, 2=, 3=]
	{%
	\def\defini{#1}%
	\def\defsubscr{#2}%
	\def\defsupscr{#3}%
	\congamstrstrauxxacb%
	}
\newcommandx{\congamstrstrauxxacb}[7][1=, 2=, 3=, 4=, 5=, 6=, 7=]
	{%
	\tuplehx{\APSet}{\AgnSet}{\ActSet}{\SttSet}{\decFun}{\trnFun}{\apFun}%
		{\isttElm}%
		[\defsubscr][\defsupscr][#1][#2][#3][#4][#5][#6][#7][\defini]%
	}
\newcommandx{\congamstrstrauxacc}[3][1=, 2=, 3=]
	{%
	\def\defini{#1}%
	\def\defsubscr{#2}%
	\def\defsupscr{#3}%
	\congamstrstrauxxacc%
	}
\newcommandx{\congamstrstrauxxacc}[8][1=, 2=, 3=, 4=, 5=, 6=, 7=, 8=]
	{%
	\tupleix{\APSet}{\AgnSet}{\ActSet}{\SttSet}{\agnFun}{\actFun}{\trnFun}%
		{\apFun}{\isttElm}%
		[\defsubscr][\defsupscr][#1][#2][#3][#4][#5][#6][#7][#8][\defini]%
	}
\newcommandx{\congamstrstrauxacd}[3][1=, 2=, 3=]
	{%
	\def\defini{#1}%
	\def\defsubscr{#2}%
	\def\defsupscr{#3}%
	\congamstrstrauxxacd%
	}
\newcommandx{\congamstrstrauxxacd}[9][1=, 2=, 3=, 4=, 5=, 6=, 7=, 8=, 9=]
	{%
	\tuplejx{\APSet}{\AgnSet}{\ActSet}{\SttSet}{\agnFun}{\actFun}{\decFun}%
		{\trnFun}{\apFun}{\isttElm}%
		[\defsubscr][\defsupscr][#1][#2][#3][#4][#5][#6][#7][#8][#9][\defini]%
	}

\newcommand{\trntabkin}{D}

\newcommand{\symset}{Sm}
\newcommandx{\SymSet}[3][1=, 2=, 3=]
	{\mthset{\symset#3}[#1][#2]}

\newcommand{\symsym}{\ell}
\newcommandx{\symSym}[3][1=, 2=, 3=]
	{\mthsym{\symsym#3}[#1][#2]}

\newcommand{\symelm}{\ell}
\newcommandx{\symElm}[3][1=, 2=, 3=]
	{\mthelm{\symelm#3}[#1][#2]}

\newcommand{\DSttSet}[1][]
	{\SttSet[\Delta#1]}

\newcommand{\ESttSet}[1][]
	{\SttSet[\exists#1]}

\newcommand{\ASttSet}[1][]
	{\SttSet[\forall#1]}

\newcommand{\trntab}{tt}
\newcommandx{\TrnTab}[5][1=, 2=, 3=, 4=, 5=]
	{\txtargname{\trntab#5{\small\argint{$[$}{#1}{$]$}}}[#2][#3]{#4}\xspace}

\newcommand{\trntabcls}{TT}
\newcommandx{\TrnTabCls}[5][1=, 2=, 3=, 4=, 5=]
	{\mthset[#5]{\trntabcls#4\text{\txtname{\small\argint{$[$}{#1}{$]$}}}}[#2]%
	[#3]}

\newcommand{\TrnTabStr}[1][]
	{%
	\IfStrEqCase{\argdef{#1}{\trntabkin}}
		{%
		{D}{\tuplecx{\SymSet}{\SttSet}{\trnFun}}%
		{N}{\tupledx{\SymSet}{\DSttSet}{\ESttSet}{\trnFun}}%
		{U}{\tupledx{\SymSet}{\DSttSet}{\ASttSet}{\trnFun}}%
		{A}{\tupleex{\SymSet}{\DSttSet}{\ESttSet}{\ASttSet}{\trnFun}}%
		}
		[\ensuremath{\clubsuit}]%
	}

\newcommandx{\PC}[5][1=, 2=, 3=, 4=, 5=]
	{\txtargname{PC#5{\small\argint{$[$}{#1}{$]$}}}[#2][#3]{#4}\xspace}

\newcommand{\Tt}
	{\mthsym{t}}

\newcommand{\Ff}
	{\mthsym{f}}

\newcommand{\bcset}{BC}
\newcommandx{\BCSet}[4][1=, 2=, 3=, 4=]
	{\mthset[3]{\bcset#4}[#1][#2]{#3}}

\newcommand{\bcelm}{\eta}
\newcommandx{\bcElm}[3][1=, 2=, 3=]
	{\mthelm{\bcelm#3}[#1][#2]}

\newcommand{\acset}{AC}
\newcommandx{\ACSet}[4][1=, 2=, 3=, 4=]
	{\mthset[3]{\acset#4}[#1][#2]{#3}}

\newcommand{\acelm}{\eta}
\newcommandx{\acElm}[3][1=, 2=, 3=]
	{\mthelm{\acelm#3}[#1][#2]}

\newcommandx{\QBF}[5][1=, 2=, 3=, 4=, 5=]
	{\txtargname{QBF#5{\small\argint{$[$}{#1}{$]$}}}[#2][#3]{#4}\xspace}

\newcommandx{\FOL}[5][1=, 2=, 3=, 4=, 5=]
	{\txtargname{FOL#5{\small\argint{$[$}{#1}{$]$}}}[#2][#3]{#4}\xspace}

\newcommand{\varset}{Vr}
\newcommandx{\VarSet}[3][1=, 2=, 3=]
	{\mthset{\varset#3}[#1][#2]}

\newcommand{\varsym}{x}
\newcommandx{\varSym}[3][1=, 2=, 3=]
	{\mthsym{\varsym#3}[#1][#2]}

\newcommand{\varelm}{x}
\newcommandx{\varElm}[3][1=, 2=, 3=]
	{\mthelm{\varelm#3}[#1][#2]}

\newcommand{\varfun}{vr}
\newcommandx{\varFun}[4][1=, 2=, 3=, 4=]
	{\mthargfun{\varfun#4}[#1][#2]{#3}}

\newcommand{\qntset}{Qn}
\newcommandx{\QntSet}[3][1=, 2=, 3=]
	{\mthset{\qntset#3}[#1][#2]}

\newcommand{\qntsym}{\wp}
\newcommandx{\qntSym}[3][1=, 2=, 3=]
	{\mthsym{\qntsym#3}[#1][#2]}

\newcommand{\qntelm}{\wp}
\newcommandx{\qntElm}[3][1=, 2=, 3=]
	{\mthelm{\qntelm#3}[#1][#2]}

\newcommand{\qntfun}{qnt}
\newcommandx{\qntFun}[4][1=, 2=, 3=, 4=]
	{\mthargfun{\qntfun#4}[#1][#2]{#3}}

\newcommand{\bndset}{Bn}
\newcommandx{\BndSet}[3][1=, 2=, 3=]
	{\mthset{\bndset#3}[#1][#2]}

\newcommand{\bndsym}{\flat}
\newcommandx{\bndSym}[3][1=, 2=, 3=]
	{\mthsym{\bndsym#3}[#1][#2]}

\newcommand{\bndelm}{\flat}
\newcommandx{\bndElm}[3][1=, 2=, 3=]
	{\mthelm{\bndelm#3}[#1][#2]}

\newcommand{\bndfun}{bnd}
\newcommandx{\bndFun}[4][1=, 2=, 3=, 4=]
	{\mthargfun{\bndfun#4}[#1][#2]{#3}}

\newcommand{\depset}{\Delta}
\newcommandx{\DepSet}[3][1=, 2=, 3=]
	{\mthset{\depset#3}[#1][#2]}

\newcommand{\denfun}{den}
\newcommandx{\denFun}[4][1=, 2=, 3=, 4=]
	{\mthargfun{\denfun#4}[#1][#2]{#3}}

\newcommand{\asgset}{Asg}
\newcommandx{\AsgSet}[3][1=, 2=, 3=]
	{\mthset{\asgset#3}[#1][#2]}

\newcommand{\asgfun}{\chi}
\newcommandx{\asgFun}[4][1=, 2=, 3=, 4=]
	{\mthargfun{\asgfun#4}[#1][#2]{#3}}

\newcommand{\smset}{SM}
\newcommandx{\SMSet}[3][1=, 2=, 3=]
	{\mthset{\smset#3}[#1][#2]}

\newcommand{\smfun}{\delta}
\newcommandx{\smFun}[4][1=, 2=, 3=, 4=]
	{\mthargfun{\smfun#4}[#1][#2]{#3}}

\newcommand{\cmset}{CM}
\newcommandx{\CMSet}[3][1=, 2=, 3=]
	{\mthset{\cmset#3}[#1][#2]}

\newcommand{\cmfun}{\gamma}
\newcommandx{\cmFun}[4][1=, 2=, 3=, 4=]
	{\mthargfun{\cmfun#4}[#1][#2]{#3}}

\newcommand{\schset}{Sch}
\newcommandx{\SchSet}[3][1=, 2=, 3=]
	{\mthset{\schset#3}[#1][#2]}

\newcommand{\schsym}{\sigma}
\newcommandx{\schSym}[3][1=, 2=, 3=]
	{\mthsym{\schsym#3}[#1][#2]}

\newcommand{\schelm}{\sigma}
\newcommandx{\schElm}[3][1=, 2=, 3=]
	{\mthelm{\schelm#3}[#1][#2]}

\newcommand{\entset}{Ent}
\newcommandx{\EntSet}[4][1=, 2=, 3=, 4=]
	{\mthset{\entset#4}[#1][#2]{#3}}

\newcommand{\entfun}{ent}
\newcommandx{\entFun}[4][1=, 2=, 3=, 4=]
	{\mthargfun{\entfun#4}[#1][#2]{#3}}

\newcommandx{\SOL}[5][1=, 2=, 3=, 4=, 5=]
	{\txtargname{SOL#5{\small\argint{$[$}{#1}{$]$}}}[#2][#3]{#4}\xspace}

\newcommandx{\TL}[5][1=, 2=, 3=, 4=, 5=]
	{\txtargname{TL#5{\small\argint{$[$}{#1}{$]$}}}[#2][#3]{#4}\xspace}

\newcommandx{\PL}[5][1=, 2=, 3=, 4=, 5=]
	{\txtargname{PL#5{\small\argint{$[$}{#1}{$]$}}}[#2][#3]{#4}\xspace}

\newcommand{\fvarset}{FVr}
\newcommandx{\FVarSet}[3][1=, 2=, 3=]
	{\mthset{\fvarset#3}[#1][#2]}

\newcommand{\fvarsym}{x}
\newcommandx{\fvarSym}[3][1=, 2=, 3=]
	{\mthsym{\fvarsym#3}[#1][#2]}

\newcommand{\fvarelm}{x}
\newcommandx{\fvarElm}[3][1=, 2=, 3=]
	{\mthelm{\fvarelm#3}[#1][#2]}

\newcommand{\fvarfun}{fvr}
\newcommandx{\fvarFun}[4][1=, 2=, 3=, 4=]
	{\mthargfun{\fvarfun#4}[#1][#2]{#3}}

\newcommand{\svarset}{SVr}
\newcommandx{\SVarSet}[3][1=, 2=, 3=]
	{\mthset{\svarset#3}[#1][#2]}

\newcommand{\svarsym}{X}
\newcommandx{\svarSym}[3][1=, 2=, 3=]
	{\mthsym{\svarsym#3}[#1][#2]}

\newcommand{\svarelm}{X}
\newcommandx{\svarElm}[3][1=, 2=, 3=]
	{\mthelm{\svarelm#3}[#1][#2]}

\newcommand{\svarfun}{svr}
\newcommandx{\svarFun}[4][1=, 2=, 3=, 4=]
	{\mthargfun{\svarfun#4}[#1][#2]{#3}}

\newcommandx{\ML}[5][1=, 2=, 3=, 4=, 5=]
	{\txtargname{ML#5{\small\argint{$[$}{#1}{$]$}}}[#2][#3]{#4}\xspace}

\newcommandx{\MC}[5][1=, 2=, 3=, 4=, 5=]
	{\txtargname{$\mu$Calculus#5{\small\argint{$[$}{#1}{$]$}}}[#2][#3]{#4}\xspace}

\newcommandx{\LTL}[5][1=, 2=, 3=, 4=, 5=]
	{\txtargname{LTL#5{\small\argint{$[$}{#1}{$]$}}}[#2][#3]{#4}\xspace}

\newcommandx{\PTL}[5][1=, 2=, 3=, 4=, 5=]
	{\txtargname{PTL#5{\small\argint{$[$}{#1}{$]$}}}[#2][#3]{#4}\xspace}

\newcommandx{\CTL}[5][1=, 2=, 3=, 4=, 5=]
	{\txtargname{CTL#5{\small\argint{$[$}{#1}{$]$}}}[#2][#3]{#4}\xspace}

\newcommandx{\CTLP}[5][1=, 2=, 3=, 4=, 5=]
	{\txtargname{CTL$^{+}$#5{\small\argint{$[$}{#1}{$]$}}}[#2][#3]{#4}\xspace}

\newcommandx{\CTLS}[5][1=, 2=, 3=, 4=, 5=]
	{\txtargname{CTL$^{\star}$#5{\small\argint{$[$}{#1}{$]$}}}[#2][#3]{#4}\xspace}

\newcommandx{\STL}[5][1=, 2=, 3=, 4=, 5=]
	{\txtargname{STL#5{\small\argint{$[$}{#1}{$]$}}}[#2][#3]{#4}\xspace}

\newcommandx{\STLP}[5][1=, 2=, 3=, 4=, 5=]
	{\txtargname{STL$^{+}$#5{\small\argint{$[$}{#1}{$]$}}}[#2][#3]{#4}\xspace}

\newcommandx{\STLS}[5][1=, 2=, 3=, 4=, 5=]
	{\txtargname{STL$^{\star}$#5{\small\argint{$[$}{#1}{$]$}}}[#2][#3]{#4}\xspace}

\newcommand{\PP}
	{\mthset[2]{P}}

\newcommandx{\ATL}[5][1=, 2=, 3=, 4=, 5=]
	{\txtargname{ATL#5{\small\argint{$[$}{#1}{$]$}}}[#2][#3]{#4}\xspace}

\newcommandx{\ATLP}[5][1=, 2=, 3=, 4=, 5=]
	{\txtargname{ATL$^{+}$#5{\small\argint{$[$}{#1}{$]$}}}[#2][#3]{#4}\xspace}

\newcommandx{\ATLS}[5][1=, 2=, 3=, 4=, 5=]
	{\txtargname{ATL$^{\star}$#5{\small\argint{$[$}{#1}{$]$}}}[#2][#3]{#4}\xspace}

\newcommandx{\SL}[5][1=, 2=, 3=, 4=, 5=]
	{\txtargname{SL#5{\small\argint{$[$}{#1}{$]$}}}[#2][#3]{#4}\xspace}

\newcommandx{\EF}[5][1=, 2=, 3=, 4=, 5=]
	{\txtargname{EF#5{\small\argint{$[$}{#1}{$]$}}}[#2][#3]{#4}\xspace}

\newcommandx{\SG}[5][1=, 2=, 3=, 4=, 5=]
	{\txtargname{SG#5{\small\argint{$[$}{#1}{$]$}}}[#2][#3]{#4}\xspace}

\newcommandx{\PG}[5][1=, 2=, 3=, 4=, 5=]
	{\txtargname{PG#5{\small\argint{$[$}{#1}{$]$}}}[#2][#3]{#4}\xspace}

\newcommandx{\LogTime}[4][1=, 2=, 3=, 4=]
	{\txtargname{LogTime#4}[#2][#3]{#1}\xspace}
\newcommandx{\LogTimeH}[4][1=, 2=, 3=, 4=]
	{\LogTime[#1][#2][#3][#4]-\HComp}
\newcommandx{\LogTimeE}[4][1=, 2=, 3=, 4=]
	{\LogTime[#1][#2][#3][#4]-\EComp}
\newcommandx{\LogTimeC}[4][1=, 2=, 3=, 4=]
	{\LogTime[#1][#2][#3][#4]-\CComp}

\newcommand{\NLogTime}
	{\txtname{N}\LogTime}
\newcommandx{\NLogTimeH}[4][1=, 2=, 3=, 4=]
	{\NLogTime[#1][#2][#3][#4]-\HComp}
\newcommandx{\NLogTimeE}[4][1=, 2=, 3=, 4=]
	{\NLogTime[#1][#2][#3][#4]-\EComp}
\newcommandx{\NLogTimeC}[4][1=, 2=, 3=, 4=]
	{\NLogTime[#1][#2][#3][#4]-\CComp}

\newcommand{\CoNLogTime}
	{\txtname{Co}\NLogTime}
\newcommandx{\CoNLogTimeH}[4][1=, 2=, 3=, 4=]
	{\CoNLogTime[#1][#2][#3][#4]-\HComp}
\newcommandx{\CoNLogTimeE}[4][1=, 2=, 3=, 4=]
	{\CoNLogTime[#1][#2][#3][#4]-\EComp}
\newcommandx{\CoNLogTimeC}[4][1=, 2=, 3=, 4=]
	{\CoNLogTime[#1][#2][#3][#4]-\CComp}

\newcommandx{\ALogTime}
	{\txtname{A}\LogTime}
\newcommandx{\ALogTimeH}[4][1=, 2=, 3=, 4=]
	{\ALogTime[#1][#2][#3][#4]-\HComp}
\newcommandx{\ALogTimeE}[4][1=, 2=, 3=, 4=]
	{\ALogTime[#1][#2][#3][#4]-\EComp}
\newcommandx{\ALogTimeC}[4][1=, 2=, 3=, 4=]
	{\ALogTime[#1][#2][#3][#4]-\CComp}

\newcommandx{\LogSpace}[4][1=, 2=, 3=, 4=]
	{\txtargname{LogSpace#4}[#2][#3]{#1}\xspace}
\newcommandx{\LogSpaceH}[4][1=, 2=, 3=, 4=]
	{\LogSpace[#1][#2][#3][#4]-\HComp}
\newcommandx{\LogSpaceE}[4][1=, 2=, 3=, 4=]
	{\LogSpace[#1][#2][#3][#4]-\EComp}
\newcommandx{\LogSpaceC}[4][1=, 2=, 3=, 4=]
	{\LogSpace[#1][#2][#3][#4]-\CComp}

\newcommandx{\NLogSpace}
	{\txtname{N}\LogSpace}
\newcommandx{\NLogSpaceH}[4][1=, 2=, 3=, 4=]
	{\NLogSpace[#1][#2][#3][#4]-\HComp}
\newcommandx{\NLogSpaceE}[4][1=, 2=, 3=, 4=]
	{\NLogSpace[#1][#2][#3][#4]-\EComp}
\newcommandx{\NLogSpaceC}[4][1=, 2=, 3=, 4=]
	{\NLogSpace[#1][#2][#3][#4]-\CComp}

\newcommandx{\CoNLogSpace}
	{\txtname{Co}\NLogSpace}
\newcommandx{\CoNLogSpaceH}[4][1=, 2=, 3=, 4=]
	{\CoNLogSpace[#1][#2][#3][#4]-\HComp}
\newcommandx{\CoNLogSpaceE}[4][1=, 2=, 3=, 4=]
	{\CoNLogSpace[#1][#2][#3][#4]-\EComp}
\newcommandx{\CoNLogSpaceC}[4][1=, 2=, 3=, 4=]
	{\CoNLogSpace[#1][#2][#3][#4]-\CComp}

\newcommandx{\ALogSpace}
	{\txtname{A}\LogSpace}
\newcommandx{\ALogSpaceH}[4][1=, 2=, 3=, 4=]
	{\ALogSpace[#1][#2][#3][#4]-\HComp}
\newcommandx{\ALogSpaceE}[4][1=, 2=, 3=, 4=]
	{\ALogSpace[#1][#2][#3][#4]-\EComp}
\newcommandx{\ALogSpaceC}[4][1=, 2=, 3=, 4=]
	{\ALogSpace[#1][#2][#3][#4]-\CComp}

\newcommandx{\PTime}[4][1=, 2=, 3=, 4=]
	{\txtargname{PTime#4}[#2][#3]{#1}\xspace}
\newcommandx{\PTimeH}[4][1=, 2=, 3=, 4=]
	{\PTime[#1][#2][#3][#4]-\HComp}
\newcommandx{\PTimeE}[4][1=, 2=, 3=, 4=]
	{\PTime[#1][#2][#3][#4]-\EComp}
\newcommandx{\PTimeC}[4][1=, 2=, 3=, 4=]
	{\PTime[#1][#2][#3][#4]-\CComp}

\newcommandx{\UPTime}
	{\txtname{U}\PTime}
\newcommandx{\UPTimeH}[4][1=, 2=, 3=, 4=]
	{\UPTime[#1][#2][#3][#4]-\HComp}
\newcommandx{\UPTimeE}[4][1=, 2=, 3=, 4=]
	{\UPTime[#1][#2][#3][#4]-\EComp}
\newcommandx{\UPTimeC}[4][1=, 2=, 3=, 4=]
	{\UPTime[#1][#2][#3][#4]-\CComp}

\newcommandx{\CoUPTime}
	{\txtname{Co}\UPTime}
\newcommandx{\CoUPTimeH}[4][1=, 2=, 3=, 4=]
	{\CoUPTime[#1][#2][#3][#4]-\HComp}
\newcommandx{\CoUPTimeE}[4][1=, 2=, 3=, 4=]
	{\CoUPTime[#1][#2][#3][#4]-\EComp}
\newcommandx{\CoUPTimeC}[4][1=, 2=, 3=, 4=]
	{\CoUPTime[#1][#2][#3][#4]-\CComp}

\newcommandx{\NPTime}
	{\txtname{N}\PTime}
\newcommandx{\NPTimeH}[4][1=, 2=, 3=, 4=]
	{\NPTime[#1][#2][#3][#4]-\HComp}
\newcommandx{\NPTimeE}[4][1=, 2=, 3=, 4=]
	{\NPTime[#1][#2][#3][#4]-\EComp}
\newcommandx{\NPTimeC}[4][1=, 2=, 3=, 4=]
	{\NPTime[#1][#2][#3][#4]-\CComp}

\newcommandx{\CoNPTime}
	{\txtname{Co}\NPTime}
\newcommandx{\CoNPTimeH}[4][1=, 2=, 3=, 4=]
	{\CoNPTime[#1][#2][#3][#4]-\HComp}
\newcommandx{\CoNPTimeE}[4][1=, 2=, 3=, 4=]
	{\CoNPTime[#1][#2][#3][#4]-\EComp}
\newcommandx{\CoNPTimeC}[4][1=, 2=, 3=, 4=]
	{\CoNPTime[#1][#2][#3][#4]-\CComp}

\newcommandx{\APTime}
	{\txtname{A}\PTime}
\newcommandx{\APTimeH}[4][1=, 2=, 3=, 4=]
	{\APTime[#1][#2][#3][#4]-\HComp}
\newcommandx{\APTimeE}[4][1=, 2=, 3=, 4=]
	{\APTime[#1][#2][#3][#4]-\EComp}
\newcommandx{\APTimeC}[4][1=, 2=, 3=, 4=]
	{\APTime[#1][#2][#3][#4]-\CComp}

\newcommandx{\PSpace}[4][1=, 2=, 3=, 4=]
	{\txtargname{PSpace#4}[#2][#3]{#1}\xspace}
\newcommandx{\PSpaceH}[4][1=, 2=, 3=, 4=]
	{\PSpace[#1][#2][#3][#4]-\HComp}
\newcommandx{\PSpaceE}[4][1=, 2=, 3=, 4=]
	{\PSpace[#1][#2][#3][#4]-\EComp}
\newcommandx{\PSpaceC}[4][1=, 2=, 3=, 4=]
	{\PSpace[#1][#2][#3][#4]-\CComp}

\newcommandx{\NPSpace}
	{\txtname{N}\PSpace}
\newcommandx{\NPSpaceH}[4][1=, 2=, 3=, 4=]
	{\NPSpace[#1][#2][#3][#4]-\HComp}
\newcommandx{\NPSpaceE}[4][1=, 2=, 3=, 4=]
	{\NPSpace[#1][#2][#3][#4]-\EComp}
\newcommandx{\NPSpaceC}[4][1=, 2=, 3=, 4=]
	{\NPSpace[#1][#2][#3][#4]-\CComp}

\newcommandx{\CoNPSpace}
	{\txtname{Co}\NPSpace}
\newcommandx{\CoNPSpaceH}[4][1=, 2=, 3=, 4=]
	{\CoNPSpace[#1][#2][#3][#4]-\HComp}
\newcommandx{\CoNPSpaceE}[4][1=, 2=, 3=, 4=]
	{\CoNPSpace[#1][#2][#3][#4]-\EComp}
\newcommandx{\CoNPSpaceC}[4][1=, 2=, 3=, 4=]
	{\CoNPSpace[#1][#2][#3][#4]-\CComp}

\newcommandx{\APSpace}
	{\txtname{A}\PSpace}
\newcommandx{\APSpaceH}[4][1=, 2=, 3=, 4=]
	{\APSpace[#1][#2][#3][#4]-\HComp}
\newcommandx{\APSpaceE}[4][1=, 2=, 3=, 4=]
	{\APSpace[#1][#2][#3][#4]-\EComp}
\newcommandx{\APSpaceC}[4][1=, 2=, 3=, 4=]
	{\APSpace[#1][#2][#3][#4]-\CComp}

\newcommandx{\ExpTime}[4][1=, 2=, 3=, 4=]
	{\txtargname{ExpTime#4}[#2][#3]{#1}\xspace}
\newcommandx{\ExpTimeH}[4][1=, 2=, 3=, 4=]
	{\ExpTime[#1][#2][#3][#4]-\HComp}
\newcommandx{\ExpTimeE}[4][1=, 2=, 3=, 4=]
	{\ExpTime[#1][#2][#3][#4]-\EComp}
\newcommandx{\ExpTimeC}[4][1=, 2=, 3=, 4=]
	{\ExpTime[#1][#2][#3][#4]-\CComp}

\newcommandx{\NExpTime}
	{\txtname{N}\ExpTime}
\newcommandx{\NExpTimeH}[4][1=, 2=, 3=, 4=]
	{\NExpTime[#1][#2][#3][#4]-\HComp}
\newcommandx{\NExpTimeE}[4][1=, 2=, 3=, 4=]
	{\NExpTime[#1][#2][#3][#4]-\EComp}
\newcommandx{\NExpTimeC}[4][1=, 2=, 3=, 4=]
	{\NExpTime[#1][#2][#3][#4]-\CComp}

\newcommandx{\CoNExpTime}
	{\txtname{Co}\NExpTime}
\newcommandx{\CoNExpTimeH}[4][1=, 2=, 3=, 4=]
	{\CoNExpTime[#1][#2][#3][#4]-\HComp}
\newcommandx{\CoNExpTimeE}[4][1=, 2=, 3=, 4=]
	{\CoNExpTime[#1][#2][#3][#4]-\EComp}
\newcommandx{\CoNExpTimeC}[4][1=, 2=, 3=, 4=]
	{\CoNExpTime[#1][#2][#3][#4]-\CComp}

\newcommandx{\AExpTime}
	{\txtname{A}\ExpTime}
\newcommandx{\AExpTimeH}[4][1=, 2=, 3=, 4=]
	{\AExpTime[#1][#2][#3][#4]-\HComp}
\newcommandx{\AExpTimeE}[4][1=, 2=, 3=, 4=]
	{\AExpTime[#1][#2][#3][#4]-\EComp}
\newcommandx{\AExpTimeC}[4][1=, 2=, 3=, 4=]
	{\AExpTime[#1][#2][#3][#4]-\CComp}

\newcommandx{\ExpSpace}[4][1=, 2=, 3=, 4=]
	{\txtargname{ExpSpace#4}[#2][#3]{#1}\xspace}
\newcommandx{\ExpSpaceH}[4][1=, 2=, 3=, 4=]
	{\ExpSpace[#1][#2][#3][#4]-\HComp}
\newcommandx{\ExpSpaceE}[4][1=, 2=, 3=, 4=]
	{\ExpSpace[#1][#2][#3][#4]-\EComp}
\newcommandx{\ExpSpaceC}[4][1=, 2=, 3=, 4=]
	{\ExpSpace[#1][#2][#3][#4]-\CComp}

\newcommandx{\NExpSpace}
	{\txtname{N}\ExpSpace}
\newcommandx{\NExpSpaceH}[4][1=, 2=, 3=, 4=]
	{\NExpSpace[#1][#2][#3][#4]-\HComp}
\newcommandx{\NExpSpaceE}[4][1=, 2=, 3=, 4=]
	{\NExpSpace[#1][#2][#3][#4]-\EComp}
\newcommandx{\NExpSpaceC}[4][1=, 2=, 3=, 4=]
	{\NExpSpace[#1][#2][#3][#4]-\CComp}

\newcommandx{\CoNExpSpace}
	{\txtname{Co}\NExpSpace}
\newcommandx{\CoNExpSpaceH}[4][1=, 2=, 3=, 4=]
	{\CoNExpSpace[#1][#2][#3][#4]-\HComp}
\newcommandx{\CoNExpSpaceE}[4][1=, 2=, 3=, 4=]
	{\CoNExpSpace[#1][#2][#3][#4]-\EComp}
\newcommandx{\CoNExpSpaceC}[4][1=, 2=, 3=, 4=]
	{\CoNExpSpace[#1][#2][#3][#4]-\CComp}

\newcommandx{\AExpSpace}
	{\txtname{A}\ExpSpace}
\newcommandx{\AExpSpaceH}[4][1=, 2=, 3=, 4=]
	{\AExpSpace[#1][#2][#3][#4]-\HComp}
\newcommandx{\AExpSpaceE}[4][1=, 2=, 3=, 4=]
	{\AExpSpace[#1][#2][#3][#4]-\EComp}
\newcommandx{\AExpSpaceC}[4][1=, 2=, 3=, 4=]
	{\AExpSpace[#1][#2][#3][#4]-\CComp}

\newcommandx{\NonElm}[4][1=, 2=, 3=, 4=]
	{\txtargname{NonElementary#4}[#2][#3]{#1}\xspace}
\newcommandx{\NonElmH}[4][1=, 2=, 3=, 4=]
	{\NonElm[#1][#2][#3][#4]-\HComp}
\newcommandx{\NonElmE}[4][1=, 2=, 3=, 4=]
	{\NonElm[#1][#2][#3][#4]-\EComp}
\newcommandx{\NonElmC}[4][1=, 2=, 3=, 4=]
	{\NonElm[#1][#2][#3][#4]-\CComp}

\newcommandx{\NonElmTime}[4][1=, 2=, 3=, 4=]
	{\txtargname{NonElementaryTime#4}[#2][#3]{#1}\xspace}
\newcommandx{\NonElmTimeH}[4][1=, 2=, 3=, 4=]
	{\NonElmTime[#1][#2][#3][#4]-\HComp}
\newcommandx{\NonElmTimeE}[4][1=, 2=, 3=, 4=]
	{\NonElmTime[#1][#2][#3][#4]-\EComp}
\newcommandx{\NonElmTimeC}[4][1=, 2=, 3=, 4=]
	{\NonElmTime[#1][#2][#3][#4]-\CComp}

\newcommandx{\NonElmSpace}[4][1=, 2=, 3=, 4=]
	{\txtargname{NonElementarySpace#4}[#2][#3]{#1}\xspace}
\newcommandx{\NonElmSpaceH}[4][1=, 2=, 3=, 4=]
	{\NonElmSpace[#1][#2][#3][#4]-\HComp}
\newcommandx{\NonElmSpaceE}[4][1=, 2=, 3=, 4=]
	{\NonElmSpace[#1][#2][#3][#4]-\EComp}
\newcommandx{\NonElmSpaceC}[4][1=, 2=, 3=, 4=]
	{\NonElmSpace[#1][#2][#3][#4]-\CComp}

\newcommandx{\DLHier}[4][2=, 3=, 4=]
	{\mthargset[0]{\Delta#4}[#1][#3]{#2}\xspace}
\newcommandx{\DLHierH}[4][2=, 3=, 4=]
	{\DLHier{#1}[#2][#3][#4]-\HComp}
\newcommandx{\DLHierE}[4][2=, 3=, 4=]
	{\DLHier{#1}[#2][#3][#4]-\EComp}
\newcommandx{\DLHierC}[4][2=, 3=, 4=]
	{\DLHier{#1}[#2][#3][#4]-\CComp}

\newcommandx{\ELHier}[4][2=, 3=, 4=]
	{\mthargset[0]{\Sigma#4}[#1][#3]{#2}\xspace}
\newcommandx{\ELHierH}[4][2=, 3=, 4=]
	{\ELHier{#1}[#2][#3][#4]-\HComp}
\newcommandx{\ELHierE}[4][2=, 3=, 4=]
	{\ELHier{#1}[#2][#3][#4]-\EComp}
\newcommandx{\ELHierC}[4][2=, 3=, 4=]
	{\ELHier{#1}[#2][#3][#4]-\CComp}

\newcommandx{\ULHier}[4][2=, 3=, 4=]
	{\mthargset[0]{\Pi#4}[#1][#3]{#2}\xspace}
\newcommandx{\ULHierH}[4][2=, 3=, 4=]
	{\ULHier{#1}[#2][#3][#4]-\HComp}
\newcommandx{\ULHierE}[4][2=, 3=, 4=]
	{\ULHier{#1}[#2][#3][#4]-\EComp}
\newcommandx{\ULHierC}[4][2=, 3=, 4=]
	{\ULHier{#1}[#2][#3][#4]-\CComp}

\newcommandx{\DBHier}[4][2=, 3=, 4=]
	{\mthargset[3]{\Delta#4}[#1][#3]{#2}\xspace}
\newcommandx{\DBHierH}[4][2=, 3=, 4=]
	{\DBHier{#1}[#2][#3][#4]-\HComp}
\newcommandx{\DBHierE}[4][2=, 3=, 4=]
	{\DBHier{#1}[#2][#3][#4]-\EComp}
\newcommandx{\DBHierC}[4][2=, 3=, 4=]
	{\DBHier{#1}[#2][#3][#4]-\CComp}

\newcommandx{\EBHier}[4][2=, 3=, 4=]
	{\mthargset[3]{\Sigma#4}[#1][#3]{#2}\xspace}
\newcommandx{\EBHierH}[4][2=, 3=, 4=]
	{\EBHier{#1}[#2][#3][#4]-\HComp}
\newcommandx{\EBHierE}[4][2=, 3=, 4=]
	{\EBHier{#1}[#2][#3][#4]-\EComp}
\newcommandx{\EBHierC}[4][2=, 3=, 4=]
	{\EBHier{#1}[#2][#3][#4]-\CComp}

\newcommandx{\UBHier}[4][2=, 3=, 4=]
	{\mthargset[3]{\Pi#4}[#1][#3]{#2}\xspace}
\newcommandx{\UBHierH}[4][2=, 3=, 4=]
	{\UBHier{#1}[#2][#3][#4]-\HComp}
\newcommandx{\UBHierE}[4][2=, 3=, 4=]
	{\UBHier{#1}[#2][#3][#4]-\EComp}
\newcommandx{\UBHierC}[4][2=, 3=, 4=]
	{\UBHier{#1}[#2][#3][#4]-\CComp}

\newcommandx{\DPolHier}[4][2=, 3=, 4=]
	{\DLHier{#1}[#2][\argb{\mathrm{P}}{#3}][#4]}
\newcommandx{\DPolHierH}[4][2=, 3=, 4=]
	{\DPolHier{#1}[#2][#3][#4]-\HComp}
\newcommandx{\DPolHierE}[4][2=, 3=, 4=]
	{\DPolHier{#1}[#2][#3][#4]-\EComp}
\newcommandx{\DPolHierC}[4][2=, 3=, 4=]
	{\DPolHier{#1}[#2][#3][#4]-\CComp}

\newcommandx{\EPolHier}[4][2=, 3=, 4=]
	{\ELHier{#1}[#2][\argb{\mathrm{P}}{#3}][#4]}
\newcommandx{\EPolHierH}[4][2=, 3=, 4=]
	{\EPolHier{#1}[#2][#3][#4]-\HComp}
\newcommandx{\EPolHierE}[4][2=, 3=, 4=]
	{\EPolHier{#1}[#2][#3][#4]-\EComp}
\newcommandx{\EPolHierC}[4][2=, 3=, 4=]
	{\EPolHier{#1}[#2][#3][#4]-\CComp}

\newcommandx{\UPolHier}[4][2=, 3=, 4=]
	{\ULHier{#1}[#2][\argb{\mathrm{P}}{#3}][#4]}
\newcommandx{\UPolHierH}[4][2=, 3=, 4=]
	{\UPolHier{#1}[#2][#3][#4]-\HComp}
\newcommandx{\UPolHierE}[4][2=, 3=, 4=]
	{\UPolHier{#1}[#2][#3][#4]-\EComp}
\newcommandx{\UPolHierC}[4][2=, 3=, 4=]
	{\UPolHier{#1}[#2][#3][#4]-\CComp}

\newcommandx{\DAriHier}[4][2=, 3=, 4=]
	{\DLHier{#1}[#2][\argb{0}{#3}][#4]}
\newcommandx{\DAriHierH}[4][2=, 3=, 4=]
	{\DAriHier{#1}[#2][#3][#4]-\HComp}
\newcommandx{\DAriHierE}[4][2=, 3=, 4=]
	{\DAriHier{#1}[#2][#3][#4]-\EComp}
\newcommandx{\DAriHierC}[4][2=, 3=, 4=]
	{\DAriHier{#1}[#2][#3][#4]-\CComp}

\newcommandx{\EAriHier}[4][2=, 3=, 4=]
	{\ELHier{#1}[#2][\argb{0}{#3}][#4]}
\newcommandx{\EAriHierH}[4][2=, 3=, 4=]
	{\EAriHier{#1}[#2][#3][#4]-\HComp}
\newcommandx{\EAriHierE}[4][2=, 3=, 4=]
	{\EAriHier{#1}[#2][#3][#4]-\EComp}
\newcommandx{\EAriHierC}[4][2=, 3=, 4=]
	{\EAriHier{#1}[#2][#3][#4]-\CComp}

\newcommandx{\UAriHier}[4][2=, 3=, 4=]
	{\ULHier{#1}[#2][\argb{0}{#3}][#4]}
\newcommandx{\UAriHierH}[4][2=, 3=, 4=]
	{\UAriHier{#1}[#2][#3][#4]-\HComp}
\newcommandx{\UAriHierE}[4][2=, 3=, 4=]
	{\UAriHier{#1}[#2][#3][#4]-\EComp}
\newcommandx{\UAriHierC}[4][2=, 3=, 4=]
	{\UAriHier{#1}[#2][#3][#4]-\CComp}

\newcommandx{\DAnaHier}[4][2=, 3=, 4=]
	{\DLHier{#1}[#2][\argb{1}{#3}][#4]}
\newcommandx{\DAnaHierH}[4][2=, 3=, 4=]
	{\DAnaHier{#1}[#2][#3][#4]-\HComp}
\newcommandx{\DAnaHierE}[4][2=, 3=, 4=]
	{\DAnaHier{#1}[#2][#3][#4]-\EComp}
\newcommandx{\DAnaHierC}[4][2=, 3=, 4=]
	{\DAnaHier{#1}[#2][#3][#4]-\CComp}

\newcommandx{\EAnaHier}[4][2=, 3=, 4=]
	{\ELHier{#1}[#2][\argb{1}{#3}][#4]}
\newcommandx{\EAnaHierH}[4][2=, 3=, 4=]
	{\EAnaHier{#1}[#2][#3][#4]-\HComp}
\newcommandx{\EAnaHierE}[4][2=, 3=, 4=]
	{\EAnaHier{#1}[#2][#3][#4]-\EComp}
\newcommandx{\EAnaHierC}[4][2=, 3=, 4=]
	{\EAnaHier{#1}[#2][#3][#4]-\CComp}

\newcommandx{\UAnaHier}[4][2=, 3=, 4=]
	{\ULHier{#1}[#2][\argb{1}{#3}][#4]}
\newcommandx{\UAnaHierH}[4][2=, 3=, 4=]
	{\UAnaHier{#1}[#2][#3][#4]-\HComp}
\newcommandx{\UAnaHierE}[4][2=, 3=, 4=]
	{\UAnaHier{#1}[#2][#3][#4]-\EComp}
\newcommandx{\UAnaHierC}[4][2=, 3=, 4=]
	{\UAnaHier{#1}[#2][#3][#4]-\CComp}

\newcommandx{\DBorHier}[4][2=, 3=, 4=]
	{\DBHier{#1}[#2][\argb{\mathrm{B}}{#3}][#4]}
\newcommandx{\DBorHierH}[4][2=, 3=, 4=]
	{\DBorHier{#1}[#2][#3][#4]-\HComp}
\newcommandx{\DBorHierE}[4][2=, 3=, 4=]
	{\DBorHier{#1}[#2][#3][#4]-\EComp}
\newcommandx{\DBorHierC}[4][2=, 3=, 4=]
	{\DBorHier{#1}[#2][#3][#4]-\CComp}

\newcommandx{\EBorHier}[4][2=, 3=, 4=]
	{\EBHier{#1}[#2][\argb{\mathrm{B}}{#3}][#4]}
\newcommandx{\EBorHierH}[4][2=, 3=, 4=]
	{\EBorHier{#1}[#2][#3][#4]-\HComp}
\newcommandx{\EBorHierE}[4][2=, 3=, 4=]
	{\EBorHier{#1}[#2][#3][#4]-\EComp}
\newcommandx{\EBorHierC}[4][2=, 3=, 4=]
	{\EBorHier{#1}[#2][#3][#4]-\CComp}

\newcommandx{\UBorHier}[4][2=, 3=, 4=]
	{\UBHier{#1}[#2][\argb{\mathrm{B}}{#3}][#4]}
\newcommandx{\UBorHierH}[4][2=, 3=, 4=]
	{\UBorHier{#1}[#2][#3][#4]-\HComp}
\newcommandx{\UBorHierE}[4][2=, 3=, 4=]
	{\UBorHier{#1}[#2][#3][#4]-\EComp}
\newcommandx{\UBorHierC}[4][2=, 3=, 4=]
	{\UBorHier{#1}[#2][#3][#4]-\CComp}

\newcommand{\HComp}
	{\txtname{hard}\xspace}

\newcommand{\EComp}
	{\txtname{easy}\xspace}

\newcommand{\CComp}
	{\txtname{complete}\xspace}

\newtheorem{definition}{Definition}[section]

\newtheorem{theorem}{Theorem}[section]

\newtheorem{proposition}{Proposition}[section]

\newcounter{flushenumerate}
  {\end{list}}

% ...

\newcommand{\prtset}{Pr}
\newcommandx{\PrtSet}[3][1=, 2=, 3=]
  {\mthset{\prtset#3}[#1][#2]}

\newcommand{\prtfun}{pr}
\newcommandx{\prtFun}[4][1=, 2=, 3=, 4=]
  {\mthargfun{\prtfun#4}[#1][#2]{#3}}

\newcommand{\prefun}{pre}
\newcommandx{\preFun}[4][1=, 2=, 3=, 4=]
  {\mthargfun{\prefun#4}[#1][#2]{#3}}

\newcommand{\escfun}{esc}
\newcommandx{\escFun}[4][1=, 2=, 3=, 4=]
  {\mthargfun{\escfun#4}[#1][#2]{#3}}

\newcommand{\intfun}{int}
\newcommandx{\intFun}[4][1=, 2=, 3=, 4=]
  {\mthargfun{\intfun#4}[#1][#2]{#3}}

\newcommand{\atrfun}{atr}
\newcommandx{\atrFun}[4][1=, 2=, 3=, 4=]
  {\mthargfun{\atrfun#4}[#1][#2]{#3}}

\newcommand{\solfun}{sol}
\newcommandx{\solFun}[4][1=, 2=, 3=, 4=]
  {\mthargfun{\solfun#4}[#1][#2]{#3}}

\newcommand{\srcfun}{src}
\newcommandx{\srcFun}[4][1=, 2=, 3=, 4=]
  {\mthargfun{\srcfun#4}[#1][#2]{#3}}

\newcommand{\regset}{Rg}
\newcommandx{\RegSet}[3][1=, 2=, 3=]
  {\mthset{\regset#3}[#1][#2]}

\newcommand{\qdset}{QD}
\newcommandx{\QDSet}[3][1=, 2=, 3=]
  {\mthset{\qdset#3}[#1][#2]}

\newcommand{\bepfun}{bep}
\newcommandx{\bepFun}[4][1=, 2=, 3=, 4=]
  {\mthargfun{\bepfun#4}[#1][#2]{#3}}

\newcommand{\brpfun}{brp}
\newcommandx{\brpFun}[4][1=, 2=, 3=, 4=]
  {\mthargfun{\brpfun#4}[#1][#2]{#3}}

\newcommand{\topelm}{\top}
\newcommandx{\topElm}[3][1=, 2=, 3=]
  {\mthelm{\topelm#3}[#1][#2]}

\newcommand{\ordrel}{\prec}
\newcommandx{\ordRel}[3][1=, 2=, 3=]
  {\mthrel{\ordrel#3}[#1][#2]}

\newcommand{\comrel}{\Yright}
\newcommandx{\comRel}[3][1=, 2=, 3=]
  {\mthrel{\comrel#3}[#1][#2]}

\newcommand{\qryfun}{\Re}
\newcommandx{\qryFun}[4][1=, 2=, 3=, 4=]
  {\mthargfun{\qryfun#4}[#1][#2]{#3}}

\newcommand{\sucfun}{\downarrow}
\newcommandx{\sucFun}[4][1=, 2=, 3=, 4=]
  {\mthargfun{\sucfun#4}[#1][#2]{#3}}

\renewcommandx{\PP}[5][1=, 2=, 3=, 4=, 5=]
  {\txtargname{PP#5{\small\argint{$[$}{#1}{$]$}}}[#2][#3]{#4}\xspace}

\newcommandx{\PPP}[5][1=, 2=, 3=, 4=, 5=]
  {\txtargname{PP+#5{\small\argint{$[$}{#1}{$]$}}}[#2][#3]{#4}\xspace}

\newcommandx{\DP}[5][1=, 2=, 3=, 4=, 5=]
  {\txtargname{DP#5{\small\argint{$[$}{#1}{$]$}}}[#2][#3]{#4}\xspace}

% ...

\usepackage{tikz}
\usetikzlibrary{arrows,shapes,calc}
\usetikzlibrary{decorations.text}

\usepackage{pgfplots}

\usepackage{wrapfig}
\usepackage[caption=false,font=footnotesize]{subfig}

\usepackage{float}

\tikzstyle{every node} =
  [draw = none, fill = white, thin]
\tikzstyle{every edge} +=
  [black, thick]

\tikzstyle{noall} =
  [draw = none, fill = none]
\tikzstyle{nodraw} =
  [draw = none, fill = white]
\tikzstyle{nofill} =
  [draw = black, fill = none]

\tikzstyle{cnode} =
  [circle, draw = black]
\tikzstyle{snode} =
  [regular polygon, regular polygon sides = 4, draw = gray!50]
\tikzstyle{lnode} =
  [diamond, draw = gray!75]
\tikzstyle{pnode} =
  [regular polygon, regular polygon sides = 5, draw = gray]

\newenvironment{customlegend}[1][]
  {%
  \begingroup
  \csname pgfplots@init@cleared@structures\endcsname
  \pgfplotsset{#1}%
  }
  {%
  \csname pgfplots@createlegend\endcsname
  \endgroup
  }
\def\addlegendimage{\csname pgfplots@addlegendimage\endcsname}

\AfterEndPreamble
  {

  \newcommand{\figexm}
    {
    \begin{center}
      \footnotesize
      \scalebox{0.80}[0.80]
        {
        \begin{tikzpicture}[node distance = 5em, bend angle = 45]

          \tikzset{every loop/.style = {max distance = 1.5em}}

          \node [snode]
                (C)
                []
                {$\cSym/4$};
          \node [snode]
                (E)
                [right of = C]
                {$\eSym/2$};
          \node [lnode]
                (J)
                [below of = E]
                {$\jSym/4$};
          \node [snode]
                (A)
                [right of = E]
                {$\aSym/6$};
          \node [snode]
                (I)
                [below of = A]
                {$\iSym/0$};
          \node [lnode]
                (D)
                [right of = A]
                {$\dSym/3$};
          \node [lnode]
                (B)
                [below of = D]
                {$\bSym/5$};
          \node [lnode]
                (G)
                [right of = D]
                {$\gSym/1$};
          \node [snode]
                (H)
                [below of = G]
                {$\hSym/1$};
          \node [snode]
                (F)
                [right of = H]
                {$\fSym/2$};

          \path[->]
            (A)   edge  []
                        (D)
                  edge  [bend right]
                        (E)
            (B)   edge  [bend right]
                        (D)
                  edge  []
                        (I)
            (C)   edge  []
                        (J)
                  edge  [bend right]
                        (E)
            (D)   edge  [bend right]
                        (B)
                  edge  [bend left]
                        (G)
            (E)   edge  [loop above]
                        ()
                  edge  [bend right]
                        (A)
                  edge  [bend right]
                        (C)
            (F)   edge  [loop above]
                        ()
                  edge  [bend left]
                        (H)
            (G)   edge  [loop above]
                        ()
                  edge  [bend left]
                        (D)
            (H)   edge  []
                        (B)
                  edge  [bend left]
                        (F)
            (I)   edge  [loop below]
                        ()
                  edge  []
                        (A)
            (J)   edge  []
                        (I)
            ;

          \begin{scope}
            [very thick, draw = gray, fill = gray, fill opacity = 0.25]

            \filldraw
              ($(E) + (0, 1)$)
                to [out = 180, in = 0]
              ($(C) + (0, 0.80)$)
                to [out = 180, in = 90]
              ($(C) + (-0.80, 0)$)
                to [out = 270, in = 180]
              ($(C) + (0, -0.80)$)
                to [out = 0, in = 180]
              ($(E) + (0, -0.80)$)
                to [out = 0, in = 270]
              ($(E) + (0.80, 0)$)
                to [out = 90, in = 0]
              ($(E) + (0, 1)$)
              ;

          \end{scope}

          \begin{scope}
            [very thick, draw = gray, fill = none, fill opacity = 0, dashed]

            \filldraw
              ($(F) + (0, 1)$)
                to [out = 180, in = 0]
              ($(H) + (0, 0.80)$)
                to [out = 180, in = 90]
              ($(H) + (-0.80, 0)$)
                to [out = 270, in = 180]
              ($(H) + (0, -0.80)$)
                to [out = 0, in = 180]
              ($(F) + (0, -0.80)$)
                to [out = 0, in = 270]
              ($(F) + (0.80, 0)$)
                to [out = 90, in = 0]
              ($(F) + (0, 1)$)
              ;

          \end{scope}

        \end{tikzpicture}
        }
      \vspace{-0.625em}
      \captionof{figure}{\label{fig:exm} Running example.}
    \end{center}
    }

  \newcommand{\figprtprmlowbnd}
    {
    \begin{wrapfigure}[8]{R}{0.250\textwidth}
      \vspace{-2.2em}
      \begin{center}
        \footnotesize
        \scalebox{0.80}[0.80]
          {
          \begin{tikzpicture}[node distance = 3.5em, bend angle = 45]

            \tikzset{every loop/.style = {max distance = 1.5em}}

            \node [lnode]
                  (0)
                  []
                  {$0$};
            \node []
                  (X)
                  [below of = 0]
                  {};
            \node [lnode]
                  (8)
                  [node distance = 1.75em, left of = X]
                  {$8$};
            \node [lnode]
                  (7)
                  [node distance = 1.75em, right of = X]
                  {$7$};
            \node [lnode]
                  (9)
                  [left of = 8]
                  {$9$};
            \node [lnode]
                  (6)
                  [right of = 7]
                  {$6$};
            \node [lnode]
                  (1)
                  [below of = 9]
                  {$1$};
            \node [snode]
                  (2)
                  [below of = 8]
                  {$2$};
            \node [lnode]
                  (3)
                  [below of = 7]
                  {$3$};
            \node [snode]
                  (4)
                  [below of = 6]
                  {$4$};

            \path[->]
              (0)   edge  [loop above]
                          ()
              (1)   edge  [loop below]
                          ()
                    edge  []
                          (9)
              (2)   edge  [loop below]
                          ()
                    edge  []
                          (8)
              (3)   edge  [loop below]
                          ()
                    edge  []
                          (7)
              (4)   edge  [loop below]
                          ()
                    edge  []
                          (6)
              (6)   edge  []
                          (0)
              (7)   edge  []
                          (0)
              (8)   edge  []
                          (0)
              (9)   edge  []
                          (0)
              ;

          \end{tikzpicture}
          }
          \vspace{-.6em}
        \caption{\label{fig:prtprmlowbnd} \small The $\GamName[4][\PPP]$ game.}
      \end{center}
      \vspace{-1em}
    \end{wrapfigure}
    }

  \newcommand{\figprf}
    {
    \begin{figure}
      \begin{center}
        \begin{tikzpicture}[every node/.style={scale=0.65}]
          \hspace{-.1cm}
          \begin{axis}
            [
            ylabel=Time (sec), xlabel=Number of nodes $/ 10^3$, title={},
            name=axis1,
            xmin=0, xmax=20,
            width=8.8cm, height=7.5cm,
            ymin=0,  ymax=35,
            ylabel near ticks,
            xlabel near ticks,
            x axis line style={-}, y axis line style={-},
            legend entries={$PP$, $PP+$, $DP$, $Rec$, $Strat$},
            legend pos=north east,
            ]
            \addplot [mark=triangle, red!60!black, dashed, mark
              options={solid}, line width=.7, mark size=1.2] table [x index=0,
              y index=1] {experiments/timesummary.plt};
            \addplot [mark=pentagon, green!60!black, solid,
              mark options={solid}, line width=.7, mark size=1.2] table
              [x index=0, y index=2] {experiments/timesummary.plt};
            \addplot [mark=o, blue!60!black, solid, mark options={solid},
              line width=.7, mark size=1.2] table [x index=0, y index=3]
              {experiments/timesummary.plt};
            \addplot [mark=x, black, solid, mark options={solid},
              line width=.7, mark size=1.2] table [x index=0, y index=4]
              {experiments/timesummary.plt};
            \addplot [mark=-, magenta, solid, mark options={solid},
              line width=.7, mark size=1.2] table [x index=0, y index=5]
              {experiments/timesummary.plt};
          \end{axis}
          \begin{axis}
            [
            xshift = 3.25in, xmode=log,
            width=8.5cm, height=7.5cm,
            ymode=log, ymin=10,
            name=axis2,
            ylabel near ticks, xlabel near ticks,
            every axis y label/.style={at={(axis description cs:-0.07,0.5)},
            rotate=90},
            every axis x label/.style={at={(axis description cs:0.5,-0.07)},
            rotate=0},
            xlabel={PP+}, ylabel={DP},
            enlargelimits=false
            ]
            \addplot[scatter/classes={0={mark=o,red, mark size=.7},
              1={mark=+,blue!70!black, mark size=1}}, scatter, only marks,
              scatter src=explicit symbolic] table [x=pppm, y=ppdp, meta=flag]
              {experiments/resultsscattertotal.plt};
            \addplot[-,mark=none, green!70!black, line width=.7, dashdotted]
              [domain=10:5000]{2*\x};
            \addplot[color=blue!70!black, line width=.7, domain=10:10000,
              mark=none] {\x};
            \addplot[color=green!70!black, line width=.7, mark=none,
              dashdotted] [domain=20:10000] {0.5*\x};
            \addplot[color=orange!70!black, line width=.7, mark=none,
              domain=40:10000, dashed] {0.25*\x};
            \addplot[color=red!70!black, line width=.7, domain=80:10000,
              mark=none, solid] {0.125*\x};
          \end{axis}

          \begin{customlegend}
            [
            legend entries={ $2\times$, $4\times$, $8\times$},
            legend style={at={(9.6,5.75)}, font=\footnotesize}
            ]
            \addlegendimage{dashdotted,green!70!black}
            \addlegendimage{dashed,orange!70!black}
            \addlegendimage{solid,red!70!black}
          \end{customlegend}

          \begin{scope}
            [every node/.style={text width=.48\textwidth,align=center}]
            \node [below=1.25em] at (axis1.south)
            {\captionof{figure}{Solution times on random games
            from~\cite{BDM16}.\label{fig:prf}}};
            \node [below=1.25em] at (axis2.south)
            {\captionof{figure}{Comparison between \PPP and \DP on random
            games with 50000 positions.\label{fig:sct}}};
          \end{scope}

        \end{tikzpicture}
      \end{center}
      \vspace{-2.5em}
    \end{figure}
    }

  }

\usepackage{multirow}

% ...

\AfterEndPreamble
  {

  \newcommand{\tabexmprtprm}
    {
    \begin{center}
      \scriptsize
      \scalebox{0.90}[0.75]
        {
        \begin{tabular}{|c||c|c|c|c|c|c|}
          \hline & & & & & & \\[-1.10em]
          & & & & & & \\[-0.85em]
          & 1 & 2 & 3 & 4 & 5 & 6 \\
          \hline \hline & & & & & & \\[-1.00em]
          & & & & & & \\[-0.80em]
          6 & $\aSym \!\downarrow$
          & $\cdots$
          & $\cdots$
          & $\cdots$
          & $\cdots$
          & $\!\!\aSym,\bSym,\dSym,\gSym,\iSym,\jSym \!\downarrow\!\!$ \\
          \hline & & & & & & \\[-1.00em]
          & & & & & & \\[-0.80em]
          5 & $\!\!\bSym,\fSym,\hSym \!\downarrow\!\!$
          & $\cdots$
          & $\cdots$
          & $\!\!\bSym,\dSym,\fSym,\gSym,\hSym \!\downarrow\!\!$
          & $\cdots$
          & \\
          \hline & & & & & & \\[-1.00em]
          & & & & & & \\[-0.80em]
          4 & $\cSym,\jSym \!\downarrow$
          & $\!\!\cSym,\eSym,\jSym \!\downarrow\!\!$
          & $\cdots$
          & $\cSym,\jSym \!\downarrow$
          & $\!\!\cSym,\eSym,\jSym \!\downarrow\!\!$
          & $\!\!\cSym,\eSym \!\uparrow_{6}\!\!$ \\
          \hline & & & & & & \\[-1.00em]
          & & & & & & \\[-0.80em]
          3 & $\dSym \!\downarrow$
          & $\dSym \!\downarrow$
          & $\!\!\dSym,\gSym \!\uparrow_{5}\!\!$
          &
          &
          & \\
          \hline & & & & & & \\[-1.00em]
          & & & & & & \\[-0.80em]
          2 & $\eSym \!\uparrow_{4}$
          &
          &
          & $\eSym \!\uparrow_{4}$
          &
          & \\
          \hline & & & & & & \\[-1.00em]
          & & & & & & \\[-0.80em]
          1 &
          & $\gSym \!\uparrow_{3}$
          &
          &
          &
          & \\
          \hline & & & & & & \\[-1.00em]
          & & & & & & \\[-0.80em]
          0 &
          &
          &
          &
          & $\iSym \!\uparrow_{6}$
          & \\
          \hline
        \end{tabular}
        }
      \vspace{-0.125em}
      \captionof{table}{\label{tab:exmprtprm} \PPP simulation.}
    \end{center}
    }

  \newcommand{\tabexmdelprm}
    {
    \begin{wraptable}[11]{R}{0.375\textwidth}
      \vspace{-1.5em}
      \begin{center}
        \scriptsize
        \scalebox{1.00}[1.00]
          {
          \begin{tabular}{|c||c|c|c|c|}
            \hline & & & & \\[-1.10em]
            & & & & \\[-0.85em]
            & 1 & 2 & 3 & 4 \\
            \hline \hline & & & & \\[-1.00em]
            & & & & \\[-0.80em]
            6 & $\aSym \!\downarrow$
            & $\cdots$
            & $\cdots$
            & $\!\!\aSym,\bSym,\dSym,\gSym,\iSym,\jSym \!\downarrow\!\!$ \\
            \hline & & & & \\[-1.00em]
            & & & & \\[-0.80em]
            5 & $\!\!\bSym,\fSym,\hSym \!\downarrow\!\!$
            & $\cdots$
            & $\cdots$
            & \\
            \hline & & & & \\[-1.00em]
            & & & & \\[-0.80em]
            4 & $\cSym,\jSym \!\downarrow$
            & $\!\!\cSym,\eSym,\jSym \!\downarrow\!\!$
            & $\cdots$
            & $\!\!\cSym,\eSym \!\uparrow_{6}\!\!$ \\
            \hline & & & & \\[-1.00em]
            & & & & \\[-0.80em]
            3 & $\dSym \!\downarrow$
            & $\dSym \!\downarrow$
            & $\!\!\dSym,\gSym \!\not\,\uparrow_{5}\!\!$
            & \\
            \hline & & & & \\[-1.00em]
            & & & & \\[-0.80em]
            2 & $\eSym \!\uparrow_{4}$
            &
            &
            & \\
            \hline & & & & \\[-1.00em]
            & & & & \\[-0.80em]
            1 &
            & $\gSym \!\uparrow_{3}$
            &
            & \\
            \hline & & & & \\[-1.00em]
            & & & & \\[-0.80em]
            0 &
            &
            & $\iSym \!\uparrow_{6}$
            & \\
            \hline
          \end{tabular}
          }
        \vspace{-0.2em}
        \caption{\label{tab:exmdelprm} \small \DP simulation.}
      \end{center}
      \vspace{-0em}
    \end{wraptable}
    }

  }

\usepackage[ruled,vlined]{algorithm2e}

\DontPrintSemicolon

\SetKw{Signature}{signature}
\SetKwFor{Function}{function}{}{}

\SetKwFor{Macro}{macro}{}{}
\SetKwIF{MIf}{MElseIf}{MElse}{\#if}{\#then}{\#else if}{\#else}{\#endif}

\SetKwFor{Let}{let}{in}{}

\newcommand{\nlr}[1]
  {
  \addtocounter{AlgoLine}{1}
  \nlset{\arabic{AlgoLine}-\addtocounter{AlgoLine}{#1}\arabic{AlgoLine}}
  }

\AfterEndPreamble
  {

  \newcommand{\algsrc}
    {
    \begin{wrapfigure}[10]{R}{0.378\textwidth}
      \vspace{-1.3em}
      \hspace{-5pt}\begin{algorithm}[H]
        \caption{\label{alg:src} The Searcher.~\cite{BDM16}}
        \Signature{$\srcFun[\DName] : \SSet[\DName] \to \QDSet[
          {\GamName[\DName]} ][+]$} \;
        \Function{$\srcFun[\DName][][\sElm]$}
          {
          \nl $(\QSet, \alpha) \gets \qryFun[\DName][][\sElm]$ \;
          \nl \eIf{$(\QSet, \alpha) \in \QDSet[ {\GamName[\DName]} ][+]$}
            {
            \nl \Return $(\QSet, \alpha)$ \;
            }
            {
            \nl \Return $\srcFun[\DName][][\sElm {\,\sucFun[\DName]} (\QSet,
                \alpha)]$\!\! \;
            }
          }
      \end{algorithm}
      \vspace{-0.00em}
    \end{wrapfigure}
    }

  \newcommand{\algqry}
    {
    \begin{wrapfigure}[9]{R}{0.375\textwidth}
      \vspace{-.10em}
      \begin{algorithm}[H]
        \caption{\label{alg:qry} Query Function.}
        \Signature{$\qryFun : \SSet \to \pow{\PosSet} \!\times\! \{ 0, 1
          \}$}\!\!\! \;
        \Function{$\qryFun[][][\sElm]$}
          {
          \Let{$(\rFun, \pElm) = \sElm$}
            {
            \nl $\alpha \gets \pElm \bmod{2}$ \;
            \nl $\RSet \gets \atrFun[ {\GamName[\sElm]}
                ][\alpha][{\rFun[][-1](\pElm)}]$ \;
            }
          \nl \Return $(\RSet, \alpha)$ \;
          }
      \end{algorithm}
      \vspace{-0.00em}
    \end{wrapfigure}
    }

  \newcommand{\algsucprtprm}
    {
    \begin{wrapfigure}[12]{R}{0.485\textwidth}
      \vspace{-1.7em}
      \SetVlineSkip{0pt}
      \begin{algorithm}[H]
        \caption{\label{alg:sucprtprm} Successor Function.}
        \Signature{$\sucFun : \comRel \to \Delta \times \PrtSet$} \;
        \Function{$\sElm \,\sucFun\, (\RSet, \alpha)$}
          {
          \Let{$(\rFun, \pElm) = \sElm$}
            {
            \nl \eIf{$(\RSet, \alpha) \in \RegSet[ {\GamName[\sElm]} ][-]$}
              {
              \nl $\rFun[][\star] \gets {\rFun}[\RSet \mapsto \pElm]$ \;
              \vspace{-.3em}\nl $\pElm[][\star] \gets \max(\rng{\rFun[][\star]^{(< \pElm)}})$
                  \;\vspace{-.3em}
              }
              {\vspace{-.3em}
              \nl $\pElm[][\star] \gets \bepFun[][\dual{\alpha}][\RSet, \rFun]$
                  \;
              \nl $\rFun[][\star] \gets \prtFun \uplus \rFun^{(\geq
                  \pElm[][\star]) \lor (\equiv_{2} \pElm[][\star])}[\RSet
                  \mapsto {\pElm[][\star]}]$\!\! \;
              }
            }
           \nl \Return $(\rFun[][\star], \pElm[][\star])$ \;
          }
      \end{algorithm}
      \vspace{-0.00em}
    \end{wrapfigure}
    }

  \newcommand{\algsucdelprm}
    {
    \SetVlineSkip{1.5pt}
    \begin{algorithm}[H]
      \caption{\label{alg:sucdelprm} Successor Function.}
      \Signature{$\sucFun : \comRel \to (\Delta \!\times\! \PrtSet) \!\times\!
        \Delta^{\pto} \!\times\! \pow{\PrtSet}$\!} \;
      \Function{$\sElm \,\sucFun\, (\RSet, \alpha)$}
        {
        \Let{$((\rFun, \pElm), \adj{\rFun}, \PSet) = \sElm$}
          {
          \nl \eIf{$(\RSet, \alpha) \in \RegSet[ {\GamName[\sElm]} ][-]$}
            {
            \nlr{3} $\mthelm{\#Assignment}(\Tt)$ \;
            }
            {
            \nl $\qElm \gets \bepFun[][\dual{\alpha}][\RSet, \rFun \uplus
                \adj{\rFun}]$ \;
            \nl \eIf{$\phi_{Lck}(\qElm, \sElm)$}
              {
              \nl $\der{\rFun} \gets {\adj{\rFun}}[\RSet \mapsto \qElm]$ \;
              \nl \eIf{$\RSet \neq \PosSet[ {\GamName[\sElm]} ]$}
                {
                \nlr{3} $\mthelm{\#Assignment}(\Ff)$ \;
                }
                {
                \nlr{3} $\mthelm{\#DelayedPromotion}$ \;
                }
              }
              {
              \nlr{3} $\mthelm{\#InstantPromotion}$ \;
              }
            }
          }
        \nl \Return $((\rFun[][\star], \pElm[][\star]), \adj{\rFun}[][\star],
            \PSet[][\star])$ \;
        }
    \end{algorithm}
    }

  \newcommand{\algmacdelprm}
    {
    \NoCaptionOfAlgo
    \begin{algorithm}[H]
      \caption{Assignment \& Promotion Macros.}
      \vspace{0.055em}
      \Macro{$\mthelm{\#Assignment(\xi)}$}
        {
        \nl $\rFun[][\star] \gets {\rFun}[\RSet \mapsto \pElm]$ \;
        \nl $\pElm[][\star] \gets \max(\rng{\rFun[][\star]^{(< \pElm)}})$ \;
        \nl $\adj{\rFun}[][\star] \gets \text{\bfseries{\#if }} \xi
            \text{\bfseries{ \#then }} \adj{\rFun} \text{\bfseries{ \#else }}
            \der{\rFun}$ \;
        \nl $\PSet[][\star] \gets \PSet$ \;
        }
      \vspace{0.051em}
      \setcounter{AlgoLine}{0}
      \Macro{$\mthelm{\#DelayedPromotion}$}
        {
        \nl $\pElm[][\star] \gets \max(\rng{\der{\rFun}})$ \;
        \nl $\rFun[][\star] \gets \prtFun \uplus (\rFun \uplus
            \der{\rFun})^{(\geq \pElm[][\star]) \lor (\equiv_{2}
            \pElm[][\star])}$ \;
        \nl $\adj{\rFun}[][\star] \gets \emptyset$ \;
        \nl $\PSet[][\star] \gets \emptyset$ \;
        }
      \vspace{0.051em}
      \setcounter{AlgoLine}{0}
      \Macro{$\mthelm{\#InstantPromotion}$}
        {
        \nl $\pElm[][\star] \gets \qElm$ \;
        \nl $\rFun[][\star] \gets \prtFun \uplus \rFun^{(\geq \pElm[][\star])
            \lor (\equiv_{2} \pElm[][\star])}[\RSet \mapsto {\pElm[][\star]}]$
            \;
        \nl $\adj{\rFun}[][\star] \gets \adj{\rFun}^{(> \pElm[][\star])}$ \;
        \nl $\PSet[][\star] \gets \PSet \cap \rng{\rFun[][\star]} \cup \{
            \pElm[][\star] \}$ \;
        }
      \vspace{0.055em}
    \end{algorithm}
    }

  }

\hypersetup
  {
  pdftitle  = {A Delayed Promotion Policy for Parity Games},
  pdfauthor = {M. Benerecetti, D. Dell'Erba, \& F. Mogavero}
  }

\begin{document}

  \title{A Delayed Promotion Policy for Parity Games}
  \def\titlerunning{A Delayed Promotion Policy for Parity Games}

  \author{Massimo Benerecetti \institute{Universit\`a degli Studi di Napoli
  Federico II} \and Daniele Dell'Erba \institute{Universit\`a degli Studi di
  Napoli Federico II} \and Fabio Mogavero \institute{Oxford University}}

  \def\authorrunning{M. Benerecetti, D. Dell'Erba, \& F. Mogavero}

  \maketitle



\begin{abstract}

  \emph{Parity games} are two-player infinite-duration games on graphs that play
  a crucial role in various fields of theoretical computer science.
  Finding efficient algorithms to solve these games in practice is widely
  acknowledged as a core problem in formal verification, as it leads to
  efficient solutions of the model-checking and satisfiability problems of
  expressive temporal logics, \eg, the modal \MC.
  Their solution can be reduced to the problem of identifying sets of positions
  of the game, called dominions, in each of which a player can force a win by
  remaining in the set forever.
  Recently, a novel technique to compute dominions, called \emph{priority
  promotion}, has been proposed, which is based on the notions of quasi
  dominion, a relaxed form of dominion, and dominion space.
  The underlying framework is general enough to accommodate different
  instantiations of the solution procedure, whose correctness is ensured by the
  nature of the space itself.
  In this paper we propose a new such instantiation, called \emph{delayed
  promotion}, that tries to reduce the possible exponential behaviours exhibited
  by the original method in the worst case.
  The resulting procedure not only often outperforms the original priority
  promotion approach, but so far no exponential worst case is known.

\end{abstract}





\begin{section}{Introduction}

  The abstract concept of \emph{game} has proved to be a fruitful metaphor in
  theoretical computer science~\cite{AG11}.
  Several \emph{decision problems} can, indeed, be encoded as \emph{path-forming
  games on graphs}, where a player willing to achieve a certain goal, usually the
  verification of some property on the plays derived from the original problem,
  has to face an opponent whose aim is to pursue the exact opposite task.
  One of the most prominent instances of this connection is represented by the
  notion of \emph{parity game}~\cite{Mos91}, a simple two-player turn-based
  perfect-information game played on directed graphs, whose nodes are labelled
  with natural numbers called \emph{priorities}.
  The goal of the first (\resp, second) player, \aka, \emph{even} (\resp,
  \emph{odd}) player, is to force a play $\pthElm$, whose maximal priority
  occurring infinitely often along $\pthElm$ is of even (\resp, odd) parity.
  The importance of these games is due to the numerous applications in the area
  of system specification, verification, and synthesis, where it is used as
  algorithmic back-end of satisfiability and model-checking procedures for
  temporal logics~\cite{EJ88,EJS93,KVW00}, and as a core for several techniques
  employed in automata theory~\cite{Mos84,EJ91,KV98,GTW02}.
  In particular, it has been proved to be linear-time interreducible with the
  model-checking problem for the \emph{modal \MC}~\cite{EJS93} and it is closely
  related to other games of infinite duration, as \emph{mean
  payoff}~\cite{EM79,GKK90}, \emph{discounted payoff}~\cite{ZP96}, \emph{simple
  stochastic}~\cite{Con92}, and \emph{energy}~\cite{CDHR10} games.
  Besides the practical importance, parity games are also interesting from a
  computational complexity point of view, since their solution problem is one of
  the few inhabitants of the \UPTime $\cap$ \CoUPTime class~\cite{Jur98}.
  That result improves the \NPTime $\cap$ \CoNPTime membership~\cite{EJS93},
  which easily follows from the property of \emph{memoryless
  determinacy}~\cite{EJ91,Mos91}.
  Still open is the question about the membership in \PTime.
  The literature on the topic is reach of algorithms for solving parity games,
  which can be mainly classified into two families.
  The first one contains the algorithms that, by employing a \divideetimpera
  approach, recursively decompose the problem into subproblems, whose solutions
  are then suitably assembled to obtain the desired result.
  In this category fall, for example, \emph{Zielonka's recursive
  algorithm}~\cite{Zie98} and its \emph{dominion decomposition}~\cite{JPZ08} and
  \emph{big step}~\cite{Sch07} improvements.
  The second family, instead, groups together those algorithms that try to
  compute a winning strategy for the two players on the entire game.
  The principal members of this category are represented by \emph{Jurdzi\'nski's
  progress measure} algorithm~\cite{Jur00} and the \emph{strategy improvement}
  approaches~\cite{VJ00,Sch08a,STV15}.

  Recently, a new \divideetimpera solution algorithm, called \emph{priority
  promotion} (\PP, for short), has been proposed in~\cite{BDM16}, which is fully
  based on the decomposition of the winning regions into \emph{dominions}.
  The idea is to find a dominion for some of the two players and then remove it
  from the game, thereby allowing for a recursive solution.
  The important difference \wrt the other two approaches~\cite{JPZ08,Sch07}
  based on the same notion is that these procedures only look for dominions of a
  certain size in order to speed up classic Zielonka's algorithm in the worst
  case.
  Consequently, they strongly rely on this algorithm for their completeness.
  On the contrary, the \PP procedure autonomously computes dominions of any
  size, by suitably composing quasi dominions, a weaker notion of dominion.
  Intuitively, a \emph{quasi dominion} $\QSet$ for player $\alpha \in \{ 0, 1
  \}$ is a set of vertices from each of which player $\alpha$ can enforce a
  winning play that never leaves the region, unless one of the following two
  conditions holds: \emph{(i)} the opponent $\dual{\alpha}$ can escape from
  $\QSet$ (\ie, there is an edge from a vertex of $\dual{\alpha}$ exiting from
  $\QSet$) or \emph{(ii)} the only choice for player $\alpha$ itself is to exit
  from $\QSet$ (\ie, no edge from a vertex of $\alpha$ remains in $\QSet$).
  A crucial feature of quasi dominion is that they can be ordered by assigning
  to each of them a priority corresponding to an under-approximation of the best
  value for $\alpha$ the opponent $\dual{\alpha}$ can be forced to visit along
  any play exiting from it.
  Indeed, under suitable and easy to check assumptions, a higher priority quasi
  $\alpha$-dominion $\QSet[1]$ and a lower priority one $\QSet[2]$, can be
  merged into a single quasi $\alpha$-dominion of the higher priority, thus
  improving the approximation for $\QSet[2]$.
  This merging operation is called a priority promotion of $\QSet[2]$ to
  $\QSet[1]$.
  The \PP solution procedure has been shown to be very effective in practice and
  to often significantly outperform all other solvers.
  Moreover, it also improves on the space complexity of the best know algorithm
  with an exponential gain \wrt the number of priorities and by a logarithmic
  factor \wrt the number of vertexes.
  Indeed, it only needs $\AOmicron{n \cdot \log{k}}$ space against the
  $\AOmicron{k \cdot n \cdot \log{n}}$ required by Jurdzi\'nski's
  approach~\cite{Jur00}, where $n$ and $k$ are, respectively, the numbers of
  vertexes and priorities of the game.
  Unfortunately, the \PP algorithm also exhibits exponential behaviours on a
  simple family of games.
  This is due to the fact that, in general, promotions to higher priorities
  requires resetting promotions previously performed at lower ones.

  In this paper, we continue the study of the priority promotion approaches
  trying to find a remedy to this problem.
  We propose a new algorithm, called \DP, built on top of a slight variation of
  \PP, called \PPP.
  The \PPP algorithm simply avoids resetting previous promotions to quasi
  dominions of the same parity.
  In this case, indeed, the relevant properties of those quasi dominions are
  still preserved.
  This variation enables the new \DP promotion policy, that delays promotions
  that require a reset and only performs those leading to the highest quasi
  dominions among the available ones.
  For the resulting algorithm no exponential worst case has been found.
  Experiments on randomly generated games also show that the new approach
  performs much better than \PP in practice, while still preserving the same
  space complexity.

\end{section}





\begin{section}{Preliminaries}
  \label{sec:prl}

  Let us first briefly recall the notation and basic definitions concerning
  parity games that expert readers can simply skip.
  We refer to~\cite{AG11}\cite{Zie98} for a comprehensive presentation of the
  subject.

  Given a partial function $\fFun : \ASet \pto \BSet$, by $\dom{\fFun} \subseteq
  \ASet$ and $\rng{\fFun} \subseteq \BSet$ we indicate the domain and range of
  $\fFun$, respectively.
  In addition, $\uplus$ denotes the \emph{completion operator} that, taken
  $\fFun$ and another partial function $\gFun : \ASet \pto \BSet$, returns the
  partial function $\fFun \uplus \gFun \defeq (\fFun \setminus \dom{\gFun}) \cup
  \gFun : \ASet \pto \BSet$, which is equal to $\gFun$ on its domain and assumes
  the same values of $\fFun$ on the remaining part of $\ASet$.

  A two-player turn-based \emph{arena} is a tuple $\ArnName =
  \tuplec{\PosSet[][0]}{\PosSet[][1]}{\MovRel}$, with $\PosSet[][0] \cap
  \PosSet[][1] = \emptyset$ and $\PosSet \defeq \PosSet[][0] \cup \PosSet[][1]$,
  such that $\tupleb{\PosSet}{\MovRel}$ is a finite directed graph.
  $\PosSet[][0]$ (\resp, $\PosSet[][1]$) is the set of positions of player $0$
  (\resp, $1$) and $\MovRel \subseteq \PosSet \times \PosSet$ is a left-total
  relation describing all possible moves.
  A \emph{path} in $\VSet \subseteq \PosSet$ is a finite or infinite sequence
  $\pthElm \in \PthSet(\VSet)$ of positions in $\VSet$ compatible with the move
  relation, \ie, $(\pthElm_{i}, \pthElm_{i + 1}) \in \MovRel$, for all $i \in
  \numco{0}{\card{\pthElm} - 1}$.
  For a finite path $\pthElm$, with $\lst{\pthElm}$ we denote the last position
  of $\pthElm$.
  A positional \emph{strategy} for player $\alpha \in \{ 0, 1 \}$ on $\VSet
  \subseteq \PosSet$ is a partial function $\strElm[\alpha] \in
  \StrSet[][\alpha](\VSet) \subseteq (\VSet \cap \PosSet[][\alpha]) \pto \VSet$,
  mapping each $\alpha$-position $\posElm \in \dom{\strElm[\alpha]}$
  to position $\strElm[\alpha](\posElm)$ compatible with the move relation,
  \ie, $(\posElm, \strElm[\alpha](\posElm)) \in \MovRel$.
  With $\StrSet[][\alpha](\VSet)$ we denote the set of all $\alpha$-strategies
  on $\VSet$.
  A \emph{play} in $\VSet \subseteq \PosSet$ from a position $\posElm \in \VSet$
  \wrt a pair of strategies $(\strElm[0], \strElm[1]) \in \StrSet[][0](\VSet)
  \times \StrSet[][1](\VSet)$, called \emph{$((\strElm[0], \strElm[1]),
  \posElm)$-play}, is a path $\pthElm \in \PthSet(\VSet)$ such that $\pthElm[0]
  = \posElm$ and, for all $i \in \numco{0}{\card{\pthElm} - 1}$, if $\pthElm_{i}
  \in \PosSet[][0]$ then $\pthElm_{i + 1} = \strElm[][0](\pthElm_{i})$ else
  $\pthElm_{i + 1} = \strElm[][1](\pthElm_{i})$.
  The \emph{play function} $\playFun : (\StrSet[][0](\VSet) \times
  \StrSet[][1](\VSet)) \times \VSet \to \PthSet(\VSet)$ returns, for each
  position $\posElm \in \VSet$ and pair of strategies $(\strElm[0], \strElm[1])
  \in \StrSet[][0](\VSet) \times \StrSet[][1](\VSet)$, the maximal
  $((\strElm[0], \strElm[1]), \posElm)$-play $\playFun((\strElm[][0],
  \strElm[][1]), \posElm)$.

  A \emph{parity game} is a tuple $\GamName =
  \tuplec{\ArnName}{\PrtSet}{\prtFun}$, where $\ArnName$ is an arena, $\PrtSet
  \subset \SetN$ is a finite set of priorities, and $\prtFun : \PosSet \to
  \PrtSet$ is a \emph{priority function} assigning a priority to each position.
  The priority function can be naturally extended to games and paths as follows:
  $\prtFun(\GamName) \defeq \max[\posElm \in \PosSet] \, \prtFun(\posElm)$; for
  a path $\pthElm \in \PthSet$, we set $\prtFun(\pthElm) \defeq \max_{i \in
  \numco{0}{\card{\pthElm}}} \, \prtFun(\pthElm_{i})$, if $\pthElm$ is finite,
  and $\prtFun(\pthElm) \defeq \limsup_{i \in \SetN} \prtFun(\pthElm_{i})$,
  otherwise.
  A set of positions $\VSet \subseteq \PosSet$ is an $\alpha$-\emph{dominion},
  with $\alpha \in \{ 0, 1 \}$, if there exists an $\alpha$-strategy
  $\strElm[\alpha] \in \StrSet[][\alpha](\VSet)$ such that, for all
  $\dual{\alpha}$-strategies $\strElm[\dual{\alpha}] \in
  \StrSet[][\dual{\alpha}](\VSet)$ and positions $\posElm \in \VSet$, the
  induced play $\pthElm = \playFun((\strElm[0], \strElm[1]), \posElm)$ is
  infinite and $\prtFun(\pthElm) \equiv_{2} \alpha$.
  In other words, $\strElm[\alpha]$ only induces on $\VSet$ infinite plays whose
  maximal priority visited infinitely often has parity $\alpha$.
  By $\GamName \!\setminus\! \VSet$ we denote the maximal subgame of $\GamName$
  with set of positions $\PosSet'$ contained in $\PosSet \!\setminus\! \VSet$
  and move relation $\MovRel'$ equal to the restriction of $\MovRel$ to
  $\PosSet'$.

  The $\alpha$-predecessor of $\VSet$, in symbols $\preFun[][\alpha](\VSet)
  \defeq \set{ \posElm \in \PosSet[][\alpha] }{ \MovRel(\posElm) \cap \VSet \neq
  \emptyset } \cup \set{ \posElm \in \PosSet[][\dual{\alpha}] }{
  \MovRel(\posElm) \subseteq \VSet }$, collects the positions from which player
  $\alpha$ can force the game to reach some position in $\VSet$ with a single
  move.
  The $\alpha$-attractor $\atrFun[][\alpha](\VSet)$ generalises the notion of
  $\alpha$-predecessor $\preFun[][\alpha](\VSet)$ to an arbitrary number of
  moves, and corresponds to the least fix-point of that operator.
  When $\VSet = \atrFun[][\alpha](\VSet)$, we say that $\VSet$ is
  $\alpha$-maximal.
  Intuitively, $\VSet$ is $\alpha$-maximal if player $\alpha$ cannot force any
  position outside $\VSet$ to enter the set.
  For such a $\VSet$, the set of positions of the subgame $\GamName \setminus
  \VSet$ is precisely $\PosSet \setminus \VSet$.
  Finally, the set
  $\escFun[][\alpha](\VSet) \defeq \preFun[][\alpha](\PosSet \setminus \VSet)
  \cap \VSet$, called the \emph{$\alpha$-escape} of $\VSet$, contains the
  positions in $\VSet$ from which $\alpha$ can leave $\VSet$ in one move.  The
  dual notion of \emph{$\alpha$-interior}, defined as
  $\intFun[][\alpha](\VSet) \defeq (\VSet \cap \PosSet[][\alpha]) \setminus
  \escFun[][\alpha][\VSet]$, contains, instead, the $\alpha$-positions from which
  $\alpha$ cannot escape with a single move.
  All the operators and sets above actually depend on the specific game
  $\GamName$ they are applied in.
  In the rest of the paper, we shall only add $\GamName$ as subscript of an
  operator, \eg, $\escFun[\GamName][\alpha][\VSet]$, when the game is not clear
  from the context.

\end{section}





\begin{section}{The Priority Promotion Approach}
  \label{sec:prtprmapr}

  The priority promotion approach proposed in~\cite{BDM16} attacks the problem
  of solving a parity game $\GamName$ by computing one of its dominions $\DSet$,
  for some player $\alpha \in \{ 0, 1 \}$, at a time.
  Indeed, once the $\alpha$-attractor $\DSet[][\star]$ of $\DSet$ is removed
  from $\GamName$, the smaller game $\GamName \setminus \DSet[][\star]$ is
  obtained, whose positions are winning for one player iff they are winning for
  the same player in the original game.
  This allows for decomposing the problem of solving a parity game to that of
  iteratively finding its dominions~\cite{JPZ08}.

  In order to solve the dominion problem, the idea is to start from a much
  weaker notion than that of dominion, called \emph{quasi dominion}.
  Intuitively, a quasi $\alpha$-dominion is a set of positions on which player
  $\alpha$ has a strategy whose induced plays either remain inside the set
  forever and are winning for $\alpha$ or can exit from it passing through a
  specific set of escape positions.

  \begin{definition}[Quasi Dominion~\cite{BDM16}]
    \label{def:qsidom}
    Let $\GamName \in \PG$ be a game and $\alpha \in \{ 0, 1 \}$ a player.
    A non-empty set of positions $\QSet \subseteq \PosSet$ is a \emph{quasi
    $\alpha$-dominion} in $\GamName$ if there exists an $\alpha$-strategy
    $\strElm[\alpha] \in \StrSet[][\alpha](\QSet)$ such that, for all
    $\dual{\alpha}$-strategies $\strElm[\dual{\alpha}] \in
    \StrSet[][\dual{\alpha}](\QSet)$, with $\intFun[][\dual{\alpha}](\QSet)
    \subseteq \dom{\strElm[\dual{\alpha}]}$, and positions $\posElm \in \QSet$,
    the induced play $\pthElm = \playFun((\strElm[0], \strElm[1]), \posElm)$
    satisfies $\prtFun(\pthElm) \equiv_{2} \alpha$, if $\pthElm$ is infinite,
    and $\lst{\pthElm} \in \escFun[][\dual{\alpha}][\QSet]$, otherwise.
  \end{definition}

  Observe that, if all the induced plays remain in the set $\QSet$ forever, this
  is actually an $\alpha$-dominion and, therefore, a subset of the winning
  region $\WinSet[\alpha]$ of $\alpha$.
  In this case, the escape set of $\QSet$ is empty, \ie,
  $\escFun[][\dual{\alpha}][\QSet] = \emptyset$, and $\QSet$ is said to be
  \emph{$\alpha$-closed}.
  In general, however, a quasi $\alpha$-dominion $\QSet$ that is not an
  $\alpha$-dominion, \ie, such that $\escFun[][\dual{\alpha}][\QSet] \neq
  \emptyset$, need not be a subset of $\WinSet[\alpha]$ and it is called
  \emph{$\alpha$-open}.
  Indeed, in this case, some induced play may not satisfy the winning condition
  for that player once exited from $\QSet$, by visiting a cycle containing a
  position with maximal priority of parity $\dual{\alpha}$.
  The set of pairs $(\QSet, \alpha) \in \pow{\PosSet} \times \{ 0, 1 \}$, where
  $\QSet$ is a quasi $\alpha$-dominion, is denoted by $\QDSet$, and is
  partitioned into the sets $\QDSet[][-]$ and $\QDSet[][+]$ of open and closed
  quasi $\alpha$-dominion pairs, respectively.

  \algsrc

  The priority promotion algorithm explores a partial order, whose elements,
  called \emph{states}, record information about the open quasi dominions
  computed along the way.
  The initial state of the search is the top element of the order, where the
  quasi dominions are initialised to the sets of positions with the same
  priority.
  At each step, a new quasi dominion is extracted from the current state, by
  means of a \emph{query} operator $\qryFun$, and used to compute a successor
  state, by means of a \emph{successor} operator $\sucFun$, if the quasi
  dominion is open.
  If, on the other hand, it is closed, the search is over.
  Algorithm~\ref{alg:src} implements the dominion search procedure
  $\srcFun[\DName]$.
  A \emph{compatibility relation} $\comRel$ connects the query and the successor
  operators.
  The relation holds between states of the partial order and the quasi dominions
  that can be extracted by the query operator.
  Such a relation defines the domain of the successor operator.
  The partial order, together with the query and successor operator and the
  compatibility relation, forms what is called a \emph{dominion space}.

  \begin{definition}[Dominion Space~\cite{BDM16}]
    \label{def:domspc}
    A \emph{dominion space} for a game $\GamName \in \PG$ is a tuple $\DName
    \defeq \tuplee{\GamName} {\SName} {\comRel} {\qryFun} {\sucFun}$, where
    \emph{(1)} $\SName \defeq \tuplec{\SSet} {\topElm} {\ordRel}$ is a
    well-founded partial order \wrt $\ordRel \:\subset \SSet \times \SSet$ with
    distinguished element $\topElm \in \SSet$, \emph{(2)} $\comRel \subseteq
    \SSet \times \QDSet[][-]$ is the \emph{compatibility relation}, \emph{(3)}
    $\qryFun : \SSet \to \QDSet$ is the \emph{query operator} mapping each
    element $\sElm \in \SSet$ to a quasi dominion pair $(\QSet, \alpha) \defeq
    \qryFun(\sElm) \in \QDSet$ such that, if $(\QSet, \alpha) \in \QDSet[][-]$
    then $\sElm \,\comRel (\QSet, \alpha)$, and \emph{(4)} $\sucFun : \comRel
    \to \SSet$ is the \emph{successor operator} mapping each pair $(\sElm,
    (\QSet, \alpha)) \in \comRel$ to the element $\sElm[][\star] \defeq \sElm
    \,\sucFun (\QSet, \alpha) \in \SSet$ with $\sElm[][\star] \ordRel \sElm$.
  \end{definition}

  The notion of dominion space is quite general and can be instantiated in
  different ways, by providing specific query and successor operators.
  In~\cite{BDM16}, indeed, it is shown that the search procedure
  $\srcFun[\DName]$ is sound and complete on any dominion space $\DName$.
  In addition, its time complexity is linear in the \emph{execution depth} of
  the dominion space, namely the length of the longest chain in the underlying
  partial order compatible with the successor operator, while its space
  complexity is only logarithmic in the space \emph{size}, since only one state
  at the time needs to be maintained.
  A specific instantiation of dominion space, called \emph{\PP dominion space},
  is the one proposed and studied in~\cite{BDM16}.
  Here we propose a different one, starting from a slight optimisation, called
  \PPP, of that original version.

  \begin{paragraph}{\PPP Dominion Space.}

    In order to instantiate a dominion space, we need to define a suitable query
    function to compute quasi dominions and a successor operator to ensure
    progress in the search for a closed dominion.
    The priority promotion algorithm proceeds as follows.
    The input game is processed in descending order of priority.
    At each step, a subgame of the entire game, obtained by removing the quasi
    domains previously computed at higher priorities, is considered.
    At each priority of parity $\alpha$, a quasi $\alpha$-domain $\QSet$ is
    extracted by the query operator from the current subgame.
    If $\QSet$ is closed in the entire game, the search stops and returns
    $\QSet$ as result.
    Otherwise, a successor state in the underlying partial order is computed by
    the successor operator, depending on whether $\QSet$ is open in the current
    subgame or not.
    In the first case, the quasi $\alpha$-dominion is removed from the current
    subgame and the search restarts on the new subgame that can only contain
    position with lower priorities.
    In the second case, $\QSet$ is merged together with some previously
    computed quasi $\alpha$-dominion with higher priority.
    Being a dominion space well-ordered, the search is guaranteed to eventually
    terminate and return a closed quasi dominion.
    The procedure requires the solution of two crucial problems: \emph{(a)}
    \emph{extracting a quasi dominion} from a subgame and \emph{(b)}
    \emph{merging together two quasi $\alpha$-dominions} to obtain a bigger,
    possibly closed, quasi $\alpha$-dominion.

    The solution of the first problem relies on the definition of a specific
    class of quasi dominions, called \emph{regions}.
    An $\alpha$-region $\RSet$ of a game $\GamName$ is a special form of quasi
    $\alpha$-dominion of $\GamName$ with the additional requirement that all the
    escape positions in $\escFun[][\dual{\alpha}](\RSet)$ have the
    maximal priority $\pElm \defeq \prtFun(\GamName) \equiv_{2} \alpha$ in
    $\GamName$.
    In this case, we say that $\alpha$-region $\RSet$ has priority $\pElm$.
    As a consequence, if the opponent $\dual{\alpha}$ can escape from the
    $\alpha$-region $\RSet$, it must visit a position with the highest priority
    in it, which is of parity $\alpha$.

    \begin{definition}[Region~\cite{BDM16}]
      \label{def:reg}
      A quasi $\alpha$-dominion $\RSet$ is an $\alpha$-region in $\GamName$ if
      $\prtFun[][][\GamName] \equiv_{2} \alpha$ and all the positions in
      $\escFun[][\dual{\alpha}][\RSet]$ have priority $\prtFun[][][\GamName]$,
      \ie
      $\escFun[][\dual{\alpha}][\RSet] \subseteq \prtFun[][-1][
      {\prtFun[][][\GamName]} ]$.
    \end{definition}

    It is important to observe that, in any parity game, an $\alpha$-region
    always exists, for some $\alpha \in \{ 0, 1 \}$.
    In particular, the set of positions of maximal priority in the game always
    forms an $\alpha$-region, with $\alpha$ equal to the parity of that maximal
    priority.
    In addition, the $\alpha$-attractor of an $\alpha$-region is always an
    ($\alpha$-maximal) $\alpha$-region.
    A closed $\alpha$-region in a game is clearly an $\alpha$-dominion in that
    game.
    These observations give us an easy and efficient way to extract a quasi
    dominion from every subgame: collect the $\alpha$-attractor of the
    positions with maximal priority $\pElm$ in the subgame, where
    $\pElm \equiv_{2} \alpha$, and assign $\pElm$ as priority of the resulting
    region $\RSet$.
    This priority, called \emph{measure} of $\RSet$, intuitively corresponds to
    an under-approximation of the best priority player $\alpha$ can force the
    opponent $\dual{\alpha}$ to visit along any play exiting from $\RSet$.

    \begin{proposition}[Region Extension~\cite{BDM16}]
      \label{prp:regext}
      Let $\GamName \in \PG$ be a game and $\RSet \subseteq \PosSet$ an
      $\alpha$-region in $\GamName$.
      Then, $\RSet[][\star] \defeq \atrFun[][\alpha][\RSet]$ is an
      $\alpha$-maximal $\alpha$-region in $\GamName$.
    \end{proposition}

    A solution to the second problem, the merging operation, is obtained as
    follows.
    Given an $\alpha$-region $\RSet$ in some game $\GamName$ and an
    $\alpha$-dominion $\DSet$ in a subgame of $\GamName$ that does not contain
    $\RSet$ itself, the two sets are merged together, if the only moves exiting
    from $\dual{\alpha}$-positions of $\DSet$ in the entire game lead to higher
    priority $\alpha$-regions and $\RSet$ has the lowest priority among them.
    The priority of $\RSet$ is called the \emph{best escape priority} of
    $\DSet$ for $\dual{\alpha}$.
    The correctness of this merging operation is established by the following
    proposition.

    \begin{proposition}[Region Merging~\cite{BDM16}]
      \label{prp:regmer}
      Let $\GamName \in \PG$ be a game, $\RSet \subseteq \PosSet$ an
      $\alpha$-region, and $\DSet \subseteq \PosSet[\GamName \setminus \RSet]$
      an $\alpha$-dominion in the subgame $\GamName \setminus \RSet$.
      Then, $\RSet[][\star] \defeq \RSet \cup \DSet$ is an $\alpha$-region in
      $\GamName$.
      Moreover, if both $\RSet$ and $\DSet$ are $\alpha$-maximal in $\GamName$
      and $\GamName \setminus \RSet$, respectively, then $\RSet[][\star]$ is
      $\alpha$-maximal in $\GamName$ as well.
    \end{proposition}

    The merging operation is implemented by promoting all the positions of
    $\alpha$-dominion $\DSet$ to the measure of $\RSet$, thus improving the
    measure of $\DSet$.
    For this reason, it is called a \emph{priority promotion}.
    In~\cite{BDM16} it is shown that, after a promotion to some measure $\pElm$,
    the regions with measure lower than $\pElm$ might need to be destroyed, by
    resetting all the contained positions to their original priority.
    This necessity derives from the fact that the new promoted region may
    attract positions from lower ones, thereby potentially invalidating their
    status as regions.
    Indeed, in some cases, the player that wins by remaining in the region may
    even change from $\alpha$ to $\dual{\alpha}$.
    As a consequence, the reset operation is, in general, unavoidable.
    The original priority promotion algorithm applies the reset operation to all
    the lower priority regions.
    However, the following property ensures that this can be limited to the
    regions belonging to the opponent player only.

    \begin{proposition}[Region Splitting]
      \label{prp:regspl}
      Let $\GamName[][\star] \in \PG$ be a game and $\RSet[][\star] \subseteq
      \PosSet[ {\GamName[][\star]} ]$ an $\alpha$-maximal $\alpha$-region in
      $\GamName[][\star]$.
      For any subgame $\GamName$ of $\GamName[][\star]$ and $\alpha$-region
      $\RSet \subseteq \PosSet$ in $\GamName$, if $\RSet[][\natural]
      \defeq \RSet \setminus \RSet[][\star] \neq \emptyset$, then
      $\RSet[][\natural]$ is an $\alpha$-region in $\GamName \setminus
      \RSet[][\star]$.
    \end{proposition}

    This proposition, together with the observation that $\dual{\alpha}$-regions
    that can be extracted from the corresponding subgames cannot attract
    positions contained in any retained $\alpha$-region, allows for preserving
    all the lower $\alpha$-regions computed so far.

    To exemplify the idea, Table~\ref{tab:exmprtprm} shows a simulation of the
    resulting procedure on the parity game of Figure~\ref{fig:exm}, where
    diamond shaped positions belong to player $0$ and square shaped ones to its
    opponent $1$.
    Player $0$ wins the entire game, hence the $0$-region containing all the
    positions is a $0$-dominion in this case.
    Each cell of the table contains a computed region.
    A downward arrow denotes a region that is open in the subgame where it is
    computed, while an upward arrow means that the region gets to be promoted to
    the priority in the subscript.
    The index of each row corresponds to the measure of the region.
    Following the idea sketched above, the first region obtained is the
    single-position $0$-region $\{ \aSym \}$ of measure 6, which is open because
    of the two moves leading to $\dSym$ and $\eSym$.
    The open $1$-region $\{ \bSym, \fSym, \hSym \}$ of measure 5 is, then,
    formed by attracting both $\fSym$ and $\hSym$ to $\bSym$, which is open in
    the subgame where $\{ \aSym \}$ is removed.
    Similarly, the $0$-region $\{ \cSym, \jSym \}$ of measure $4$ and the
    $1$-region $\{ \dSym \}$ of measure $3$ are open, once removed $\{ \aSym,
    \bSym, \fSym, \hSym \}$ and $\{ \aSym, \bSym, \cSym, \fSym, \hSym \}$,
    respectively, from the game.
    At priority $2$, the $0$-region $\{ \eSym \}$ is closed in the corresponding
    subgame.
    However, it is not closed in the whole game, because of the move leading to
    $\cSym$, \ie, to the region of measure $4$.
    Proposition~\ref{prp:regmer} can now be applied and a promotion of $\{ \eSym
    \}$ to $4$ is performed, resulting in the new $0$-region $\{ \cSym,
    \eSym, \jSym \}$ that resets $1$-region $\{ \dSym \}$.
    The search resumes at the corresponding priority and, after computing the
    extension of such a region via the attractor, we obtain that it is still
    open in the corresponding subgame.
    Consequently, the $1$-region $\{ \dSym \}$ of measure $3$ is recomputed
    and, then, priority $1$ is processed to build the $1$-region $\{ \gSym \}$.
    The latter is closed in the associated subgame, but not in the original
    game, because of a move leading to position $\dSym$.
    Hence, another promotion is performed, leading to the closed region of
    measure 3 at Column~3, which in turn triggers a promotion to $5$.
    When the promotion of $0$-region $\{ \iSym \}$ to priority $6$ is performed,
    however, $0$-region $\{ \cSym, \eSym, \jSym \}$ of measure $4$ is not reset.
    This leads directly to the configuration in Column~6, after the maximisation
    of $0$-region $6$, which attracts $\bSym$, $\dSym$, $\gSym$, and $\jSym$.
    Notice that, as prescribed by Proposition~\ref{prp:regspl}, the set $\{
    \cSym, \eSym \}$, highlighted by the grey area, is still a $0$-region.
    On the other hand, the set $\{ \fSym, \hSym \}$, highlighted by the dashed
    line and originally included in $1$-region $\{ \bSym, \dSym, \fSym, \gSym,
    \hSym \}$ of priority $5$, needs to be reset, since it is not a $1$-region
    any more.
    It is, actually, an open $0$-region instead.
    Now, $0$-region $4$ is closed in its subgame and it is promoted to $6$.
    As result of this promotion, we obtain the closed $0$-region
    $\{ \aSym, \bSym, \cSym, \dSym, \eSym, \gSym, \iSym, \jSym \}$, which is a
    dominion for player $0$.

    \vspace{-0.5em}
    \noindent
    \begin{minipage}[t]{0.50\textwidth}
      \null
      \vspace{0.25em}
      \figexm
    \end{minipage}
    \begin{minipage}[t]{0.50\textwidth}
      \null
      \tabexmprtprm
    \end{minipage}
    \vspace{-0.1em}

    We can now provide the formal account of the \PPP dominion space.
    We shall denote with $\RegSet$ the set of region pairs in $\GamName$ and
    with $\RegSet[][-]$ and $\RegSet[][+]$ the sets of open and closed region
    pairs, respectively.

    Similarly to the priority promotion algorithm, during the search for a
    dominion, the computed regions, together with their current measure, are
    kept track of by means of an auxiliary priority function $\rFun \in \Delta
    \defeq \PosSet \to \PrtSet$, called \emph{region function}.
    Given a priority $\pElm \in \PrtSet$, we denote by $\rFun^{(\geq \pElm)}$
    (\resp, $\rFun^{(> \pElm)}$, $\rFun^{(< \pElm)}$, and $\rFun^{(\equiv_{2}
    \pElm)}$) the function obtained by restricting the domain of $\rFun$ to the
    positions with measure greater than or equal to $\pElm$ (\resp, greater
    than, lower than, and congruent modulo $2$ to $\pElm$).
    Formally, $\rFun^{(\sim \pElm)} \defeq \rFun \rst \set{ \posElm \in \PosSet
    }{ \rFun(\posElm) \sim \pElm }$, for $\sim \,\in \{ \geq, >, <, \equiv_{2}
    \}$.
    By $\GamName[\rFun][\leq \pElm] \defeq \GamName \setminus \dom{\rFun^{(>
    \pElm)}}$, we denote the largest subgame obtained by removing from
    $\GamName$ all the positions in the domain of $\rFun^{(> \pElm)}$.
    The \emph{maximisation} of a priority function $\rFun \in \Delta$ is the
    unique priority function $\mFun \in \Delta$ such that $\mFun[][-1](\qElm) =
    \atrFun[ {\GamName[\mFun][\leq \qElm]} ][\alpha][ {\rFun[][-1](\qElm) \cap
    \PosSet[ {\GamName[\mFun][\leq \qElm]} ]} ]$, for all priorities $\qElm \in
    \rng{\rFun}$ with $\alpha \defeq \qElm \bmod{2}$.
    In addition, we say that $\rFun$ is \emph{maximal} above $\pElm \in \PrtSet$
    iff $\rFun^{(> \pElm)} = \mFun^{(> \pElm)}$.

    As opposed to the \PP approach, where a promotion to $\pElm \equiv_{2}
    \alpha$ resets all the regions lower than $\pElm$, here we need to take into
    account the fact that the regions of the opponent $\dual{\alpha}$ are reset,
    while the ones of player $\alpha$ are retained.
    In particular, we need to ensure that, as the search proceeds from $\pElm$
    downward to any priority $\qElm < \pElm$, the maximisation of the regions
    contained at priorities higher than $\qElm$ can never make the region
    recorded in $\rFun$ at $\qElm$ invalid.
    To this end, we consider only priority functions $\rFun$ that satisfy the
    requirement that, at all priorities, they contain regions \wrt the subgames
    induced by their maximisations $\mFun$.
    Formally, $\rFun \in \SetR \subseteq \Delta$ is a \emph{region function}
    iff, for all priorities $\qElm \in \rng{\mFun}$ with $\alpha \defeq \qElm
    \bmod{2}$, it holds that $\rFun[][-1](\qElm) \cap \PosSet[
    {\GamName[\mFun][\leq \qElm]} ]$ is an $\alpha$-region in the subgame
    $\GamName[\mFun][\leq \qElm]$, where $\mFun$ is the maximisation of $\rFun$.

    The status of the search of a dominion is encoded by the notion of
    \emph{state} $\sElm$ of the dominion space, which contains the current
    region function $\rFun$ and the current priority $\pElm$ reached by the
    search in $\GamName$.
    Initially, $\rFun$ coincides with the priority function $\prtFun$ of the
    entire game $\GamName$, while $\pElm$ is set to the maximal priority
    $\prtFun(\GamName)$ available in the game.
    To each of such states $\sElm \defeq (\rFun, \pElm)$, we then associate the
    \emph{subgame at $\sElm$} defined as $\GamName[\sElm] \defeq
    \GamName[\rFun][\leq \pElm]$, representing the portion of the original game
    that still has to be processed.

    The following state space specifies the configurations in which the \PPP
    procedure can reside and the relative order that the successor function must
    satisfy.

    \begin{definition}[State Space for \PPP]
      \label{def:sttprtprm}
      A \emph{\PPP state space} is a tuple $\SName \defeq \tuplec{\SSet}
      {\topElm} {\ordRel}$, where:
      \begin{enumerate}
        \item\label{def:sttprtprm(stt)}
          $\SSet \subseteq \SetR \times \PrtSet$ is the set of all pairs $\sElm
          \defeq (\rFun, \pElm)$, called \emph{states}, composed of a region
          function $\rFun \in \SetR$ and a priority $\pElm \in \PrtSet$ such
          that \emph{(a)}~$\rFun$ is maximal above $\pElm$ and \emph{(b)}~$\pElm
          \in \rng{\rFun}$;
        \item\label{def:sttprtprm(top)}
          $\topElm \defeq (\prtFun, \prtFun[][][\GamName])$;
        \item\label{def:sttprtprm(ord)}
          two states $\sElm[1] \defeq (\rFun[1], \pElm[1]), \sElm[2] \defeq
          (\rFun[2], \pElm[2]) \in \SSet$ satisfy $\sElm[1] \ordRel \sElm[2]$
          iff either \emph{(a)}~$\rFun[1]^{(> \qElm)} = \rFun[2]^{(> \qElm)}$
          and $\rFun[2][-1](\qElm) \subset \rFun[1][-1](\qElm)$, for some
          priority $\qElm \in \rng{\rFun[1]}$ with $\qElm \geq \pElm[1]$, or
          \emph{(b)}~both $\rFun[1] = \rFun[2]$ and $\pElm[1] < \pElm[2]$ hold.
      \end{enumerate}
    \end{definition}

    Condition~\ref{def:sttprtprm(stt)} requires that every region
    $\rFun[][-1](\qElm)$ with measure $\qElm > \pElm$ be $\alpha$-maximal, where
    $\alpha = \qElm \bmod{2}$.
    This implies that $\rFun[][-1](\qElm) \subseteq \PosSet[
    {\GamName[\rFun][\leq \qElm]} ]$.
    Moreover, the current priority $\pElm$ of the state must be one of the
    measures recorded in $\rFun$.
    In addition, Condition~\ref{def:sttprtprm(top)} specifies the initial state,
    while Condition~\ref{def:sttprtprm(ord)} defines the ordering relation among
    states, which the successor operation has to comply with.
    It asserts that a state $\sElm[1]$ is strictly smaller than another state
    $\sElm[2]$ if either there is a region recorded in $\sElm[1]$ with some
    higher measure $\qElm$ that strictly contains the corresponding one in
    $\sElm[2]$ and all regions with measure grater than $\qElm$ are equal in
    the two states, or state $\sElm[1]$ is currently processing a lower priority
    than the one of $\sElm[2]$.

    A region pair $(\RSet, \alpha)$ is compatible with a state $\sElm \defeq
    (\rFun, \pElm)$ if it is an $\alpha$-region in the current subgame
    $\GamName[\sElm]$.
    Moreover, if such a region is $\alpha$-open in that game, it has to be
    $\alpha$-maximal and needs to necessarily contain the current region
    $\rFun[][-1](\pElm)$ of priority $\pElm$ in $\rFun$.

    \begin{definition}[Compatibility Relation]
      \label{def:comprtprm}
      An open quasi dominion pair $(\RSet, \allowbreak \alpha) \in \QDSet[][-]$
      is \emph{compatible} with a state $\sElm \defeq (\rFun, \pElm) \in \SSet$,
      in symbols $\sElm \,\comRel (\RSet, \alpha)$, iff \emph{(1)}~$(\RSet,
      \alpha) \in \RegSet[ {\GamName[\sElm]} ]$ and \emph{(2)}~if $\RSet$ is
      $\alpha$-open in $\GamName[\sElm]$ then $\RSet = \atrFun[
      {\GamName[\sElm]} ][\alpha][ {\rFun[][-1](\pElm)} ]$.
    \end{definition}

    \algqry

    Algorithm~\ref{alg:qry} provides the implementation for the query function
    compatible with the priority-promotion mechanism.
    Line~1 simply computes the parity $\alpha$ of the priority to process in the
    state $\sElm \defeq (\rFun, \pElm)$.
    Line~2, instead, computes the attractor \wrt player $\alpha$ in subgame
    $\GamName[\sElm]$ of the region contained in $\rFun$ at the current priority
    $\pElm$.
    The resulting set $\RSet$ is, according to Proposition~\ref{prp:regext}, an
    $\alpha$-maximal $\alpha$-region of $\GamName[\sElm]$ containing
    $\rFun[][-1](\pElm)$.

    The promotion operation is based on the notion of best escape priority
    mentioned above, namely the priority of the lowest $\alpha$-region in
    $\rFun$ that has an incoming move coming from the $\alpha$-region, closed in
    the current subgame, that needs to be promoted.
    This concept is formally defined as follows.
    Let $\IRel \defeq \MovRel \cap ((\RSet \cap \PosSet[][\dual{\alpha}]) \times
    (\dom{\rFun} \!\setminus\! \RSet))$ be the \emph{interface relation} between
    $\RSet$ and $\rFun$, \ie, the set of $\dual{\alpha}$-moves exiting from
    $\RSet$ and reaching some position within a region recorded in $\rFun$.
    Then, $\bepFun[][\dual{\alpha}][\RSet, \rFun]$ is set to the minimal
    measure of those regions that contain positions reachable by a move in
    $\ISet$.
    Formally, $\bepFun[][\dual{\alpha}][\RSet, \rFun] \defeq \min(\rng{\rFun
    \rst \rng{\IRel}})$.
    Such a value represents the best priority associated with an $\alpha$-region
    contained in $\rFun$ and reachable by $\dual{\alpha}$ when escaping from
    $\RFun$.
    Note that, if $\RSet$ is a closed $\alpha$-region in $\GamName[\sttElm]$,
    then $\bepFun[][\dual{\alpha}][\RSet, \rFun]$ is necessarily of parity
    $\alpha$ and greater than the measure $\pElm$ of $\RSet$.
    This property immediately follows from the maximality of $\rFun$ above
    $\pElm$.
    Indeed, no move of an $\dual{\alpha}$-position can lead to a
    $\dual{\alpha}$-maximal $\dual{\alpha}$-region.
    For instance, for $0$-region $\RSet = \{ \eSym \}$ with measure $2$ in
    Column~1 of Figure~\ref{fig:exm}, we have that $\IRel = \{ (\eSym, \aSym),
    (\eSym, \cSym)\}$ and $\rFun \rst \rng{\IRel} = \{ (\aSym, 6), (\cSym, 4)
    \}$.
    Hence, $\bepFun[][1][\RSet, \rFun] = 4$.

    \algsucprtprm

    Algorithm~\ref{alg:sucprtprm} reports the pseudo-code of the successor
    function, which differs from the one proposed in~\cite{BDM16} only in
    Line~5, where Proposition~\ref{prp:regspl} is applied.
    Given the current state $\sElm$ and a compatible region pair $(\RSet,
    \alpha)$ open in the whole game as inputs, it produces a successor state
    $\sElm[][\star] \defeq (\rFun[][\star], \pElm[][\star])$ in the dominion
    space.
    It first checks whether $\RSet$ is open also in the subgame
    $\GamName[\sElm]$ (Line~1).
    If this is the case, it assigns the measure $\pElm$ to region $\RSet$ and
    stores it in the new region function $\rFun[][\star]$ (Line~2).
    The new current priority $\pElm[][\star]$ is, then, computed as the highest
    priority lower than $\pElm$ in $\rFun[][\star]$ (Line~3).
    If, on the other hand, $\RSet$ is closed in $\GamName[\sElm]$, a promotion,
    merging $\RSet$ with some other $\alpha$-region contained in $\rFun$, is
    required.
    The next priority $\pElm[][\star]$ is set to the $\bepFun$ of $\RSet$ for
    player $\dual{\alpha}$ in the entire game $\GamName$ \wrt $\rFun$ (Line~4).
    Region $\RSet$ is, then, promoted to priority $\pElm[][\star]$ and all and
    only the regions of the opponent $\dual{\alpha}$ with lower measure than
    $\pElm[][\star]$ in the region function $\rFun$ are reset by means
    of the completion operator defined in Section~\ref{sec:prl} (Line~5).

    The following theorem asserts that the \PPP state space, together with the
    same query function of \PP and the successor function of
    Algorithm~\ref{alg:sucprtprm} is a dominion space.

    \begin{theorem}[\PPP Dominion Space]
      \label{thm:prtprmdomspc}
      For a game $\GamName$, the \emph{\PPP structure} $\DName \defeq
      \tuplee{\GamName} {\SName}{\comRel} {\qryFun} {\sucFun}$, where $\SName$
      is given in Definition~\ref{def:sttprtprm}, $\comRel$ is the relation of
      Definition~\ref{def:comprtprm}, and $\qryFun$ and $\sucFun$ are the
      functions computed by Algorithms~\ref{alg:qry} and~\ref{alg:sucprtprm} is
      a dominion space.
    \end{theorem}

    The \PPP procedure does reduce, \wrt \PP, the number of reset needed to
    solve a game and the exponential worst-case game presented in~\cite{BDM16}
    does not work any more.
    However, a worst-case, which is a slight modification of the one for \PP,
    does exist for this procedure as well.
    Consider the game $\GamName[h]$ containing $h$ chains of length $2$ that
    converge into a single position of priority $0$ with a self loop.
    The $i$-th chain has a head of priority $2(h + 1) - i$ and a body composed
    of a single position with priority $i$ and a self loop.
    An instance of this game with $h = 4$ is depicted in
    Figure~\ref{fig:prtprmlowbnd}.
    The labels of the positions correspond to the associated priorities.
    Intuitively, the execution depth of the \PPP dominion space for this game is
    exponential, since the consecutive promotion operations performed on each
    chain can simulate the increments of a partial form of binary counter, some
    of whose configurations are missing.
    As a result, the number of configurations of the counter follows a
    Fibonacci-like sequence of the form $F(h) = F(h-1) + F(h-2) + 1$, with
    $F(0)=1$ and $F(1)=2$.

    \figprtprmlowbnd

    \noindent
    The search procedure on $\GamName[4]$ starts by building the following four
    open regions: the $1$-region $\{ 9 \}$, the $0$-region $\{ 8 \}$, the
    $1$-region $\{ 7 \}$, and $0$-region $\{ 6 \}$.
    This state represents the configuration of the counter, where all four
    digits are set to $0$.
    The closed $0$-region $\{ 4 \}$ is then found and promoted to $6$.
    Now, the counter is set to $0001$.
    After that, the closed $1$-region $\{ 3 \}$ is computed that is promoted to
    $7$.
    Due to the promotion to $7$, the positions in the $0$-region with priority
    $6$ are reset to their original priority, as they belong to the opponent
    player.
    This releases the chain with head $6$, which corresponds to the reset of the
    least significant digit of the counter caused by the increment of the second
    one, \ie, the counter displays $0010$.
    The search resumes at priority $6$ and the $0$-regions $\{ 6 \}$ and $\{ 4
    \}$ are computed once again.
    A second promotion of $\{ 4 \}$ to $6$ is performed, resulting in the
    counter assuming value $0011$.
    When the closed $0$-region $\{ 2 \}$ is promoted to $8$, however, only the
    $1-$region $\{ 7, 3 \}$ is reset, leading to configuration $0101$ of the
    counter.
    Hence, configuration $0100$ is skipped.
    Similarly, when, the counter reaches configuration $0111$ and $1$-region
    $\{ 1 \}$ is promoted to $9$, the $0$-regions $\{ 8, 2 \}$ and $\{ 6, 4 \}$
    are reset, leaving $1$-region $\{ 7, 3 \}$ intact.
    This leads directly to configuration $1010$ of the counter, skipping
    configuration $1000$.

    An estimate of the depth of the \PPP dominion space on the game
    $\GamName[h]$ is given by the following theorem.

    \begin{theorem}[Execution-Depth Lower Bound]
      \label{thm:prtprmlowbnd}
      For all $h \in \SetN$, there exists a \PPP dominion space
      $\DName[h][\PPP]$ with $k = 2h + 1$ positions and priorities, whose
      execution depth is $\mthfun{Fib}(2(h + 4)) / \mthfun{Fib}(h + 4) - (h +
      6) = \ATheta{((1 + \sqrt{5}) / 2)^{k / 2}}$.
    \end{theorem}

  \end{paragraph}

\end{section}





\begin{section}{Delayed Promotion Policy}
  \label{sec:delprmpol}

  At the beginning of the previous section, we have observed that the time
  complexity of the dominion search procedure $\srcFun[\DName]$ linearly depends
  on the execution depth of the underlying dominion space $\DName$.
  This, in turn, depends on the number of promotions performed by the associated
  successor function and is tightly connected with the reset mechanism applied
  to the regions with measure lower than the one of the target region of the
  promotion.
  In fact, it can be proved that, when no resets are performed, the number of
  possible promotions is bounded by a polynomial in the number of priorities and
  positions of the game under analysis.
  Consequently, the exponential behaviours exhibited by the \PP algorithm and
  its enhanced version \PPP are strictly connected with the particular reset
  mechanism employed to ensure the soundness of the promotion approach.
  The correctness of the \PPP method shows that the reset can be restricted to
  only the regions of opposite parity \wrt the one of the promotion and, as we
  shall show in the next section, this enhancement is also relatively effective
  in practice.
  However, we have already noticed that this improvement does not suffices in
  avoiding some pathological cases and we do not have any finer criteria to
  avoid the reset of the opponent regions.
  Therefore, to further reduce such resets, in this section we propose a finer
  promotion policy that tries in advance to minimise the necessity of
  application of the reset mechanism.
  The new solution procedure is based on delaying the promotions of regions,
  called \emph{locked promotions}, that require the reset of previously
  performed promotions of the opponent parity, until a complete knowledge of the
  current search phase is reached.
  Once only locked promotions are left, the search phase terminates by choosing
  the highest measure $\pElm[][\star]$ among those associated with the locked
  promotions and performing all the postponed ones of the same parity as
  $\pElm[][\star]$ altogether.
  In order to distinguish between locked and unlocked promotions, the
  corresponding target priorities of the performed ones, called \emph{instant
  promotions}, are recorded in a supplementary set $\PSet$.
  Moreover, to keep track of the locked promotions, a supplementary partial
  priority function $\adj{\rFun}$ is used.
  In more detail, the new procedure evolves exactly as the \PPP algorithm, as
  long as open regions are discovered.
  When a closed one with measure $\pElm$ is provided by the query function,
  two cases may arise.
  If the corresponding promotion is not locked, the destination priority $\qElm$
  is recorded in the set $\PSet$ and the instant promotion is performed similarly
  to the case of \PPP.
  Otherwise, the promotion is not performed.
  Instead, it is recorded in the supplementary function $\adj{\rFun}$, by
  assigning to $\RSet$ in $\adj{\rFun}$ the target priority $\qElm$ of that
  promotion and in $\rFun$ its current measure $\pElm$.
  Then, the positions in $\RSet$ are removed from the subgame and the search
  proceeds at the highest remaining priority, as in the case $\RSet$ was open in
  the subgame.
  In case the region $\RSet$ covers the entire subgame, all priorities available
  in the original game have been processed and, therefore, there is no further
  subgame to analyse.
  At this point, the delayed promotion to the highest priority $\pElm[][\star]$
  recorded in $\adj{\rFun}$ is selected and all promotions of the same parity
  are applied at once.
  This is done by first moving all regions from $\adj{\rFun}$ into $\rFun$ and
  then removing from the resulting function the regions of opposite parity \wrt
  $\pElm[][\star]$, exactly as done by \PPP.
  The search, then, resumes at priority $\pElm[][\star]$.
  Intuitively, a promotion is considered as locked if its target priority is
  either \emph{(a)} greater than some priority in $\PSet$ of opposite parity,
  which would be otherwise reset, or \emph{(b)} lower than the target of some
  previously delayed promotion recorded in $\adj{\rFun}$, but greater than the
  corresponding priority set in $\rFun$.
  The latter condition is required to ensure that the union of a region in
  $\rFun$ together with the corresponding recorded region in $\adj{\rFun}$ is
  still a region.
  Observe that the whole approach is crucially based on the fact that when a
  promotion is performed all the regions having lower measure but same parity
  are preserved.
  If this was not the case, we would have no criteria to determine which
  promotions need to be locked and which, instead, can be freely performed.

  \tabexmdelprm

  This idea is summarized by Table~\ref{tab:exmdelprm}, which
  contains the execution of the new algorithm on the example in
  Figure~\ref{fig:exm}.
  The computation proceeds as for \PPP, until the promotion of the $1$-region
  $\{ \gSym \}$ shown in Column~2, occurs.
  This is an instant promotion to $3$, since the only other promotion already
  computed and recorded in $\PSet$ has value $4$.
  Hence, it can be performed and saved in $\PSet$ as well.
  Starting from priority $3$ in Column~3, the closed $1$-region $\{ \dSym,
  \gSym \}$ could be promoted to $5$.
  However, since its target is greater than $4 \in \PSet$, it is delayed and
  recorded in $\adj{\rFun}$, where it is assigned priority $5$.
  At priority $0$, a delayed promotion of $0$-region $\{ \iSym \}$ to priority
  $6$ is encountered and registered, since it would overtake priority $3 \in
  \PSet$.
  Now, the resulting subgame is empty.
  Since the highest delayed promotion is the one to priority $6$ and no other
  promotion of the same parity was delayed, $0$-region $\{ \iSym \}$ is promoted
  and both the auxiliary priority function $\adj{\rFun}$ and the set of
  performed promotions $\PSet$ are emptied.
  Previously computed $0$-region $\{ \cSym, \eSym, \jSym \}$ has the same parity
  and, therefore, it is not reset, while the positions in both $1$-regions $\{
  \bSym, \fSym, \hSym \}$ and $\{ \dSym, \gSym \}$ are reset to their original
  priorities.
  After maximisation of the newly created $0$-region $\{ \aSym, \iSym \}$,
  positions $\bSym$, $\dSym$, $\gSym$, and $\jSym$ get attracted as well.
  This leads to the first cell of Column~4, where $0$-region $\{ \aSym, \bSym,
  \dSym, \gSym, \iSym, \jSym \}$ is open.
  The next priority to process is then $4$, where $0$-region $\{ \cSym, \eSym
  \}$, the previous $0$-region $\{ \cSym, \eSym, \jSym \}$ purged of position
  $\jSym$, is now closed in the corresponding subgame and gets promoted to $6$.
  This results in a $0$-region closed in the entire game, hence, a dominion
  for player $0$ has been found.
  Note that postponing the promotion of $1$-region $\{ \dSym, \gSym \}$ allowed
  a reduction in the number of operations.
  Indeed, the redundant maximisation of $1$-region $\{ \bSym, \fSym, \hSym \}$
  is avoided.

  It is worth noting that this procedure only requires a linear number of
  promotions, precisely $\floor{\frac{h + 1}{2}}$, on the lower bound game
  $\GamName[h][\PPP]$ for \PPP.
  This is due to the fact that all resets are performed on regions that are not
  destination of any promotion.
  Also, the procedure appears to be much more robust, in terms of preventing
  resets, than the \PPP technique alone, to the point that it does not seem
  obvious whether an exponential lower bound even exists.
  Further investigation is, however, needed for a definite answer on its actual
  time complexity.

  \begin{paragraph}{\DP Dominion Space.}

    As it should be clear from the above informal description, the delayed
    promotion mechanism is essentially a refinement of the one employed in \PPP.
    Indeed, the two approaches share all the requirements on the corresponding
    components of a state, on the orderings, and on compatibility relations.
    However, \DP introduces in the state two supplementary elements: a partial
    priority function $\adj{\rFun}$, which collects the delayed promotions that
    were not performed on the region function $\rFun$, and a set of priorities
    $\PSet$, which collects the targets of the instant promotions performed.
    Hence, in order to formally define the corresponding dominion space, we need
    to provide suitable constraints connecting them with the other components of
    the search space.
    The role of function $\adj{\rFun}$ is to record the delayed promotions
    obtained by moving the corresponding positions from their priority $\pElm$
    in $\rFun$ to the new measure $\qElm$ in $\adj{\rFun}$.
    Therefore, as dictated by Proposition~\ref{prp:regmer}, the union of
    $\rFun^{-1}(\qElm)$ and $\adj{\rFun}^{-1}(\qElm)$ must always be a region in
    the subgame $\GamName[\rFun][\leq \qElm]$.
    In addition, $\adj{\rFun}^{-1}(q)$ can only contain positions whose measure
    in $\rFun$ is of the same parity as $\qElm$ and recorded in $\rFun$ at some
    lower priority greater than the current one $\pElm$.
    Formally, we say that $\adj{\rFun} \in \Delta^{\pto} \defeq \PosSet \pto
    \PrtSet$ is \emph{aligned} with a region function $\rFun$ \wrt $\pElm$ if,
    for all priorities $\qElm \in \rng{\adj{\rFun}}$, it holds that
    \emph{(a)}~$\adj{\rFun}[][-1](\qElm) \subseteq \dom{\rFun^{(> \pElm) \land
    (< \qElm) \land (\equiv_{2} \qElm)}}$ and \emph{(b)}~$\rFun^{-1}(\qElm) \cup
    \adj{\rFun}^{-1}(\qElm)$ is an $\alpha$-region in $\GamName[\rFun][\leq
    \qElm]$ with $\alpha \equiv_{2} \qElm$.
    The state space for \DP is, therefore, defined as follows.

    \begin{definition}[State Space for \DP]
      \label{def:sttdelprm}
      A \emph{\DP state space} is a tuple $\SName \defeq \tuplec{\SSet}
      {\topElm} {\ordRel}$, where:
      \begin{enumerate}
        \item\label{def:sttdelprm(stt)}
          $\SSet \subseteq \SSet[][\PPP] \times \Delta^{\pto} \times
          \pow{\PrtSet}$ is the set of all triples $\sElm \defeq ((\rFun,
          \pElm), \adj{\rFun}, \PSet)$, called \emph{states}, composed by a \PPP
          state $(\rFun, \pElm) \in \SSet[][\PPP]$, a partial priority function
          $\adj{\rFun} \in \Delta^{\pto}$ aligned with $\rFun$ \wrt $\pElm$, and
          a set of priorities $\PSet \subseteq \PrtSet$.
        \item\label{def:sttdelprm(top)}
          $\topElm \defeq (\topElm[][\PPP], \emptyfun, \emptyset)$;
        \item\label{def:sttdelprm(ord)}
          $\sElm[1] \ordRel \sElm[2]$ iff $\der{\sElm}[1] \ordRel[][\PPP]
          \der{\sElm}[2]$, for any two states $\sElm[1] \defeq (\der{\sElm}[1],
          \_, \_), \sElm[2] \defeq (\der{\sElm}[2], \_, \_) \in \SSet$.
      \end{enumerate}
    \end{definition}

    The second property we need to enforce is expressed in the compatibility
    relation connecting the query and successor functions for \DP and regards
    the closed region pairs that are locked \wrt the current state.
    As stated above, a promotion is considered locked if its target priority is
    either \emph{(a)} greater than some priority in $\PSet$ of opposite parity
    or \emph{(b)} lower than the target of some previously delayed promotion
    recorded in $\adj{\rFun}$, but greater than the corresponding priority set
    in $\rFun$.
    Condition~\emph{(a)} is the one characterising the delayed promotion
    approach, as it reduces the number of resets of previously promoted regions.
    The two conditions are expressed by the following two formulas,
    respectively, where $\qElm$ is the target priority of the blocked promotion.
    \vspace{-0.5em}
    \[
      \phi_{a}(\qElm, \PSet) \defeq \exists \lElm \in \PSet \,.\, \lElm
      \not\equiv_{2} \qElm \land \lElm < \qElm
      \hspace{2em}
      \phi_{b}(\qElm, \rFun, \adj{\rFun}) \defeq \exists \posElm \in
      \dom{\adj{\rFun}} \,.\, \rFun(\posElm) < \qElm \leq  \adj{\rFun}(\posElm)
      \vspace{-0.5em}
    \]
    Hence, an $\alpha$-region $\RSet$ is called \emph{$\alpha$-locked} \wrt a
    state $\sElm \defeq ((\rFun, \pElm), \adj{\rFun}, \PSet)$ if the predicate
    $\phi_{Lck}(\qElm, \sElm) \defeq \phi_{a}(\qElm, \PSet) \vee \phi_{b}(\qElm,
    \rFun, \adj{\rFun})$ is satisfied, where $\qElm =
    \bepFun[][\dual{\alpha}][\RSet, \rFun \uplus \adj{\rFun}]$.
    In addition to the compatibility constraints for \PPP, the compatibility
    relation for \DP requires that any $\alpha$-locked region, possibly returned
    by the query function, be maximal and contain the region
    $\rFun[][-1](\pElm)$ associated to the priority $\pElm$ of the current
    state.

    \begin{definition}[Compatibility Relation for \DP]
      \label{def:comdelprm}
      An open quasi dominion pair $(\RSet, \alpha) \in \QDSet[][-]$ is
      \emph{compatible} with a state $\sElm \defeq ((\rFun, \pElm), \_, \_) \in
      \SSet$, in symbols $\sElm \,\comRel (\RSet, \alpha)$, iff
      \emph{(1)}~$(\RSet, \alpha) \in \RegSet[ {\GamName[\sElm]} ]$ and
      \emph{(2)}~if $\RSet$ is $\alpha$-open in $\GamName[\sElm]$ or it is
      $\alpha$-locked \wrt $\sElm$ then $\RSet = \atrFun[ {\GamName[\sElm]}
      ][\alpha][ {\rFun[][-1](\pElm)} ]$.
    \end{definition}

    Algorithm~\ref{alg:sucdelprm} implements the successor function for \DP.
    The pseudo-code on the right-hand side consists of three macros, namely
    $\mthelm{\#Assignment}$, $\mthelm{\#InstantPromotion}$, and
    $\mthelm{\#DelayedPromotion}$, used by the algorithm.
    Macro $\mthelm{\#Assignment}(\xi)$ performs the insertion of a new region
    $\RSet$ into the region function $\rFun$.
    In presence of a blocked promotion, \ie, when the parameter $\xi$ is set to
    $\Ff$, the region is also recorded in $\adj{\rFun}$ at the target priority
    of the promotion.
    Macro $\mthelm{\#InstantPromotion}$ corresponds to the \DP version of the
    standard promotion operation of \PPP.
    The only difference is that it must also take care of updating the
    supplementary elements $\adj{\rFun}$ and $\PSet$.
    Macro $\mthelm{\#DelayedPromotion}$, instead, is responsible for the delayed
    promotion operation specific to \DP.\vspace{-.5em}

    \noindent
    \begin{minipage}[t]{0.550\textwidth}
      \null
      \algsucdelprm
    \end{minipage}
    \begin{minipage}[t]{0.440\textwidth}
      \null
      \algmacdelprm
    \end{minipage}
    \vspace{.5em}

    If the current region $\RSet$ is open in the subgame $\GamName[\sttElm]$,
    the main algorithm proceeds, similarly to Algorithm~\ref{alg:sucprtprm}, at
    assigning to it the current priority $\pElm$ in $\rFun$.
    This is done by calling macro $\mthelm{\#Assignment}$ with parameter $\Tt$.
    Otherwise, the region is closed and a promotion should be performed at
    priority $\qElm$, corresponding to the $\bepFun$ of that region \wrt the
    composed region function $\rFun \uplus \adj{\rFun}$.
    In this case, the algorithm first checks whether such promotion is locked
    \wrt $\sttElm$ at Line~7.
    If this is not the case, then the promotion is performed as in \PPP, by
    executing $\mthelm{\#InstantPromotion}$, and the target is kept track of in
    the set $\PSet$.
    If, instead, the promotion to $\qElm$ is locked, but some portion of the
    game still has to be processed, the region is assigned its measure $\pElm$
    in $\rFun$ and the promotion to $\qElm$ is delayed and stored in
    $\adj{\rFun}$.
    This is done by executing $\mthelm{\#Assignment}$ with parameter $\Ff$.
    Finally, in case the entire game has been processed, the delayed promotion
    to the highest priority recorded in $\adj{\rFun}$ is selected and applied.
    Macro $\mthelm{\#DelayedPromotion}$ is executed, thus merging $\rFun$ with
    $\adj{\rFun}$.
    Function $\adj{\rFun}$ and set $\PSet$ are, then, erased, in order to begin
    a new round of the search.
    Observe that, when a promotion is performed, whether instant or delayed, we
    always preserve the underlying regions of the same parity, as done by the
    \PPP algorithm.
    This is a crucial step in order to avoid the pathological exponential worst
    case for the original \PP procedure.

    The soundness of the solution procedure relies on the following theorem.

    \begin{theorem}[\DP Dominion Space]
      \label{thm:delprmdomspc}
      For a game $\GamName$, the \emph{\DP structure} $\DName \defeq
      \tuplee{\GamName} {\SName}{\comRel} {\qryFun} {\sucFun}$, where $\SName$
      is given in Definition~\ref{def:sttdelprm}, $\comRel$ is the relation of
      Definition~\ref{def:comdelprm}, and $\qryFun$ and $\sucFun$ are the
      functions computed by Algorithms~\ref{alg:qry} and~\ref{alg:sucdelprm},
      where in the former the assumption "\emph{\textbf{let}} $(\rFun, \pElm) =
      \sElm$" is replaced by "\emph{\textbf{let}} $((\rFun, \pElm), \_, \_) =
      \sElm$" is a dominion space.
    \end{theorem}

    It is immediate to observe that the following mapping $\hFun : ((\rFun,
    \pElm), \_, \_) \in \SSet[][\DP] \mapsto (\rFun, \pElm) \in \SSet[][\PPP]$,
    which takes \DP states to \PPP states by simply forgetting the additional
    elements $\adj{\rFun}$ and $\PSet$, is a homomorphism.
    This, together with a trivial calculation of the number of possible states,
    leads to the following theorem.

    \begin{theorem}[\DP Size \& Depth Upper Bounds]
      \label{thm:delprmuppbnd}
      The size of the \DP dominion space $\DName[\GamName]$ for a game $\GamName
      \in \PG$ with $n \in \SetN[+]$ positions and $k \in \numcc{1}{n}$
      priorities is bounded by $2^{k}k^{2n}$.
      Moreover, its depth is not greater then the one of the \PPP dominion space
      $\DName[\GamName][\PPP]$ for the same game.
    \end{theorem}

  \end{paragraph}

\end{section}





\vspace{-0.75em}
\begin{section}{Experimental Evaluation}
  \label{sec:expevl}

  The technique proposed in the paper has been implemented in the tool
  \textsc{PGSolver}~\cite{FL09}, which collects implementations of several
  parity game solvers proposed in the literature and provides benchmarking tools
  that can be used to evaluate the solver performances.\footnote{All the
  experiments were carried out on a 64-bit 3.1GHz \textsc{Intel\textregistered}
  quad-core machine, with i5-2400 processor and 8GB of RAM, running
  \textsc{Ubuntu}~12.04 with \textsc{Linux} kernel version~3.2.0.
  \textsc{PGSolver} was compiled with OCaml version 2.12.1.}

  Figure~\ref{fig:prf} compares the running times of the new algorithms, \PPP
  and \DP, against the original version \PP\footnote{The version of \PP used in
  the experiments is actually an improved implementation of the one described
  in~\cite{BDM16}.} and the well-known solvers \emph{Rec} and \emph{Str},
  implementing the recursive algorithm~\cite{Zie98} and the strategy improvement
  technique~\cite{VJ00}, respectively.
  This first pool of benchmarks is taken from~\cite{BDM16} and involves $2000$
  random games of size ranging from $1000$ to $20000$ positions and $2$ outgoing
  moves per position.
  Interestingly, random games with very few moves prove to be much more
  challenging for the priority promotion based approaches than those with a
  higher number of moves per position, and often require a much higher number of
  promotions.
  Since the behaviour of the solvers is typically highly variable, even on games
  of the same size and priorities, to summarise the results we took the average
  running time on clusters of games.
  Therefore, each point in the graph shows the average time over a cluster of
  $100$ different games of the same size: for each size value $n$, we chose the
  numbers $k = n \cdot i / 10$ of priorities, with $i \in \numcc{1}{10}$, and
  $10$ random games were generated for each pair $n$ and $k$.
  We set a time-out to 180 seconds (3 minutes).
  Solver \PPP performs slightly better than \PP, while \DP shows a much more
  convincing improvement on the average time.
  \figprf
  All the other solvers provided in \textsc{PGSolver}, including the Dominion
  Decomposition~\cite{JPZ08} and the Big Step~\cite{Sch07} algorithms, perform
  quite poorly on those games, hitting the time-out already for very small
  instances.
  Figure~\ref{fig:prf} shows only the best performing ones on the considered
  games, namely \emph{Rec} and \emph{Str}.
  Similar experiments were also conducted on random games with a higher number
  of moves per position and up to $100000$ positions.
  The resulting games turn out to be very easy to solve by all the priority
  promotion based approaches.
  The reason seems to be that the higher number of moves significantly increases
  the dimension of the computed regions and, consequently, also the chances to
  find a closed one.
  Indeed, the number of promotions required by \PPP and \DP on all those games
  is typically zero, and the whole solution time is due exclusively to a very
  limited number of attractors needed to compute the few regions contained in
  the games.
  We reserve the presentation of the results for the extended version.

  To further stress the \DP technique in comparison with \PP and \PPP, we also
  generated a second pool of much harder benchmarks, containing more than 500
  games, each with 50000 positions, 12000 priorities and 2 moves per positions.
  We selected as benchmarks only random games whose solution requires \PPP
  between 30 and 6000 seconds.  The results comparing \PPP and \DP are reported
  in Figure~\ref{fig:sct} on a logarithmic scale.  The figure shows that, in few
  cases, \PPP actually performs better than \DP.  This is due to the fact that
  the two algorithms follow different solution paths within the dominion space
  and that delaying promotions may also defer the discovery of a closed
  dominion.  Nonetheless, the \DP policy does pay off significantly on the vast
  majority of the benchmarks, often solving a game between two to eight times
  faster than \PPP, as witnessed by the points below the dash-dotted line
  labeled $2\times$ in Figure~\ref{fig:sct}.

  In~\cite{BDM16} it is shown that \PP solves all the known exponential worst
  cases for the other solvers without promotions and, clearly, the same holds of
  \DP as well.
  As a consequence, \DP only requires polynomial time on those games and the
  experimental results coincide with the ones for \PP.

\end{section}





\vspace{-0.75em}
\begin{section}{Discussion}

  Devising efficient algorithms that can solve parity games well in practice is
  a crucial endeavour towards enabling formal verification techniques, such as
  model checking of expressive temporal logics and automatic synthesis, in
  practical contexts.
  To this end, a promising new solution technique, called \emph{priority
  promotion}, was recently proposed in~\cite{BDM16}.
  While the technique seems very effective in practice, the approach still
  admits exponential behaviours.
  This is due to the fact that, to ensure correctness, it needs to forget
  previously computed partial results after each promotion.
  In this work we presented a new promotion policy that delays promotions as
  much as possible, in the attempt to reduce the need to partially reset the
  state of the search.
  Not only the new technique, like the original one, solves in polynomial time
  all the exponential worst cases known for other solvers, but requires
  polynomial time for the worst cases of the priority promotion approach as
  well.
  The actual complexity of the algorithm is, however, currently unknown.
  Experiments on randomly generated games also show that the new technique often
  outperforms the original priority promotion technique, as well as the
  state-of-the-art solvers proposed in the literature.

\end{section}



  \bibliographystyle{eptcs}
  \bibliography{References}

\end{document}